\documentclass[10pt,twocolumn,showpacs,preprintnumbers,amsmath,amssymb,aps,prb,longbibliography,superscriptaddress]{revtex4-2}
\usepackage[utf8]{inputenc}
\usepackage{color,array,verbatim,multirow,dsfont,graphicx,esint,titlesec,longtable,bbm,listings,braket,siunitx,mathtools}
\usepackage[shortlabels]{enumitem}
\usepackage[normalem]{ulem}

\usepackage[unicode=true,
 bookmarks=true,bookmarksnumbered=true,bookmarksopen=false,
 breaklinks=true,pdfborder={0 0 0},backref=false,colorlinks=true]{hyperref}
\hypersetup{citecolor={blue},urlcolor={magenta}}

\usepackage{cleveref}
\crefname{appendix}{App.}{Apps.}
\crefname{equation}{Eq.}{Eqs.}
\crefname{figure}{Fig.}{Figs.}
\crefname{table}{Tab.}{Tabs.}
\crefname{section}{Sec.}{Secs.}

\newcommand{\Z}{\mathbb{Z}}
\newcommand{\R}{\mathbb{R}}
\newcommand{\T}{\mathbb{T}}
\newcommand{\br}{\mathbf{r}}
\newcommand{\bk}{\mathbf{k}}
\newcommand{\ba}{\mathbf{a}}
\newcommand{\bb}{\mathbf{b}}
\newcommand{\btau}{\boldsymbol\tau}
\newcommand{\bK}{\mathbf{K}}
\newcommand{\bM}{\mathbf{M}}
\newcommand{\bd}{\mathbf{d}}
\newcommand{\bG}{\mathbf{G}}
\newcommand{\bJ}{\mathbf{J}}
\newcommand{\bQ}{\mathbf{Q}}
\newcommand{\bp}{\mathbf{p}}
\newcommand{\bq}{\mathbf{q}}
\newcommand{\bg}{\mathbf{g}}
\newcommand{\bsigma}{\boldsymbol\sigma}
\newcommand{\bGamma}{\boldsymbol\Gamma}
\newcommand{\bhatx}{\mathbf{\hat{x}}}
\newcommand{\bhaty}{\mathbf{\hat{y}}}
\newcommand{\bhatz}{\mathbf{\hat{z}}}
\newcommand{\bu}{\mathbf{u}}
\newcommand{\bv}{\mathbf{v}}
\newcommand{\bzero}{\mathbf{0}}
\newcommand{\bx}{\mathbf{x}}
\newcommand{\by}{\mathbf{y}}

\newcommand{\bX}{\mathbf{X}}

\begin{document}
\title{Magic angles in twisted bilayer graphene near commensuration: Towards a hypermagic regime}
\author{Michael G. Scheer}
\author{Kaiyuan Gu}
\author{Biao Lian}
\affiliation{Department of Physics, Princeton University, Princeton, New Jersey 08544, USA}
\date{\today}

\begin{abstract}
The Bistritzer-MacDonald continuum model (BM model) describes the low-energy moir\'e bands for twisted bilayer graphene (TBG) at small twist angles. We derive a generalized continuum model for TBG near any commensurate twist angle, which is characterized by complex interlayer hoppings at commensurate $AA$ stackings (rather than the real hoppings in the BM model), a real interlayer hopping at commensurate $AB/BA$ stackings, and a global energy shift. The complex phases of the $AA$ stacking hoppings and the twist angle together define a single angle parameter $\phi_0$. We compute the model parameters for the first six distinct commensurate TBG configurations, among which the $38.2^\circ$ configuration may be within experimentally observable energy scales. We identify the first magic angle for any $\phi_0$ at a condition similar to that of the BM model. At this angle, the lowest two moir\'e bands at charge neutrality become flat except near the $\bGamma_M$ point and retain fragile topology but lose particle-hole symmetry. We further identify a hypermagic parameter regime centered at $\phi_0 = \pm\pi/2$ where many moir\'e bands around charge neutrality (often $8$ or more) become flat simultaneously. Many of these flat bands resemble those in the kagome lattice and $p_x$, $p_y$ 2-orbital honeycomb lattice tight-binding models.
\end{abstract}

\maketitle

\section{Introduction}
At certain discrete commensurate twist angles $\theta_0$, the honeycomb lattices of two graphene layers align to form a perfectly periodic superlattice \cite{Shallcross2008,Shallcross2010}. The simplest such commensurate configuration is $\theta_0 = 0$ in which two layers of graphene are aligned with no twist. Bistritzer and MacDonald demonstrated that if two layers of graphene are twisted by a small angle relative to this $\theta_0 = 0$ configuration, forming twisted bilayer graphene (TBG), a moir\'e superlattice emerges, and the low energy single particle physics can be described by a continuum model \cite{Bistritzer2011}. Furthermore, at the so called magic angle, $\theta\approx 1.05^\circ$, this model predicts that the lowest two moir\'e bands (i.e. the first conduction and valence bands) at charge neutrality become approximately flat. Moreover, it has been shown that the two flat bands carry a fragile topology \cite{po_origin_2018,song_all_2019,po_faithful_2019,ahn_failure_2019,lian2020}, obstructing the construction of maximally localized symmetric Wannier orbitals \cite{kang_symmetry_2018,koshino_maximally_2018,liu2019pseudo,Tarnopolsky2019,song2022,Wang2021}. In this flat band regime, the physics is dominated by interactions. Interacting electronic states such as correlated insulators, superconductors, and Chern insulators have been observed \cite{Kim2017,Cao2018,Cao2018a,Lu2019,Yankowitz2019,Xie2019,Choi2019,Jiang2019,Kerelsky2019,Sharpe2019,Cao2020,Serlin2020,wong_cascade_2020,nuckolls_chern_2020,choi2021correlation,saito2020,das2020symmetry,wu_chern_2020,oh_unconventional_2021}, the mechanisms of which have been studied extensively \cite{ochi_possible_2018, guinea2018, venderbos2018, Lian2019TBG,Wu2018TBG-BCS, isobe2018unconventional, thomson2018triangular,  dodaro2018phases, gonzalez2019kohn, yuan2018model,kang_strong_2019,bultinck_ground_2020,seo_ferro_2019, Hejazi2021, khalaf_charged_2020,xie_superfluid_2020,julku_superfluid_2020, hu2019_superfluid, pixley2019, xie_HF_2020,zhang_HF_2020, fernandes_nematic_2020,Bernevig2021b,lian_tbg4_2021,bernevig_tbg5_2021,xie_tbg6_2021}. Flat bands and interacting electronic states are also present in other two dimensional moir\'e materials such as twisted double bilayer graphene \cite{Burg2019,Shen2020,Liu2020,Cao2020a}, twisted trilayer and multilayer graphene \cite{khalaf2019,mora2019,hao2021,park_trilayer_2021,zhang_ascendance_2021,Park2022}, ABC trilayer graphene \cite{Chen2019,Chen2019a,Chen2020,zhou2021}, and twisted transition metal dichalcogenides \cite{Wang2020,Regan2020,wangp2021}. For the purpose of exploring interacting states, the search for more flat band moir\'e platforms is important.

In this paper, we search for flat moir\'e bands in TBG twisted by a small angle relative to an arbitrary commensurate configuration. That is to say, we consider a twist angle $\theta = \theta_0 + \delta\theta$ where $\theta_0$ is a commensurate angle and $\delta\theta$ is small. Without loss of generality, we can choose $0 \leq \theta_0 < \pi/3$ because of the crystalline symmetries of TBG. When $\theta_0  = 0$, the interlayer hopping couples states near the top and bottom layer $\bK$ points only among themselves. This allows one to explicitly derive the form of the Bistritzer-MacDonald continuum model (BM model) by computing the interlayer hopping in reciprocal space and making a few well-justified approximations \cite{Bistritzer2011}. However, for all other commensurate configurations, the calculations are more complicated since states near the top and bottom $\bK$ points are coupled to many other states. The origin of this complication is the fact that the commensurate unit cell contains $4N$ atoms, where the integer $N$ is $1$ when $\theta_0 = 0$ but is $7$ or greater for all other commensurate configurations \cite{Shallcross2010}.

Assuming that the interlayer hopping is not too strong, the states far from the top and bottom $\bK$ points influence the low energy physics perturbatively. Rather than explicitly applying perturbation theory, we take an approach based on symmetry and parameter determination from a microscopic tight-binding model. We first show, based on an analysis of the magnitudes of the hopping terms, that the system is approximated by a continuum model of a certain general form. We then use the exact unitary and anti-unitary crystalline symmetries of TBG to constrain the coefficients of this general model. Near a commensurate twist angle $\theta_0$, we arrive at a TBG continuum model containing four real parameters $\chi_0$, $w_0$, $w_1$, and $w_2$, which are ultimately determined by the microscopic hopping parameters. We show that $w_1$ controls the interlayer hopping at the commensurate $AB$ and $BA$ stacking configurations while $w_0e^{i\chi_0}$ and $w_0e^{-i\chi_0}$ control the interlayer hoppings at the commensurate $AA$ stacking configuration. $w_2$ is simply a global energy shift. When $\theta_0 = 0$, the value of $\chi_0$ is negligible because of an approximate mirror symmetry, and we recover the BM model.

In order to determine the model parameters near general commensurate configurations, we consider the geometry of TBG in real space. The key observation is that a small relative rotation $\delta\theta$ of the two graphene layers can be locally approximated by an interlayer translation \cite{Moon2013}. By carefully taking the limit $\delta\theta \to 0$, we derive the model for commensurate twist angle $\theta_0$ and interlayer displacement $\bd$ from the model for twist angle $\theta = \theta_0 + \delta\theta$. We then determine the continuum model parameters from a numerical computation of the microscopic tight-binding model (without lattice relaxation or corrugation \cite{Nam2017}) at commensurate angle $\theta_0$ with two values of $\bd$ corresponding to $AA$ and $AB$ stacking configurations. For the case $\theta_0 = 0$, we recover $w_0 \approx w_1 \approx \SI{110}{\milli\electronvolt}$, $\chi_0 =0$, $w_2 = 0$ in agreement with the BM model. We additionally provide numerical values of the model parameters for the next five commensurate configurations in order of the number of atoms per commensurate unit cell. When determining the continuum model parameters, we only use numerical tight-binding results at a single crystal momentum (the commensurate $\bK$ point) and two $\bd$ vectors. However, we find that the continuum model matches the tight-binding model with high accuracy for all crystal momenta in the commensurate $\bK$ valley and all $\bd$ vectors. It is worth noting that in the first five non-trivial (i.e. $\theta_0\neq 0$) commensurate configurations, the new parameters $\chi_0$ and $w_2$ are non-negligible. Although we do not consider lattice relaxation or corrugation, we note that these effects can alter the values of the model parameters, but not the general form of the continuum model (assuming these effects preserve the moir\'e lattice symmetries) \cite{Nam2017,koshino_maximally_2018,Carr2019}. We note also the possibility that the model parameters can be altered by the effects of higher graphene bands which we do not consider.

Next, we compute the moir\'e band structures of TBG with twist angle near the first six commensurate configurations. By both the ``tripod model" approximation \cite{Bernevig2021a,Bistritzer2011} and accurate numerical computations, we identify the condition for the \emph{first magic angle} in any nearly commensurate TBG system (\cref{eq:alpha-magic,eq:theta-magic}). This condition is similar to that of the original BM model. A further simplification of the generic TBG continuum model indicates that the moir\'e band structure only depends on a single angle variable $\phi_0=\chi_0+\theta_0/2$. At the first magic angle, the lowest two bands at charge neutrality in the nearly commensurate TBG model with $\phi_0\neq 0$ are flat in most of the moir\'e Brillouin zone except in the vicinity of the $\bGamma_M$ point. These bands are no longer particle-hole symmetric, though they do retain fragile topology. According to our model, the first magic angle near any nonzero commensurate twist angle $\theta_0$ (e.g. the magic angle $0.004^\circ$ near $\theta_0 \approx 38.2^\circ$) may be too small to be realized experimentally. However, it is possible that spontaneous commensurate atomic structural reconstructions (e.g. charge density wave orders), lattice relaxation or corrugation, or effective couplings mediated by higher graphene bands may enhance the moir\'e potential and enlarge the magic angles.

Finally, we reveal the existence of a \emph{hypermagic regime} centered at $\phi_0 = \pm\pi/2$ where several moir\'e bands (often $8$ or more) near charge neutrality become extremely flat simultaneously. The second and third magic angles in the chiral limit \cite{Tarnopolsky2019} are contained in the hypermagic regime, and for these parameters the lowest two bands at charge neutrality have fragile topology \cite{po_origin_2018,song_all_2019,po_faithful_2019,ahn_failure_2019,lian2020}. On the other hand, for many parameters in the hypermagic regime the lowest bands at charge neutrality have trivial topology. In such cases, we expect that the strongly interacting physics may be similar to that of the Hubbard model with trivial bands and may host anti-ferromagnetic states. Interestingly, many of the flat bands in the hypermagic regime resemble those of the kagome lattice and $p_x$, $p_y$ 2-orbital honeycomb lattice tight-binding models, which are known to exhibit flat bands \cite{Wu2007,Bergman2008}.

The rest of this paper is organized as follows. \cref{sec:generic-model} derives the generic form of the low energy TBG continuum model near commensuration from a microscopic graphene Hamiltonian. \cref{sec:symmetry-and-model-parameters} further restricts the form of the TBG continuum model using crystalline symmetries, and gives the model parameters for the first six commensurate configurations. In \cref{sec:commensurate-bands}, we discuss the low energy bands (namely the first two conduction and valence bands) of commensurate TBG. Then in \cref{sec:realistic-moire}, we compute the moir\'e band structure near several commensurate configurations with the actual model parameters and give the condition for the first magic angle. In \cref{sec:hypermagic}, we further explore the parameter space of the nearly commensurate TBG continuum model, reveal the hypermagic regime, and investigate the topology of the moir\'e bands. Finally, we give a high level discussion in \cref{sec:discussion}.

\section{Derivation of the generic continuum model}\label{sec:generic-model}

\subsection{Microscopic Hamiltonian}\label{sec:microscopic-Hamiltonian}
\begin{figure}
	\centering
	\includegraphics{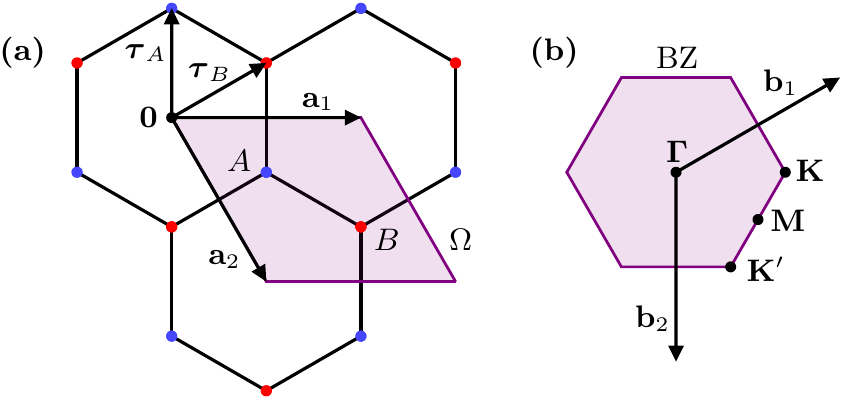}
	\caption{Illustration of the definitions of quantities in \cref{sec:microscopic-Hamiltonian}. \textbf{(a)} The graphene lattice and its primitive unit cell. \textbf{(b)} The reciprocal lattice primitive vectors, Brillouin zone, and high-symmetry crystal momenta.}
	\label{fig:lattice-definitions}
\end{figure}

The honeycomb lattice of monolayer graphene consists of two sublattices $A$ and $B$. We will often make the identifications $A = 1$ and $B = -1$ when using $A$ and $B$ in equations. As shown in \cref{fig:lattice-definitions}\textbf{(a)}, the positions of the carbon atoms in sublattice $\alpha$ are given by $\br + \btau_\alpha$ for $\alpha \in \{A, B\}$, $\br$ in a triangular Bravais lattice $L$, and constant vectors $\btau_\alpha$. It is convenient to choose the primitive vectors $(\ba_1, \ba_2)$ for $L$ where $\ba_1 = a_0\sqrt{3}\bhatx$, $\ba_2 = R_{-\pi/3}\ba_1$, $a_0\approx\SI{0.142}{\nano\meter}$ is the interatomic distance, and $R_\phi$ denotes rotation by angle $\phi$ about the $\bhatz$ axis. Additionally, we choose $\btau_A = a_0\bhaty$ and $\btau_B = R_{-\pi/3}\btau_A$ so that the origin $\bzero$ is in the center of a hexagon. We define $\Omega$ to be the primitive unit cell of $L$ and $|\Omega|$ to be its area.

The Bravais lattice $P$ that is reciprocal to $L$ has primitive vectors $(\bb_1, \bb_2)$ with $\bb_1 = R_{2\pi/3} \bb_2$ and $\bb_2 = -4\pi\bhaty/(3a_0)$ so that $\bb_j \cdot \ba_k = 2\pi \delta_{j,k}$. Explicitly, the lattices $L$ and $P$ are given by
\begin{equation}
\begin{split}
L &= \{n_1 \ba_1 + n_2 \ba_2 | n_1, n_2 \in \Z\}\\
P &= \{n_1 \bb_1 + n_2 \bb_2 | n_1, n_2 \in \Z\}.
\end{split}
\end{equation}
We define the Brillouin zone $\text{BZ}$ to be the Wigner-Seitz unit cell of $P$ and $|\text{BZ}|$ to be its area. Note that $|\Omega||\text{BZ}| = (2\pi)^2$. We additionally define the high-symmetry crystal momenta
\begin{equation}\label{eq:define-high-symmetry-points}
\begin{split}
\bGamma &= \bzero\\
\bK &= \frac{2}{3}\bb_1 + \frac{1}{3}\bb_2 = \frac{4\pi\sqrt{3}}{9a_0}\bhatx\\
\bK' &= \frac{1}{3}\bb_1 + \frac{2}{3}\bb_2 = R_{-\pi/3}\bK\\
\bM &= \frac{1}{2}\bb_1 + \frac{1}{2}\bb_2 = \frac{1}{2}\bK + \frac{1}{2}\bK'
\end{split}
\end{equation}
which are shown in \cref{fig:lattice-definitions}\textbf{(b)}.

We consider a system consisting of two stacked graphene layers denoted by $l \in \{+, -\}$. We rotate layer $l$ by the angle $-l\theta/2$ about the origin $\bzero$ and then translate it by an in-plane vector $-l\bd/2$, so that $\theta$ and $\bd$ are the relative rotation and translation of the two layers. We show in \cref{sec:equivalent-configurations} that when $\theta$ is not a commensurate angle, a change in the translation vector $\bd$ is equivalent to a unitary change of basis, but this is not generally the case when $\theta$ is a commensurate angle.

Let $L_l$, $P_l$, and $\text{BZ}_l$ be the real space lattice, reciprocal lattice, and graphene Brillouin zone of layer $l$. Explicitly, $L_l = R_{-l\theta/2} L$, $P_l = R_{-l\theta/2}P$, and $\text{BZ}_l = R_{-l\theta/2}\text{BZ}$ where we use the notation $R S = \{R s | s \in S\}$ for a set $S$ of vectors and an operator or number $R$. We will additionally use the notations $S_1 \cap S_2$ and $S_1 \cup S_2$ for the intersection and union of sets $S_1$, $S_2$, as well as the notations $S_1 + S_2 = \{s_1 + s_2 | s_1 \in S_1, s_2 \in S_2\}$ and $s_1 + S_2 = \{s_1 + s_2 | s_2 \in S_2\}$ where $s_1$ is a vector and $S_1$, $S_2$ are sets of vectors.

We neglect electron spin when describing the single-particle model because of the weak spin-orbit coupling in graphene \cite{Sichau2019}. The spinless $p_z$ orbitals $\ket{\br, l, \alpha}$ for $\br \in L_l$, $l \in \{+, -\}$, and $\alpha \in \{A, B\}$ form an orthonormal basis for the Hilbert space. The orbital $\ket{\br, l, \alpha}$ is localized at position $\br + \btau^l_\alpha$ where $\btau^l_\alpha = R_{-l\theta/2}\btau_\alpha - l\bd/2$. Note that $\bd$ enters the formalism only through the definition of $\btau_\alpha^l$.

The Bloch states are defined by
\begin{equation}\label{eq:Bloch-definition}
\ket{\bk, l, \alpha} = \frac{1}{\sqrt{|\text{BZ}|}} \sum_{\br \in L_l} e^{i\bk\cdot(\br  + \btau^l_\alpha)}\ket{\br, l, \alpha}
\end{equation}
for crystal momentum vectors $\bk \in \R^2$, and satisfy the normalization condition
\begin{equation}\label{eq:Bloch-normalization}
\begin{split}
&\braket{\bk', l', \alpha' | \bk, l, \alpha}\\
&=\delta_{l',l}\delta_{\alpha',\alpha}\sum_{\bG_l \in P_l}\delta^2(\bk'-\bk - \bG_l)e^{-i\btau_\alpha^l \cdot \bG_l}.
\end{split}
\end{equation}
Note that the origin for crystal momenta is $\bGamma$, defined in \cref{eq:define-high-symmetry-points} and shown in \cref{fig:lattice-definitions}. The Bloch states $\ket{\bk, l, \alpha}$ with $\bk \in \text{BZ}_l$ form a continuous basis for the Hilbert space. However, for convenience we will sometimes use the overcomplete set formed by all Bloch states $\ket{\bk, l, \alpha}$ for $\bk \in \R^2$.

We consider a microscopic single-particle Hamiltonian $H$ with matrix elements
\begin{equation}\label{eq:Hamiltonian-matrix-elements}
\begin{split}
\braket{\br', l', \alpha' | H | \br, l, \alpha} &= t_{l'\cdot l}(\br' + \btau_{\alpha'}^{l'} - \br - \btau_\alpha^l)\\
&- \mu \delta_{\br',\br}\delta_{l',l}\delta_{\alpha',\alpha}
\end{split}
\end{equation}
where $\mu$ is a chemical potential and $t_\pm : \R^2 \to \R$ are rotationally symmetric functions (i.e. $t_\pm(\br)$ depends only on $|\br|$) determining the intra- and interlayer hoppings. We allow the functions $t_\pm(\br)$ to remain unspecified for now. The intra-layer matrix elements are given by
\begin{equation}\label{eq:intra-matrix-element}
\begin{split}
&\braket{\bk', l, \alpha' | H | \bk, l, \alpha} = \braket{\bk', l, \alpha' | \bk, l, \alpha'}\\
&\times \left(-\mu + \sum_{\br \in L + \btau_{\alpha'} - \btau_\alpha} e^{-i(R_{l\theta/2}\bk) \cdot \br} t_+(\br)\right)
\end{split}
\end{equation}
(see \cref{sec:hamiltonian-matrix-elements}). If the value of $\mu$ is chosen appropriately, then for crystal momenta near $\bK_l = R_{-l\theta/2}\bK$, this matrix element can be approximated by a Dirac cone
\begin{equation}\label{eq:dirac-cone}
\begin{split}
&\braket{\bK_l + \bp', l, \alpha' | H | \bK_l + \bp, l, \alpha}\\
&= (\hbar v_F (\bsigma_{l\theta/2} \cdot \bp)_{\alpha',\alpha} + O(|\bp|^2))\delta^2(\bp'-\bp).
\end{split}
\end{equation}
Here, $\bsigma_\phi = e^{-i(\phi/2)\sigma_z}(\sigma_x \bhatx + \sigma_y \bhaty)e^{i(\phi/2)\sigma_z}$ is a vector of rotated Pauli matrices satisfying
\begin{equation}\label{eq:define-bsigma}
\bsigma_\phi \cdot \bp = \bsigma_0 \cdot (R_\phi \bp) = \begin{pmatrix}
0 & h.c.\\
e^{i\phi}(p_x+ip_y) & 0
\end{pmatrix}
\end{equation}
and $v_F$ is the Fermi velocity, which depends on the function $t_+(\br)$. We make the assumption throughout the paper that $v_F > 0$. See \cref{sec:dirac-cones} for a derivation of \cref{eq:dirac-cone} based on symmetry. The matrix elements for crystal momenta near the other Brillouin zone corners $R_{n\pi/3} \bK_l$ for $1 \leq n \leq 5$ are given by similar Dirac cone Hamiltonians.

The interlayer matrix elements are given by
\begin{equation}\label{eq:inter-matrix-element}
\begin{split}
&\braket{\bk', -l, \alpha' | H | \bk, l, \alpha} = \sum_{\bG_- \in P_-} \sum_{\bG_+ \in P_+} \frac{\hat{t}_-(\bk+\bG_l)}{|\Omega|}\\
&e^{i\btau^{-l}_{\alpha'}\cdot \bG_{-l}} e^{-i\btau^l_\alpha \cdot \bG_l}
\delta^2(\bk + \bG_l - \bk'- \bG_{-l})
\end{split}
\end{equation}
where the hatted functions $\hat{t}_\pm(\bk)$ are the two dimensional Fourier transforms of $t_\pm(\br)$ (see \cref{sec:hamiltonian-matrix-elements}). We see that $H$ is block diagonal: the Bloch states $\ket{\bk', l', \alpha'}$ and $\ket{\bk, l, \alpha}$ are in the same Hamiltonian block if and only if
\begin{equation}\label{eq:block-diagonal-microscopic}
\bk' - \bk \in P_- + P_+.
\end{equation}

\subsection{Commensurate configurations}\label{sec:commensurate-case}
\begin{figure}
	\centering
	\includegraphics{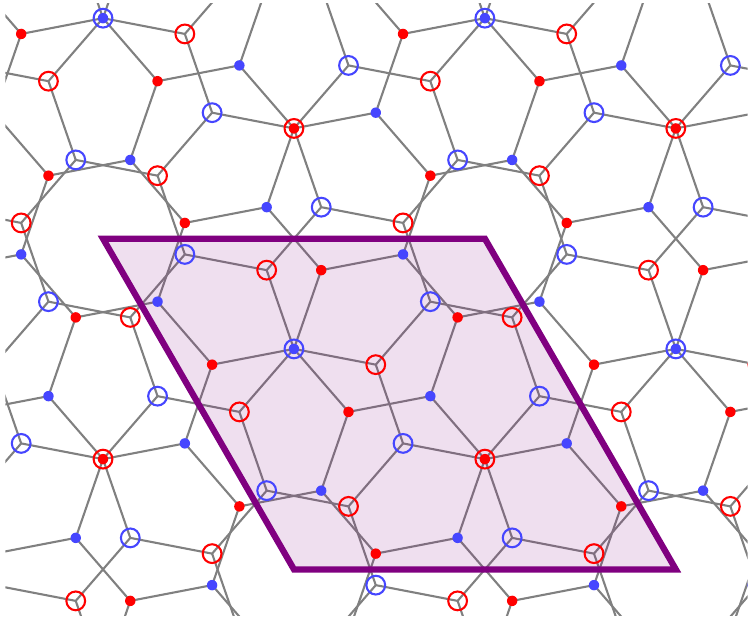}
	\caption{The real space structure of commensurate TBG with $(m, n) = (5, 3)$ and $\bd = \bzero$, in which case $\theta_0 \approx 38.2^{\circ}$ and $N = 7$. The top (bottom) atoms are represented by dots (circles) and the $A$ ($B$) sublattices in each layer are colored blue (red). The purple rhombus is an example of a primitive unit cell for the commensuration superlattice. This unit cell contains $4N = 28$ atoms. Note that this configuration has $AA$ stacking as described in \cref{sec:stacking}.}
	\label{fig:commensurate-lattice-AA}
\end{figure}

Since layer $l$ is invariant under translation by elements of the graphene Bravais lattice $L_l$, the bilayer system is invariant under translations by elements of $L_- \cap L_+$. Commensurate configurations are those for which $L_- \cap L_+ \neq \{\bzero\}$, in which case $L_- \cap L_+$ forms a Bravais lattice called the commensuration superlattice. Let $\theta = \theta_0$ be a commensurate angle, by which we mean the twist angle for a commensurate configuration.

We show in \cref{sec:equivalent-configurations} that the crystalline symmetries of TBG allow us to restrict our attention to configurations with $\theta_0 \in [0, \pi/3)$. These configurations can be enumerated by a pair of relatively prime integers $m > n \geq 0$ with
\begin{equation}\label{eq:theta0}
\theta_0 = \cos^{-1}\left(\frac{3m^2 - n^2}{3m^2 + n^2}\right)
\end{equation}
(see \cref{sec:enumeration-commensurate}). The commensurate configuration corresponding to the pair $(m, n)$ has $4N$ atoms per unit cell where the integer $N \geq 1$ is given in \cref{eq:define-N} as a function of $m$ and $n$.

As shown in \cref{sec:K-K'-equivalences}, if $3 | n$ (i.e. $3$ divides $n$) we have
\begin{equation}\label{eq:K_+ - K_-}
\bK_+ - \bK_-, \bK'_+ - \bK'_- \in P_- + P_+
\end{equation}
and otherwise
\begin{equation}\label{eq:K_+ - K'_-}
\bK_+ - \bK'_-, \bK'_+ - \bK_- \in P_- + P_+
\end{equation}
where $\bK_l = R_{-l\theta/2}\bK$ and $\bK'_l = R_{-l\theta/2}\bK'$. Additionally, in either case we have
\begin{equation}\label{eq:K_l - K'_l}
\bK_+ - \bK'_+, \bK_- - \bK'_- \not \in P_- + P_+.
\end{equation}
If $\theta_0$ is a commensurate angle then so is $\pi/3 - \theta_0$, and the Hamiltonians for these two configurations are unitarily equivalent (see \cref{sec:equivalent-configurations}). Furthermore, we show in \cref{sec:pi/3-complementarity} that among the two configurations corresponding to $\theta_0$ and $\pi/3-\theta_0$, one must satisfy $3 | n$ while the other does not. As a result, we assume without loss of generality that $3 | n$ and \cref{eq:K_+ - K_-} holds. From here on, we will always assume $3|n$ unless we explicitly state otherwise. \cref{tbl:w-parameter-table} lists properties of the first six commensurate configurations in increasing order of $N$. \cref{fig:commensurate-lattice-AA} illustrates the locations of the atoms in real space for a particular commensurate configuration.

\begin{figure}
	\centering
	\includegraphics{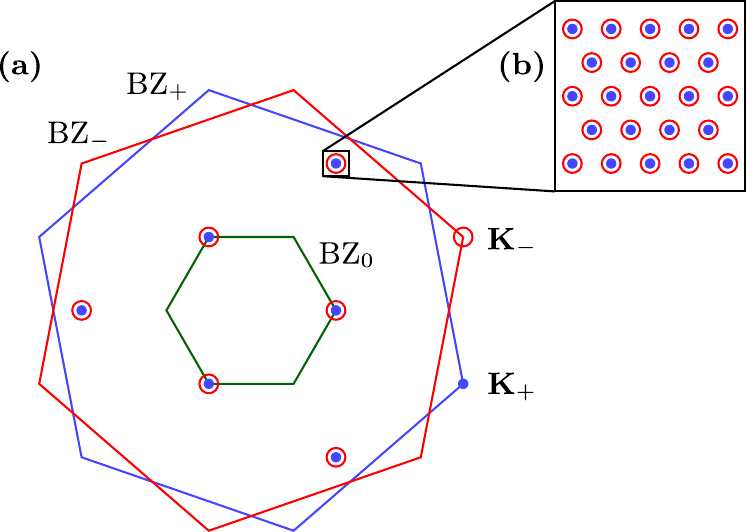}
	\caption{\textbf{(a)} Illustration of the crystal momenta involved in the Hamiltonian block containing $\bK_+$ and $\bK_-$ for the commensurate configuration with $(m, n) = (5, 3)$. The top (bottom) Brillouin zone boundaries are shown in blue (red) and the boundary of the Brillouin zone $\text{BZ}_0$ corresponding to the commensuration superlattice is shown in green. The $N=7$ top (bottom) layer momenta are marked with blue dots (red circles). All shown crystal momenta in a given layer differ by elements of $P_- + P_+$ and are contained in the given layer's Brillouin zone. \textbf{(b)} Illustration of the set $B_{l, q}$ defined in \cref{sec:low-energy-model} for $l=+$ or $l=-$, some large value $q$, and some small value $\delta\theta$. There is a lattice of crystal momenta near each point in \textbf{(a)}.}
	\label{fig:brillouin-zones}
\end{figure}

We saw in \cref{eq:block-diagonal-microscopic} that the microscopic Hamiltonian is block diagonal in accordance with the lattice $P_- + P_+$. We show in \cref{sec:commensuration-lattices} that when the system is commensurate, $P_- + P_+$ is the reciprocal lattice of the commensuration superlattice $L_- \cap L_+$. We see that the block diagonality can be attributed in this case to translation symmetry with respect to the commensuration superlattice. Each Hamiltonian block has a basis consisting of Bloch states with $N$ non-equivalent crystal momenta in each layer, for a total dimension of $4N$. As an example, \cref{fig:brillouin-zones}\textbf{(a)} illustrates the crystal momenta involved in the Hamiltonian block containing $\bK_+$ and $\bK_-$ for a particular commensurate configuration. We show in \cref{sec:minimal-norm} that $L_- \cap L_+ = \sqrt{N}L$ and $P_- + P_+ = P/\sqrt{N}$ so that the Brillouin zone $\text{BZ}_0$ corresponding to the commensuration superlattice is a regular hexagon.

\subsection{Incommensurate configurations}\label{sec:incommensurate-case}
We now consider an incommensurate twist angle $\theta$. We show in \cref{sec:density-proof} that in this case $P_- + P_+$ is a dense subset of $\R^2$. As a result, the block diagonality of $H$ given by \cref{eq:block-diagonal-microscopic} cannot be directly used to define a band structure. In this section, we will construct a notion of distance between Bloch states that can be used in place of block diagonality to analyze $H$.

We show in \cref{sec:enumeration-commensurate} that since $\theta$ is incommensurate, we have $P_- \cap P_+ = \{\bzero\}$. It follows that for any $l \in \{+, -\}$ and crystal momentum vectors $\bk$, $\bk'$ with $\bk' - \bk \in P_- + P_+$ there are unique vectors $\bG_- \in P_-$, $\bG_+ \in P_+$ such that
\begin{equation}\label{eq:k-G-k'-G}
\bk + \bG_l = \bk' + \bG_{-l}.
\end{equation}
This pair of vectors $\bG_-, \bG_+$ determines the interlayer matrix element in \cref{eq:inter-matrix-element}. Since $\hat{t}_-(\bk)$ depends only on $|\bk|$, the magnitude of $\braket{\bk', -l, \alpha' | H | \bk, l, \alpha}$ depends only on $|\bk + \bG_l|$. We assume that $\hat{t}_-(\bk)$ monotonically decreases with $|\bk|$, so that interlayer matrix elements with large magnitude correspond to small values of $|\bk + \bG_l|$. Similarly, the intralayer matrix element in \cref{eq:intra-matrix-element} is zero unless $\bk' - \bk \in P_l$. As a result, $\braket{\bk', l, \alpha' | H | \bk, l, \alpha}$ is only nonzero when $\bk$ and $\bk'$ are related as in \cref{eq:k-G-k'-G} with $|\bG_{-l}| = 0$.

With this motivation, we define a function $d$ that quantifies the magnitude of the matrix elements of $H$
\begin{equation}\label{eq:define-distance}
d(\bk, l, \bk', l') = \begin{cases}
\infty & \text{if } \bk' - \bk \not \in P_- + P_+\\
|\bk + \bG_l| & \text{if } l' = -l \text{ and \cref{eq:k-G-k'-G}}\\
|\bG_{-l}| & \text{if } l' = l \text{ and \cref{eq:k-G-k'-G}}.
\end{cases}
\end{equation}
We show in \cref{sec:distance-function} that $d$ satisfies
\begin{enumerate}
\item $d(\bk, l, \bk, l) = 0$\label{item:distance-0}
\item $d(\bk, l, \bk', l') = d(\bk', l', \bk, l)$\label{item:distance-symmetry}
\item $d(\bk, l, \bk'', l'') \leq d(\bk, l, \bk', l') + d(\bk', l', \bk'', l'')$\label{item:distance-triangle-inequality}
\end{enumerate}
so that $d$ defines a notion of distance on the set $\R^2 \times \{+, -\}$ \footnote{Technically, $d$ is not a metric because it assumes the value $\infty$ and $d(\bk, l, \bk', l') = 0$ whenever $l'= -l$, $\bk \in P_l$, $\bk' \in P_{-l}$ or $l'=l$, $\bk' - \bk \in P_l$. Nonetheless, it is useful to think of $d$ as a distance function.}. Suppose we define the distance between Bloch states $\ket{\bk, l, \alpha}$, $\ket{\bk', l', \alpha'}$ to be $d(\bk, l, \bk', l')$. Then by construction, the microscopic Hamiltonian $H$ described by \cref{eq:intra-matrix-element,eq:inter-matrix-element} is local with respect to this notion of distance.

\subsection{Continuum model for incommensurate configurations}\label{sec:low-energy-model}
We now take
\begin{equation}\label{eq:theta}
\theta = \theta_0 + \delta\theta
\end{equation}
where $\theta_0$ is a commensurate angle as in \cref{eq:theta0} and $\delta\theta$ is small. We assume that $\theta$ is an incommensurate angle so that the distance function $d$ from \cref{sec:incommensurate-case} is defined. We are interested in the single particle physics of $H$ near the Fermi level at charge neutrality, as this determines the low energy excitations of the many-body Hamiltonian. In this section, we will derive a continuum model that approximates the relevant energies and eigenvectors of $H$.

We will make use of the following characterization of the distance function $d$ that applies when $\theta = \theta_0 + \delta\theta$. Let $L_l^0 = R_{-l\theta_0/2}L$, $P_l^0 = R_{-l\theta_0/2}P$, and recall from \cref{sec:commensurate-case} that $L_-^0 \cap L_+^0$ is the commensuration superlattice corresponding to twist angle $\theta_0$ and $P_-^0 + P_+^0$ is its reciprocal lattice. Define the set
\begin{equation}
\mathcal{Q}(\bk, l, \bk', l') = - \delta_{l',l}\bk'  + (\bk + P^0_l) \cap (\bk' + P^0_{-l})
\end{equation}
and the operator
\begin{equation}\label{eq:define-D-delta-theta}
D(\delta\theta) = R_{\delta\theta/2} - R_{-\delta\theta/2} = 2\sin(\delta\theta/2) R_{\pi/2}.
\end{equation}
Let $\bk \in \R^2$, $l \in \{+, -\}$, and define $\bk_0 = R_{l\delta\theta/2}\bk$. Then for any pair $(\bk', l')$ with $d(\bk, l, \bk', l') < \infty$, there are unique vectors $\bk'_0 \in \bk_0 + P^0_- + P^0_+$ and $\bQ \in \mathcal{Q}(\bk_0, l, \bk'_0, l')$ such that
\begin{equation}\label{eq:k'-decomposition}
\bk' = R_{-l'\delta\theta/2} \bk'_0 - lD(\delta\theta)\bQ.
\end{equation}
Additionally, we have $|\bQ| = d(\bk, l, \bk', l')$ so that
\begin{equation}\label{eq:distance-lattice-relation}
|R_{l'\delta\theta/2}\bk' - \bk'_0| = 2|\sin(\delta\theta/2)|d(\bk, l, \bk', l').
\end{equation}
Conversely, if $\bk'$ is given by \cref{eq:k'-decomposition} for some $\bk'_0 \in \bk_0 + P^0_- + P^0_+$ and $\bQ \in \mathcal{Q}(\bk_0, l, \bk'_0, l')$ then $d(\bk, l, \bk', l') = |\bQ|$. These claims are proved in \cref{sec:distance-level-sets}.

Since monolayer graphene has Dirac cones at the $\bK$ and $\bK'$ points (i.e. graphene has two valleys), the single particle physics of $H$ near the Fermi level at charge neutrality is dominated by Bloch states with crystal momenta near $\bK_\pm$ or $\bK'_\pm$. Consider two momenta $\bk = \bK_l$, $\bk' = \bK'_{l'}$ from opposite graphene valleys. Then $\bk_0 = \bK^0_l$ and $R_{l'\delta\theta/2}\bk' = \bK^{\prime 0}_{l'}$ where $\bK^0_\pm = R_{\mp \theta_0/2}\bK$ and $\bK^{\prime 0}_\pm = R_{\mp \theta_0/2}\bK'$. By \cref{eq:K_+ - K_-,eq:K_l - K'_l}, $\bK^{\prime 0}_{l'} \not\in \bK^0_l + P^0_- + P^0_+$ so there is some minimal value $\kappa > 0$ taken by the quantity $|\bK^{\prime 0}_{l'} - \bk'_0|$ for $\bk'_0 \in \bk_0 + P^0_- + P^0_+$. By \cref{eq:distance-lattice-relation},
\begin{equation}
d(\bK_l, l, \bK'_{l'}, l') \geq \frac{\kappa}{2|\sin(\delta\theta/2)|}
\end{equation}
which diverges as $\delta\theta \to 0$. This implies that for small $\delta\theta$, the spectrum of $H$ splits into two nearly uncoupled valleys corresponding to $\bK$ and $\bK'$. We will neglect intervalley coupling and focus on the $\bK$ valley, noting that time-reversal symmetry interchanges the valleys (see \cref{sec:symmetry-representations}).

For any $q > 0$, define $U(\bk, l, q)$ to be the subspace generated by all Bloch states $\ket{\bk', l', \alpha'}$ with $d(\bk, l, \bk', l') < q$, and note that $U(\bk, l, q)$ is finite dimensional. To compute the eigenstates and energies of $H$ in the $\bK$ valley, we consider the projection of $H$ into $U(\bK_l + \bp, l, q)$ for a small vector $\bp$ and a large value $q$. Let $B_{l, q}$ be the set of pairs $(\bk', l')$ such that $\bk'$ is given by \cref{eq:k'-decomposition} with $\bk = \bK_l$, $\bk'_0 \in \text{BZ}^0_{l'} = R_{-l'\theta_0/2}\text{BZ}$, and $|\bQ| < q$. Then for all vectors $\bp$ small enough, the set of Bloch states $\ket{\bk' + \bp, l', \alpha'}$ with $(\bk', l') \in B_{l, q}$ forms a basis for $U(\bK_l + \bp, l, q)$. The set $B_{l, q}$ is illustrated in \cref{fig:brillouin-zones}\textbf{(b)}.

Recall from \cref{sec:commensurate-case} that we can write $L_-^0 \cap L_+^0 = \sqrt{N}L$ and $P_-^0 + P_+^0 = P/\sqrt{N}$ where $4N$ is the number of atoms in the primitive unit cell of $L_-^0 \cap L_+^0$. When $N > 1$ and $q$ is large enough, there are elements $(\bk', l') \in B_{l, q}$ for which the value of $\bk'_0$ is not $\bK^0_{l'}$ (e.g. the points shown in \cref{fig:brillouin-zones}\textbf{(b)}). The corresponding Bloch states in $U(\bK_l + \bp, l, q)$ have expected energies with respect to the intralayer Hamiltonian that are far from the Fermi level at charge neutrality. Assuming that $|\hat{t}_-(\bK)|$ is not too large, these Bloch states can be treated perturbatively. There is then some effective Hamiltonian supported only on the subspace generated by Bloch states $\ket{\bk' + \bp, l', \alpha'}$ such that $(\bk', l') \in B_{l, q}$ and the value of $\bk'_0$ is $\bK^0_{l'}$. Note that these conditions are equivalent to $\bk' + \bp = \bK_{l'} + \bp'$ where
\begin{equation}\label{eq:momentum-hopping-derived}
\bp' = \bp - lD(\delta\theta)\bQ
\end{equation}
and $\bQ \in \mathcal{Q}(\bK^0_l, l, \bK^0_{l'}, l')$ with $|\bQ| < q$. For convenience, we define
\begin{align}
\begin{split}
\mathcal{Q}_+ &= \mathcal{Q}(\bK^0_+, +, \bK^0_-, -)\\
&= \mathcal{Q}(\bK^0_-, -, \bK^0_+, +)\\
&= (\bK^0_- + P^0_-) \cap (\bK^0_+ + P^0_+)
\end{split}\\
\mathcal{Q}_- &= -\mathcal{Q}_+\\
\begin{split}
\mathcal{Q}_0 &= \mathcal{Q}(\bK^0_+, +, \bK^0_+, +)\\
&= \mathcal{Q}(\bK^0_-, -, \bK^0_-, -)\\
&= P^0_- \cap P^0_+.
\end{split}
\end{align}

We will now describe a class of continuum models that approximate these effective Hamiltonians. We introduce continuum states $\ket{\bp, l, \alpha}_c$ for $\bp \in \R^2$, $l \in \{+, -\}$, $\alpha \in \{A, B\}$ in a new Hilbert space, satisfying the normalization condition
\begin{equation}\label{eq:simpler-normalization}
\braket{\bp', l', \alpha' |_c \bp, l, \alpha}_c = \delta_{l',l}\delta_{\alpha',\alpha}\delta^2(\bp'-\bp).
\end{equation}
Although $\bp$ is allowed to range over all of $\R^2$, $\ket{\bp, l, \alpha}_c$ represents the Bloch state $\ket{\bK_l + \bp, l, \alpha}$ when $\bp$ is small. When $\bp$ is large, these states cannot be identified because they satisfy different normalization conditions, namely \cref{eq:Bloch-normalization,eq:simpler-normalization}. Because of \cref{eq:dirac-cone}, we take the part of the continuum Hamiltonian due to intralayer coupling to be $\tilde{H}_{\text{intra}} = \int d^2\bp \boldsymbol{\ket{\bp}_c} \mathcal{H}_{\text{intra}}(\bp) \boldsymbol{\bra{\bp}_c}$ where
\begin{equation}\label{eq:effective-intra-hamiltonian}
\mathcal{H}_{\text{intra}}(\bp) = \hbar v_F \begin{pmatrix}
\bsigma_{\theta/2} \cdot \bp & 0\\
0 & \bsigma_{-\theta/2} \cdot \bp
\end{pmatrix}
\end{equation}
and
\begin{equation}\label{eq:define-p-row-vector}
\boldsymbol{\ket{\bp}_c} = \begin{pmatrix}
\ket{\bp, +, A}_c & \ket{\bp, +, B}_c & \ket{\bp, -, A}_c & \ket{\bp, -, B}_c
\end{pmatrix}
\end{equation}
is a row vector of states. Because of \cref{eq:momentum-hopping-derived}, we take the part of the continuum Hamiltonian due to interlayer coupling to be $\tilde{H}_{\text{inter}} = \int d^2\bp' d^2\bp \boldsymbol{\ket{\bp'}_c} \mathcal{H}_{\text{inter}}(\bp', \bp) \boldsymbol{\bra{\bp}_c}$ where
\begin{equation}\label{eq:effective-inter-hamiltonian}
\begin{split}
&\mathcal{H}_{\text{inter}}(\bp', \bp) =\\
&\sum_{\bQ \in \mathcal{Q}_0}
\begin{pmatrix}
S_\bQ^+  & 0\\
0 & S_\bQ^-
\end{pmatrix}
\delta^2(\bp' - \bp - D(\delta\theta)\bQ)\\
&+\sum_{\bQ \in \mathcal{Q}_+} \begin{pmatrix}
0 & T_\bQ\\
0 & 0
\end{pmatrix}
\delta^2(\bp' - \bp - D(\delta\theta)\bQ)\\
&+ \sum_{\bQ \in \mathcal{Q}_-} \begin{pmatrix}
0 & 0\\
T_\bQ & 0
\end{pmatrix}
\delta^2(\bp' - \bp - D(\delta\theta)\bQ).
\end{split}
\end{equation}
Here, $T_\bQ$ and $S^l_\bQ$ denote complex $2 \times 2$ matrices which are functions of $\delta\theta$ and the translation parameter $\bd$. Note that since $\tilde{H}_{\text{inter}}$ is Hermitian, we have 
\begin{equation}\label{eq:T-S-hermiticity}
T^\dagger_\bQ = T_{-\bQ},\ (S^l_\bQ)^\dagger = S^l_\bQ.
\end{equation}
The full continuum Hamiltonian is given by $\tilde{H} = \tilde{H}_{\text{intra}} + \tilde{H}_{\text{inter}}$.

\begin{figure}
	\centering
	\includegraphics{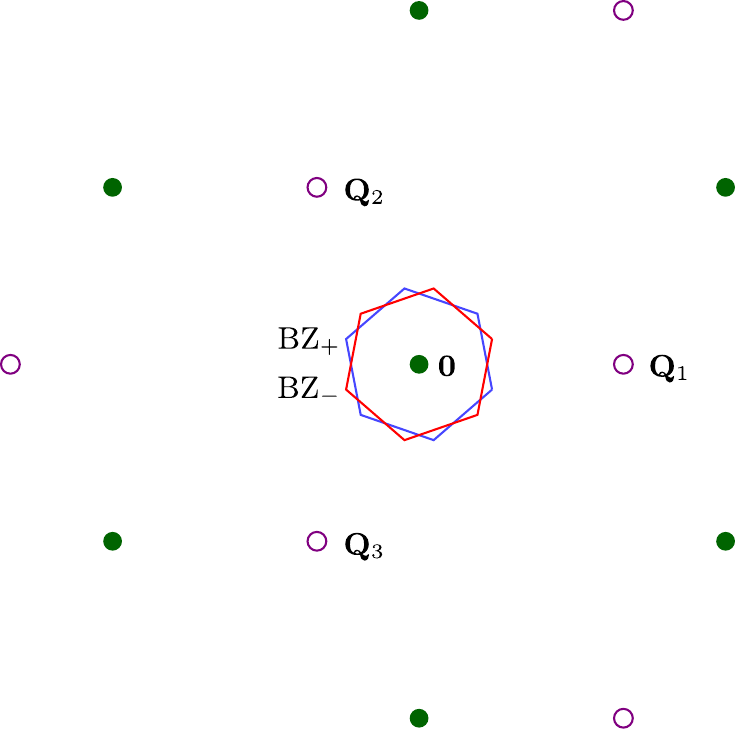}
	\caption{The $\mathcal{Q}_+$ and $\mathcal{Q}_0$ lattices for the commensurate configuration with $(m, n) = (5, 3)$, in which case $s = 1$. The elements of $\mathcal{Q}_+$ ($\mathcal{Q}_0$) are denoted by purple circles (green dots) and the elements of both lattices with minimal norm are labeled. The top (bottom) Brillouin zone boundaries are shown in blue (red).}
	\label{fig:Q_lattices}
\end{figure}

We show in \cref{sec:minimal-norm} that
\begin{equation}\label{eq:Q-lattices-form}
\begin{split}
\mathcal{Q}_+ &= s\sqrt{N}\bK + \sqrt{N}P\\
\mathcal{Q}_0 &= \sqrt{N}P
\end{split}
\end{equation}
where $s = \pm 1$ is given by \cref{eq:determine-s}. Furthermore, the elements of $\mathcal{Q}_+$ with minimal norm are
\begin{equation}\label{eq:define-Q1-Q2-Q3}
\bQ_1 = s\sqrt{N}\bK,\ \bQ_2 = R_{2\pi/3}\bQ_1,\ \bQ_3 = R_{4\pi/3}\bQ_1.
\end{equation}
The lattices $\mathcal{Q}_+$ and $\mathcal{Q}_0$ and the vectors $\bQ_1$, $\bQ_2$, and $\bQ_3$ are shown in \cref{fig:Q_lattices}.

We now observe that $\tilde{H}$ is block diagonal: the states $\ket{\bp', l', \alpha'}_c$ and $\ket{\bp, l, \alpha}_c$ are in the same Hamiltonian block if and only if
\begin{equation}\label{eq:block-diagonal-continuum}
(\bp' + l' \bq_1) - (\bp + l \bq_1) \in D(\delta\theta)\mathcal{Q}_0
\end{equation}
where
\begin{equation}\label{eq:define-q_j}
\bq_j = D(\delta\theta)\bQ_j \text{ for } j \in  \{1, 2, 3\}.
\end{equation}
More explicitly, we have
\begin{equation}\label{eq:explicit-q_j}
\begin{split}
\bq_1 &= 2\sin(\delta\theta/2)s\sqrt{N}|\bK|\bhaty,\\
\bq_2 &= R_{2\pi/3}\bq_1,\ \bq_3 = R_{4\pi/3}\bq_1.
\end{split}
\end{equation}
We refer to $D(\delta\theta)\mathcal{Q}_0$ as the moir\'e reciprocal lattice and $\bp + l \bq_1$ as the moir\'e quasi-momentum for $\ket{\bp, l, \alpha}_c$. The Wigner-Seitz unit cell of the moir\'e reciprocal lattice is $\text{BZ}_M = D(\delta\theta)\sqrt{N}\text{BZ}$ and it is called the moir\'e Brillouin zone. Additionally, we define the high-symmetry moir\'e quasi-momenta
\begin{equation}\label{eq:define-high-symmetry-points-moire}
\bX_M = D(\delta\theta)s\sqrt{N}\bX
\end{equation}
for $\bX \in \{\bGamma, \bK, \bK', \bM\}$ and note that
\begin{equation}
\bGamma_M = \bzero,\qquad \bK_M = \bq_1
\end{equation}
and
\begin{equation}\label{eq:norm-K_M}
|\bK_M| = 2|\sin(\delta\theta/2)|\sqrt{N}|\bK|.
\end{equation}

To further explicate the moir\'e translation symmetry, we transform to real space. We define states
\begin{equation}\label{eq:define-real-continuum-states}
\ket{\br, l, \alpha}_c = \frac{1}{2\pi}\int d^2\bp e^{-i\bp\cdot \br}\ket{\bp, l, \alpha}_c
\end{equation}
which satisfy the normalization condition
\begin{equation}
\braket{\br', l', \alpha'|_c \br, l, \alpha}_c = \delta_{l',l}\delta_{\alpha',\alpha}\delta^2(\br'-\br).
\end{equation}
Defining the row vector of states
\begin{equation}\label{eq:define-r-row-vector}
\boldsymbol{\ket{\br}_c} = \begin{pmatrix}
\ket{\br, +, A}_c & \ket{\br, +, B}_c & \ket{\br, -, A}_c & \ket{\br, -, B}_c
\end{pmatrix},
\end{equation}
we can rewrite the Hamiltonian in the form $\tilde{H}_{\text{intra}} = \int d^2\br \boldsymbol{\ket{\br}_c} \mathcal{H}_{\text{intra}}(\br) \boldsymbol{\bra{\br}_c}$ and $\tilde{H}_{\text{inter}} = \int d^2\br \boldsymbol{\ket{\br}_c} \mathcal{H}_{\text{inter}}(\br) \boldsymbol{\bra{\br}_c}$ where
\begin{equation}\label{eq:continuum-hamiltonian-real-space}
\begin{split}
\mathcal{H}_{\text{intra}}(\br) &= -i\hbar v_F \begin{pmatrix}
\bsigma_{\theta/2} \cdot \nabla & 0\\
0 & \bsigma_{-\theta/2} \cdot \nabla
\end{pmatrix}\\
\mathcal{H}_{\text{inter}}(\br) &= \begin{pmatrix}
S^+(\br) & T(\br)\\
T^\dagger(\br) & S^-(\br)
\end{pmatrix}
\end{split}
\end{equation}
and
\begin{equation}\label{eq:potentials-real-space}
\begin{split}
T(\br) &= \sum_{\bQ \in \mathcal{Q}_+} T_\bQ e^{i\br \cdot D(\delta\theta)\bQ}\\
S^l(\br) &= \sum_{\bQ \in \mathcal{Q}_0} S^l_\bQ e^{i\br \cdot D(\delta\theta)\bQ}.
\end{split}
\end{equation}
We interpret $\tilde{H}$ as the Hamiltonian for a system of Dirac electrons moving through the spatially varying potentials $T(\br)$, $S^+(\br)$, and $S^-(\br)$. Note that these potentials are periodic (up to a phase) with respect to the moir\'e superlattice $D(\delta\theta)^{-1}L/\sqrt{N}$ which is reciprocal to $D(\delta\theta)\mathcal{Q}_0$.

\subsection{Continuum model for commensurate configurations}\label{sec:real-space}
As in \cref{sec:low-energy-model} we take $\theta = \theta_0 + \delta\theta$ where $\theta_0$ is a commensurate twist angle, $\delta\theta$ is small, and $\theta$ is an incommensurate angle. Since the microscopic Hamiltonian is continuous with respect to twist angle, we can take the limit $\delta\theta \to 0$ to derive a continuum model for the commensurate configuration with twist angle $\theta_0$.

In this section, we use $T_\bQ(\delta\theta, \bd)$, $S^l_\bQ(\delta\theta, \bd)$, $T(\br,\delta\theta,\bd)$, and $S^l(\br,\delta\theta,\bd)$, to denote the $T_\bQ$ and $S^l_\bQ$ matrices and the $T(\br)$ and $S^l(\br)$ potentials with twist angle $\theta = \theta_0 + \delta\theta$ and translation vector $\bd$. To determine the correct definition of $T_\bQ(0,\bd)$, note that
\begin{equation}\label{eq:rotation-translation-equiv-simple}
\begin{split}
R_{-l\delta\theta/2}\br &= \br - l\delta\theta R_{\pi/2}\br/2 + O(\delta\theta^2)\\
&= \br - l D(\delta\theta)\br/2 + O(\delta\theta^2).
\end{split}
\end{equation}
This implies that the pattern of atoms near position $\br$ with $\theta = \theta_0 + \delta\theta$ and $\bd = \bzero$ is the same to first order in $\delta\theta$ as the pattern with $\theta = \theta_0$ and
\begin{equation}
\bd = D(\delta\theta)\br = 2\sin(\delta\theta/2) R_{\pi/2}\br.
\end{equation}
Taking into account the phase shift accrued by the continuum momentum states when the translation vector $\bd$ is changed (see \cref{sec:rotation-translation}), we must then have
\begin{equation}\label{eq:T-rotation-translation}
\begin{split}
&e^{i\cos(\theta/2) \bK \cdot D(\delta\theta)\br} T(\br,\delta\theta, \bzero)\\
&=T(\br, 0, D(\delta\theta)\br) + O(\delta\theta^2)
\end{split}
\end{equation}
and
\begin{equation}\label{eq:S-rotation-translation}
S^l(\br,\delta\theta, \bzero) = S^l(\br, 0, D(\delta\theta)\br) + O(\delta\theta^2).
\end{equation}
It follows that
\begin{equation}
\begin{split}
&e^{i\cos(\theta/2) \bK \cdot D(\delta\theta)\br}\sum_{\bQ \in \mathcal{Q}_+} T_\bQ(\delta\theta, \bzero)e^{i\br \cdot D(\delta\theta)\bQ} \\
=&\sum_{\bQ \in \mathcal{Q}_+} T_\bQ(0, D(\delta\theta)\br) + O(\delta\theta^2)
\end{split}
\end{equation}
and
\begin{equation}
\begin{split}
&\sum_{\bQ \in \mathcal{Q}_0} S^l_\bQ(\delta\theta, \bzero)e^{i\br \cdot D(\delta\theta)\bQ} \\
=&\sum_{\bQ \in \mathcal{Q}_0} S^l_\bQ(0, D(\delta\theta)\br) + O(\delta\theta^2).
\end{split}
\end{equation}
Taking $\br = D(\delta\theta)^{-1}\bd$ and then taking the limit as $\delta\theta \to 0$ with $\bd$ fixed, we find
\begin{equation}
\begin{split}
\sum_{\bQ \in \mathcal{Q}_+} T_\bQ(0, \bzero) e^{i\bd\cdot (\cos(\theta_0/2)\bK - \bQ)} &= \sum_{\bQ \in \mathcal{Q}_+} T_\bQ(0, \bd)\\
\sum_{\bQ \in \mathcal{Q}_0} S^l_\bQ(0, \bzero) e^{-i\bd\cdot \bQ} &= \sum_{\bQ \in \mathcal{Q}_0} S^l_\bQ(0, \bd).
\end{split}
\end{equation}
Taking $\delta\theta \to 0$ in \cref{eq:effective-inter-hamiltonian} then gives $\mathcal{H}_{\text{inter}}(\bp', \bp) = \mathcal{H}^0_{\text{inter}} \delta^2(\bp' - \bp)$ where
\begin{equation}\label{eq:effective-inter-hamiltonian-commensurate}
\mathcal{H}^0_{\text{inter}} = \begin{pmatrix}
S^+_0(\bd) & T_0(\bd)\\
T^\dagger_0(\bd) & S^-_0(\bd)
\end{pmatrix}\\
\end{equation}
and
\begin{equation}\label{eq:T_0-S^l_0}
\begin{split}
T_0(\bd) &= \sum_{\bQ \in \mathcal{Q}_+} T_\bQ(0, \bzero) e^{i\bd\cdot (\cos(\theta_0/2)\bK - \bQ)}\\
S^l_0(\bd) &= \sum_{\bQ \in \mathcal{Q}_0} S^l_\bQ(0, \bzero) e^{-i\bd\cdot \bQ}.
\end{split}
\end{equation}
We see that in the commensurate case, the continuum Hamiltonian describes four energy bands, approximating the bands nearest the Fermi level at charge neutrality.

Note that $T_0(\bd)$ and $S^l_0(\bd)$ are periodic (up to a phase) with respect to the lattice $L_-^0 + L_+^0 = L/\sqrt{N}$ which is reciprocal to $\mathcal{Q}_0$ (see \cref{sec:commensuration-lattices,sec:minimal-norm}). As a result, for $\theta = \theta_0$ the continuum Hamiltonian $\tilde{H}$ is periodic in $\bd$ (up to unitary equivalence) with respect to $L^0_- + L^0_+$. It is worthwhile to note that the microscopic Hamiltonian $H$ has the exact same periodicity in $\bd$ (see \cref{sec:equivalent-configurations}).

\section{Symmetry constraints and model parameters}\label{sec:symmetry-and-model-parameters}
We now consider the constraints that can be put on the TBG continuum model at twist angle $\theta=\theta_0+\delta\theta$ based on the symmetries of TBG, and explicitly determine the parameters of the TBG continuum model near various commensurate angles.

Note that the continuum model is fully determined by the $T_\bQ$ and $S^l_\bQ$ matrices with $\bd = \bzero$ in both the commensurate ($\delta\theta=0$) and incommensurate ($\delta\theta\neq0$) cases. We therefore make the assumption that $\bd = \bzero$ throughout this section. For $\theta \neq 0$, the valley preserving symmetries of the microscopic Hamiltonian $H$ are generated by the unitary operators $C_{3z}$ (rotation by $2\pi/3$ about $\bhatz$), $C_{2x}$ (rotation by $\pi$ about $\bhatx$), and the anti-unitary operator $C_{2z}\mathcal{T}$ (time-reversal followed by rotation by $\pi$ about $\bhatz$). The representations of these symmetry operators on the $\ket{\bk, l, \alpha}$ and $\ket{\bp, l, \alpha}_c$ states are given in \cref{sec:symmetry-representations}.

We require that $\tilde{H}$ commutes with these symmetry operators. $\tilde{H}_{\text{intra}}$ commutes with the symmetry operators identically so we need only consider $\tilde{H}_{\text{inter}}$. Assuming $\delta\theta \neq 0$, the symmetry constraint $[C_{2z}\mathcal{T}, \tilde{H}_{\text{inter}}] = 0$ is equivalent to
\begin{equation}\label{eq:C_2zT}
\sigma_x \overline{T_\bQ} \sigma_x = T_\bQ,\qquad\sigma_x \overline{S^l_\bQ} \sigma_x = S^l_{\bQ},
\end{equation}
$[C_{3z}, \tilde{H}_{\text{inter}}] = 0$ is equivalent to
\begin{equation}\label{eq:C_3z}
\begin{split}
e^{i(2\pi/3)\sigma_z} T_\bQ e^{-i(2\pi/3)\sigma_z} &= T_{R_{2\pi/3}\bQ}\\
e^{i(2\pi/3)\sigma_z} S^l_\bQ e^{-i(2\pi/3)\sigma_z} &= S^l_{R_{2\pi/3}\bQ},
\end{split}
\end{equation}
and $[C_{2x}, \tilde{H}_{\text{inter}}] = 0$ is equivalent to
\begin{equation}\label{eq:C_2x}
\begin{split}
\sigma_x T^\dagger_\bQ \sigma_x = T_{R^x\bQ},\qquad
\sigma_x S^{-l}_\bQ \sigma_x = S^l_{-R^x\bQ},
\end{split}
\end{equation}
where we use the notation $\overline{M}$ for the complex conjugate of a matrix $M$. By continuity, these equations also hold for $\delta\theta = 0$.

Since $\hat{t}_-(\bk)$ monotonically decreases with $|\bk|$, we expect that the magnitudes of $T_\bQ$ and $S^l_\bQ$ decay rapidly with $|\bQ|$. We therefore neglect $T_\bQ$ and $S^l_\bQ$ for all $\bQ$ with non-minimal norm. Recall that the elements of $\mathcal{Q}_+$ of minimal norm are $\bQ_1$, $\bQ_2$, and $\bQ_3$ which are given in \cref{eq:define-Q1-Q2-Q3}. The elements of $\mathcal{Q}_-$ with minimal norm are $-\bQ_1$, $-\bQ_2$, and $-\bQ_3$, and the only element of $\mathcal{Q}_0$ of minimal norm is $\bzero$. See \cref{fig:Q_lattices} for an illustration of the $\bQ_1$, $\bQ_2$, and $\bQ_3$ vectors.

By \cref{eq:T-S-hermiticity}, it suffices to determine the matrices $T_{\bQ_1}$, $T_{\bQ_2}$, $T_{\bQ_3}$, $S^+_\bzero$, $S^-_\bzero$ which correspond to minimal norm momenta. By expanding these matrices in the Pauli basis and applying \cref{eq:C_2zT,eq:C_3z,eq:C_2x} we find
\begin{equation}\label{eq:T-S-matrices}
\begin{split}
T_{\bQ_j}&= w_0 e^{i\chi_0 \sigma_z} +w_1\left(\sigma_x\cos\zeta_j + \sigma_y\sin\zeta_j\right),\\
S_\bzero^+ &= S_\bzero^- = w_2 \sigma_0
\end{split}
\end{equation}
for real model parameters $\chi_0$, $w_0$, $w_1$, and $w_2$ with $w_0 \geq 0$ and $\chi_0 \in [0, 2\pi)$. Here, we have used $\zeta_j=\frac{2\pi(j-1)}{3}$ for $j\in\{1, 2, 3\}$ and $\sigma_0$ for the $2\times2$ identity matrix. Note that the model parameters $\chi_0$, $w_0$, $w_1$, and $w_2$ depend on $\theta_0$ and $\delta\theta$ but not on $\bd$.

In the special case $\theta = 0$ (i.e. no twist) there is an additional valley preserving unitary mirror symmetry $M_y$ (reflection across the $xz$ plane). The symmetry constraint $[M_y, \tilde{H}_{\text{inter}}] = 0$ is equivalent to
\begin{equation}\label{eq:M_xz-T-S}
\sum_{\bQ \in \mathcal{Q}_+} [T_\bQ, \sigma_x] = \sum_{\bQ \in \mathcal{Q}_0} [S^l_\bQ, \sigma_x] = 0
\end{equation}
(see \cref{sec:symmetry-representations}). When $\theta = 0$, \cref{eq:M_xz-T-S} implies $\chi_0 = 0$. Therefore, if the twist angle is near $0$ (i.e. $\theta_0 = 0$, $\theta = \delta\theta \approx 0$) one will find $\chi_0\approx 0$ because of the approximate $M_y$ symmetry. This agrees with the Bistritzer-MacDonald model for small angle TBG \cite{Bistritzer2011}.

In \cref{sec:model-parameters}, we show that when $\delta\theta = 0$, the model parameters can be determined from numerical computations of the Hamiltonian block containing $\bK_\pm$ using \cref{eq:intra-matrix-element,eq:inter-matrix-element}. Additionally, \cref{sec:stacking,sec:model-parameters} show that the $\chi_0$, $w_0$, and $w_2$ parameters determine the band structure of $AA$ stacked commensurate configurations, while the $w_1$ and $w_2$ parameters determine the band structures of $AB$ and $BA$ stacked commensurate configurations. For numerical computations, we choose the $t_\pm(\br)$ functions in \cref{eq:Hamiltonian-matrix-elements} to be those used in Refs. \cite{Moon2013,Nam2017,Slater1954} and described in \cref{sec:t-functions}. \cref{tbl:w-parameter-table} shows approximate values of the model parameters derived from these functions for the first six commensurate configurations in order of the number of atoms per unit cell. Appendix \cref{tbl:accurate-parameters} lists these parameters with more significant figures. Appendix \cref{fig:relative-error} shows that the continuum models with parameters in Appendix \cref{tbl:accurate-parameters} are accurate low energy approximations of the microscopic Hamiltonian for all $\bd$ vectors. Additionally, Appendix \cref{fig:additional-commensurate-bandstructures} compares the band structures for each commensurate configuration in \cref{tbl:w-parameter-table} with the band structure derived from the microscopic Hamiltonian, and we see very good agreement. We note that we do not include any lattice relaxation or corrugation effects here in the microscopic model, nor do we include coupling mediated by higher graphene bands. Such effects may alter the true model parameters.

\begin{table}[h]
	\centering
	\bgroup
	\def\arraystretch{1.2}
	\begin{tabular}{|c|c|c|c|c|c|c|}
		\hline $(m, n)$ & $\theta_0$ & $N$ & $s$ & $\chi_0$ & $(w_0, w_1)$ & $w_2$\\
		\hline $(1, 0)$ & $0^{\circ}$ & $1$ & $1$ & $0.00^{\circ}$ & $(113, 113)\si{\milli\electronvolt}$ & $0.0\si{\milli\electronvolt}$\\
		\hline $(5, 3)$ & $38.2^{\circ}$ & $7$ & $1$ & $-3.10^{\circ}$ & $(959, 1050)\si{\micro\electronvolt}$ & $-4.44\si{\milli\electronvolt}$\\
		\hline $(7, 3)$ & $27.8^{\circ}$ & $13$ & $-1$ & $125^{\circ}$ & $(5.50, 3.62)\si{\micro\electronvolt}$ & $-4.43\si{\milli\electronvolt}$\\
		\hline $(4, 3)$ & $46.8^{\circ}$ & $19$ & $1$ & $-0.994^{\circ}$ & $(33.2, 33.2)\si{\micro\electronvolt}$ & $-4.32\si{\milli\electronvolt}$\\
		\hline $(11, 3)$ & $17.9^{\circ}$ & $31$ & $1$ & $1.24^{\circ}$ & $(653, 653)\si{\nano\electronvolt}$ & $-4.43\si{\milli\electronvolt}$\\
		\hline $(11, 9)$ & $50.6^{\circ}$ & $37$ & $1$ & $-0.862^{\circ}$ & $(1300, 1300)\si{\nano\electronvolt}$ & $-4.03\si{\milli\electronvolt}$\\
		\hline
	\end{tabular}
	\egroup
	\caption{Numerically determined model parameters reported with three significant figures. For the more accurate parameters used in \cref{fig:commensurate-band-structures,fig:velocity-and-bandwidth,fig:moire-band-structures,fig:relative-error,fig:additional-commensurate-bandstructures,fig:additional-moire-band-structures-near-commensuration}, see Appendix \cref{tbl:accurate-parameters}.}
	\label{tbl:w-parameter-table}
\end{table}

\section{Commensurate models: band structures}\label{sec:commensurate-bands}
\begin{figure}
	\centering
	\includegraphics{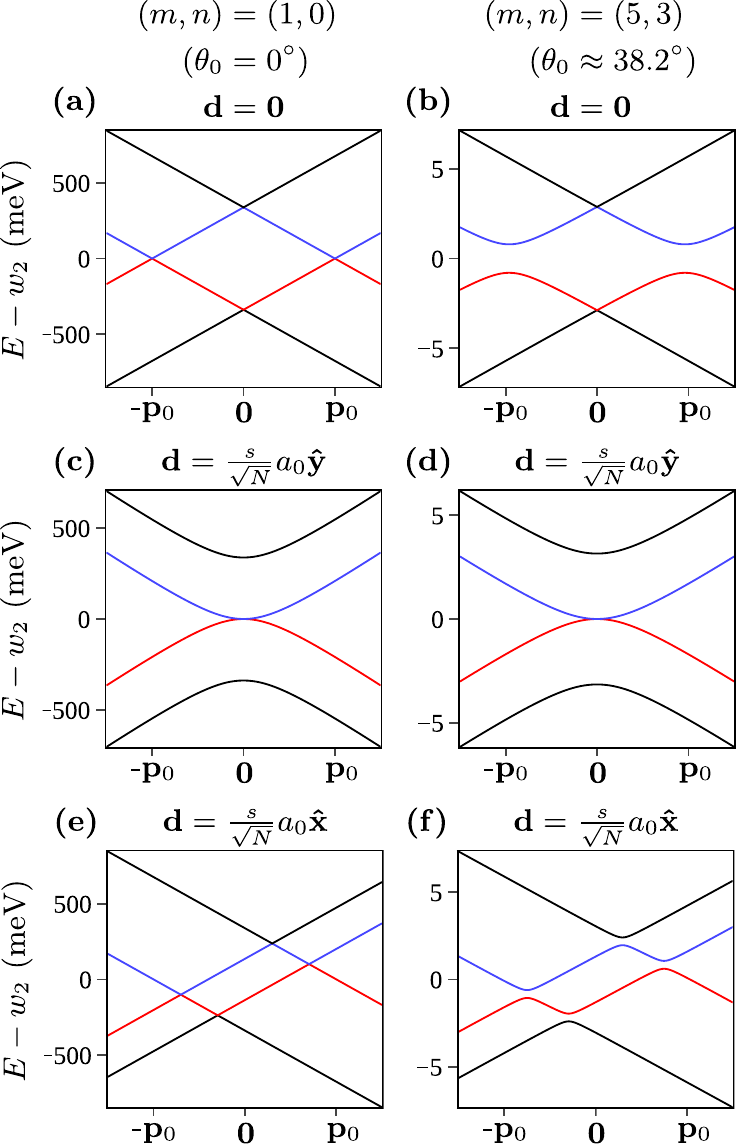}
	\caption{Commensurate band structures using the model in \cref{eq:commensurate-continuum-model,eq:truncated-T_0} with parameters in Appendix \cref{tbl:accurate-parameters}. The vector $\bp$ ranges linearly from $-3\bp_0/2$ to $3\bp_0/2$ where $\hbar v_F \bp_0 = 3|w_0|\bhatx$. The first and second columns correspond to commensurate configurations with $(m, n) = (1, 0)$ ($\theta_0=0^\circ$) and $(m, n) = (5, 3)$ ($\theta_0\approx38.2^\circ$), and the first and second rows correspond to $AA$ and $AB$ stackings, respectively (see \cref{sec:stacking}).}
	\label{fig:commensurate-band-structures}
\end{figure}

By \cref{eq:effective-intra-hamiltonian,eq:effective-inter-hamiltonian-commensurate,eq:T_0-S^l_0}, the continuum model corresponding to commensurate twist angle $\theta_0$ and translation vector $\bd$ is $\tilde{H} = \int d^2\bp \boldsymbol{\ket{\bp}_c} \mathcal{H}_0(\bp) \boldsymbol{\bra{\bp}_c}$, where the explicit Hamiltonian matrix is
\begin{align}
\mathcal{H}_0(\bp) &= w_2 I + \begin{pmatrix}
\hbar v_F \bsigma_{\theta_0/2} \cdot \bp & T_0(\bd)\\
T_0^\dagger(\bd) & \hbar v_F \bsigma_{-\theta_0/2} \cdot \bp
\end{pmatrix}\label{eq:commensurate-continuum-model}\\
T_0(\bd) &= \sum_{j=1}^3 T_{\bQ_j}e^{i\bd \cdot (\cos(\theta_0/2)\bK -\bQ_j)}.\label{eq:truncated-T_0}
\end{align}
The matrices $T_{\bQ_j}$ are given in \cref{eq:T-S-matrices} and $I$ is the $4\times4$ identity matrix. Recall that $\bsigma_\phi$ is the Pauli matrix vector defined in \cref{eq:define-bsigma}, $\bQ_j$ is defined in \cref{eq:define-Q1-Q2-Q3}, and the momentum space basis $\boldsymbol{\ket{\bp}_c}$ is defined in \cref{eq:define-p-row-vector}. Using \cref{eq:continuum-hamiltonian-real-space,eq:potentials-real-space} we can also describe this model in real space as $\tilde{H} = \int d^2\br \boldsymbol{\ket{\br}_c} \mathcal{H}_0(\br) \boldsymbol{\bra{\br}_c}$, where the Hamiltonian matrix takes the form
\begin{equation}
\mathcal{H}_0(\br) = w_2 I + \begin{pmatrix}
-i\hbar v_F \bsigma_{\theta_0/2}\cdot \nabla & T_0(\bd)\\
T^\dagger_0(\bd) & -i\hbar v_F \bsigma_{-\theta_0/2}\cdot \nabla
\end{pmatrix}
\end{equation}
and the real space basis $\boldsymbol{\ket{\br}_c}$ is defined in \cref{eq:define-r-row-vector}.

\cref{fig:commensurate-band-structures} shows the low energy band structures of the model in \cref{eq:commensurate-continuum-model,eq:truncated-T_0} for the first two commensurate configurations in \cref{tbl:w-parameter-table}, namely $(m, n) = (1, 0)$ (the untwisted configuration with $\theta_0 = 0$) and $(m, n) = (5, 3)$ ($\theta_0 \approx 38.2^\circ$). For both configurations, we show three translation vectors $\bd$, and use the parameters in Appendix \cref{tbl:accurate-parameters}. Similar band structures for the other commensurate configurations in \cref{tbl:w-parameter-table} are shown in Appendix \cref{fig:additional-commensurate-bandstructures}. We compare the band structures of untwisted bilayer graphene and commensurate TBG in the following cases:
\begin{enumerate}
\item At $AA$ stacking where $\bd=\bzero$. In this case, untwisted bilayer graphene is gapless at momentum $|\bp|=|\bp_0|=3|w_0|/(\hbar v_F)$ at charge neutrality as in \cref{fig:commensurate-band-structures}\textbf{(a)}. In contrast, commensurate TBG develops a gap at $|\bp|=|\bp_0|$ at charge neutrality as in \cref{fig:commensurate-band-structures}\textbf{(b)}, due to the relative rotation angle between the Dirac fermions in different layers and the nonzero value of $\chi_0$. Specifically, the gap at $|\bp| = |\bp_0|$ is given in general by
\begin{equation}\label{eq:comm-gap}
12|w_0|\min(|\cos(\phi_0/2)|, |\sin(\phi_0/2)|)
\end{equation}
where $\phi_0 = \chi_0 + \theta_0/2$. In the $\theta_0 \approx 38.2^\circ$ commensurate configuration, the charge neutrality gap in \cref{fig:commensurate-band-structures}\textbf{(b)} is approximately $\SI{1.6}{\milli\electronvolt}$, which should be experimentally measurable.

\item At $AB$ stacking where $\bd=\frac{s}{\sqrt{N}}a_0\hat{\by}$ (recall that $s=\pm1$ was introduced in \cref{eq:Q-lattices-form}). In this case, both untwisted (Bernal) bilayer graphene, shown in \cref{fig:commensurate-band-structures}\textbf{(c)}, and commensurate TBG, shown in \cref{fig:commensurate-band-structures}\textbf{(d)}, have gapless quadratic Dirac band touchings \cite{McCann2006} at charge neutrality.

\item At generic asymmetric stackings such as $\bd=\frac{s}{\sqrt{N}}a_0\hat{\bx}$. Untwisted bilayer graphene remains gapless as in \cref{fig:commensurate-band-structures}\textbf{(e)}. In contrast, commensurate TBG has a tilted band gap at charge neutrality as in \cref{fig:commensurate-band-structures}\textbf{(f)}, but there may not be an indirect gap.
\end{enumerate}

Although the above observations are made at exactly commensurate angles, they may also hold for local measurements (e.g. scanning tunneling microscopy experiments) near the corresponding stackings if the angle $\theta$ is close enough to a commensurate angle $\theta_0$. In particular, when $\theta_0$ is significantly far from zero, one expects to observe a local charge neutrality gap at $AA$ stacking positions (e.g. a $\SI{1.6}{\milli\electronvolt}$ gap at $\theta_0 \approx 38.2^\circ$). However, we note that the local charge neutrality at $AA$ stacking is generically different from the global charge neutrality of an incommensurate angle, due to local charge transfers between $AA$ stacking regions and $AB$ stacking regions. This can be seen in \cref{fig:moire-band-structures}\textbf{(c)}, by noting that the moir\'e bands at global charge neutrality are close to the conduction band energy at $AA$ stacking in \cref{fig:commensurate-band-structures}\textbf{(b)}.

\section{Continuum models near commensuration: moir\'e band structures and magic angles}\label{sec:realistic-moire}
\begin{figure}
	\centering
	\includegraphics{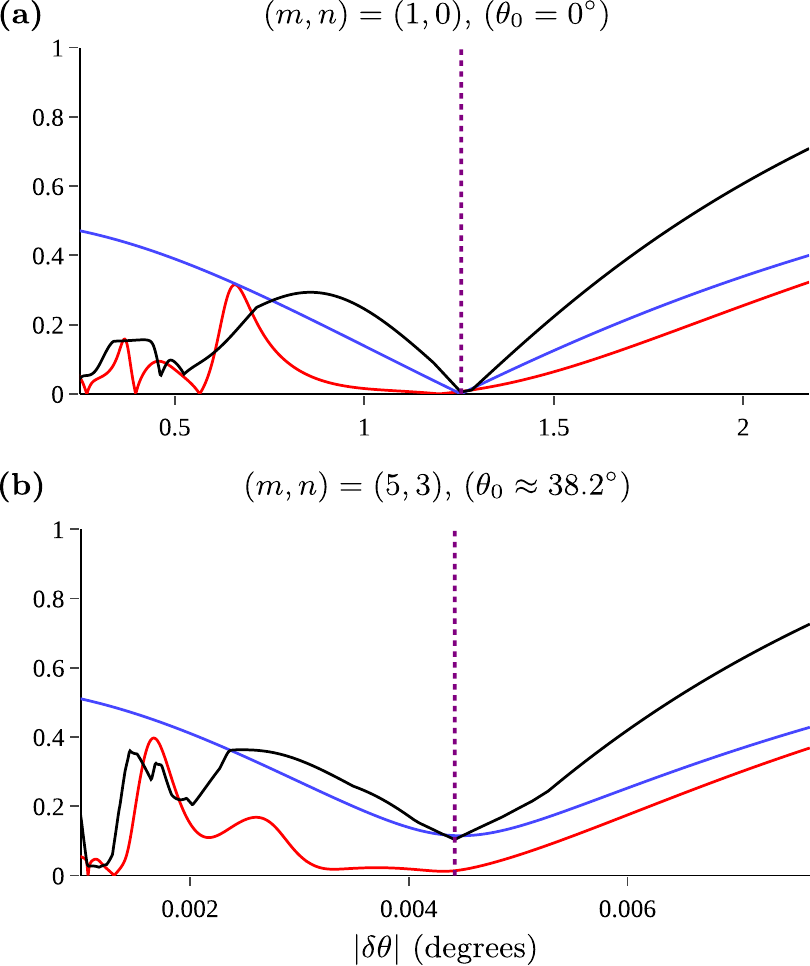}
	\caption{The red, blue, and black curves show properties of the spectrum of the continuum Hamiltonian in \cref{eq:continuum-hamiltonian-complete} as a function of $\delta\theta$. The red and blue curves show $v_M / v_F$ ($v_M$ is the Dirac velocity at the $\bK_M$ point at charge neutrality), while the black curve shows the bandwidth (in units of $\hbar v_F |\bK_M|$) of the two lowest bands at charge neutrality. The purple dashed lines indicate $\delta\theta = \delta\theta_{\text{magic}}$, at which point the bandwidth is minimized. The blue curves use the $8$ band tripod model analyzed in \cref{sec:tripod-analysis} while the red and black curves use the more accurate $768$ band model illustrated in Appendix \cref{fig:moire-hopping-lattice}. The bandwidth shown in the black curve is the difference between the highest conduction energy and the lowest valence energy among the points $\bGamma_M, \bK_M, \bM_M, \bK_M/2, \bM_M/2, -\bM_M/2$ in $\text{BZ}_M$. Panels \textbf{(a)} and \textbf{(b)} correspond to the commensurate configurations with $(m, n) = (1, 0)$ ($\theta_0=0^\circ$) and $(m, n) = (5, 3)$ ($\theta_0\approx38.2^\circ$), respectively and use the parameters in Appendix \cref{tbl:accurate-parameters}.}
	\label{fig:velocity-and-bandwidth}
\end{figure}

\begin{figure*}
	\centering
	\includegraphics{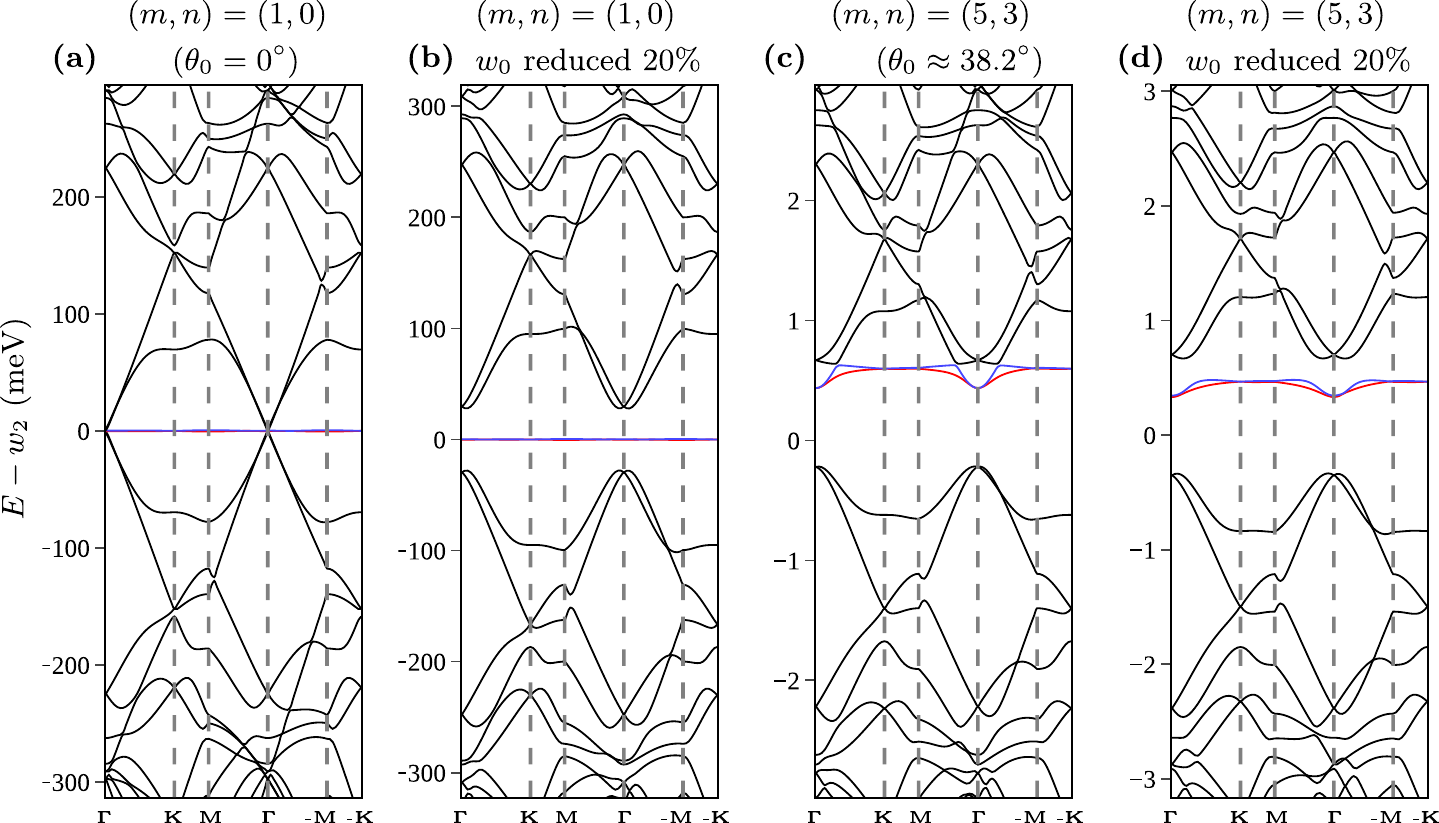}
	\caption{Moir\'e band structures using the model in \cref{eq:continuum-hamiltonian-complete} with $\delta\theta =\theta-\theta_0= \delta\theta_{\text{magic}}$ (where the bandwidth is minimal in \cref{fig:velocity-and-bandwidth}) and the quasi-momentum truncation illustrated in Appendix \cref{fig:moire-hopping-lattice}. The horizontal axes follow the moir\'e quasi-momentum trajectory $\bGamma_M \to \bK_M \to \bM_M \to \bGamma_M \to -\bM_M \to -\bK_M$. The two bands nearest charge neutrality are shown in blue and red while all other bands are shown in black. Panels \textbf{(a)} and \textbf{(b)} correspond to the commensurate configurations with $(m, n) = (1, 0)$ ($\theta_0=0^\circ$) and $(m, n) = (5, 3)$ ($\theta_0\approx38.2^\circ$), respectively and use the parameters in Appendix \cref{tbl:accurate-parameters}. The parameters for panels \textbf{(b)} and \textbf{(d)} are the same as those for \textbf{(a)} and \textbf{(c)} except with the values of $w_0$ reduced by $20\%$. Similar plots for the other commensurate configurations in \cref{tbl:w-parameter-table} are shown in Appendix \cref{fig:additional-moire-band-structures-near-commensuration}.}
	\label{fig:moire-band-structures}
\end{figure*}

The continuum model corresponding to twist angle $\theta = \theta_0 + \delta\theta$ and translation vector $\bd = \bzero$ is described by \cref{eq:effective-intra-hamiltonian,eq:effective-inter-hamiltonian,eq:T-S-matrices}. Note that when $\delta\theta \neq 0$, the microscopic Hamiltonians for different choices of translation vector $\bd$ differ only by a unitary transformation (see \cref{sec:equivalent-configurations}) so it is sufficient to consider the case $\bd = \bzero$. In this section, we further develop the continuum model Hamiltonian and investigate its moir\'e band structures and magic angles using the parameters determined in \cref{sec:symmetry-and-model-parameters}.

Since $\delta\theta$ is small, we approximate the rotation angles $\pm\theta/2$ of the Dirac cones in \cref{eq:effective-intra-hamiltonian} by $\pm\theta_0/2$. This is a common approximation in the literature \cite{Bistritzer2011}. Additionally, we approximate the $\chi_0$, $w_0$, $w_1$, and $w_2$ parameters by their values at angle $\theta_0$ (i.e. with $\delta\theta = 0$), which can be determined using the method described in \cref{sec:symmetry-and-model-parameters}. The continuum model then becomes $\tilde{H} = \int d^2\bp' d^2\bp \boldsymbol{\ket{\bp'}_c} \mathcal{H}(\bp', \bp) \boldsymbol{\bra{\bp}_c}$, where the Hamiltonian matrix is
\begin{equation}\label{eq:continuum-hamiltonian-complete}
\begin{split}
\mathcal{H}(\bp', \bp)& = w_2 I \delta^2(\bp' -\bp)\\
&+ \hbar v_F\begin{pmatrix}
\bsigma_{\theta_0/2} \cdot \bp & 0\\
0 & \bsigma_{-\theta_0/2}\cdot \bp \\
\end{pmatrix}\delta^2(\bp'-\bp)\\
&+ \sum_{j=1}^3 \begin{pmatrix}
0 & T_{\bQ_j}\\
0 & 0
\end{pmatrix}\delta^2(\bp' -\bp - \bq_j)\\
&+ \sum_{j=1}^3 \begin{pmatrix}
0 & 0\\
T_{\bQ_j}^\dagger & 0
\end{pmatrix}\delta^2(\bp' -\bp + \bq_j)\\
\end{split}
\end{equation}
and the matrices $T_{\bQ_j}$ are defined in \cref{eq:T-S-matrices}. Recall that $\bsigma_\phi$ is the Pauli matrix vector defined in \cref{eq:define-bsigma}, $\bq_j$ is defined in \cref{eq:define-q_j}, and the momentum space basis $\boldsymbol{\ket{\bp}_c}$ is defined in \cref{eq:define-p-row-vector}.  Note that $w_2$ only provides a constant energy shift.

Using \cref{eq:continuum-hamiltonian-real-space,eq:potentials-real-space} we can also describe this model in real space by $\tilde{H} = \int d^2\br \boldsymbol{\ket{\br}_c} \mathcal{H}(\br) \boldsymbol{\bra{\br}_c}$, where
\begin{equation}
\begin{split}
\mathcal{H}(\br) &= w_2 I + \begin{pmatrix}
-i\hbar v_F \bsigma_{\theta_0/2} \cdot \nabla & T(\br)\\
T^\dagger(\br) & -i\hbar v_F \bsigma_{-\theta_0/2} \cdot \nabla
\end{pmatrix},\\
T(\br) &= \sum_{j=1}^3 T_{\bQ_j} e^{i \br \cdot \bq_j},
\end{split}
\end{equation}
and the real space basis $\boldsymbol{\ket{\br}_c}$ is defined in \cref{eq:define-r-row-vector}.

Following Refs. \cite{Bistritzer2011, Tarnopolsky2019}, we introduce the dimensionless parameter
\begin{equation}\label{eq:alpha-def}
\alpha = \frac{|w_1|}{\hbar v_F |\bK_M|} = \frac{|w_1|}{2|\sin(\delta\theta/2)|\hbar v_F \sqrt{N} |\bK|}.
\end{equation}
Recall that $4N$ is the number of atoms in each commensurate unit cell at twist angle $\theta_0$. Note that $\alpha^{-1}\propto|\delta \theta|$ when $\delta\theta$ is small.

As a first step in the search for magic angles, we cut off the continuum model in \cref{eq:continuum-hamiltonian-complete} to a subspace of four quasi-momenta, namely $\bp$ and $\bp-\bq_j$ for $j \in \{1, 2, 3\}$. This truncation is known as the tripod model approximation \cite{Bistritzer2011,Bernevig2021a} and it yields an approximate $\bk\cdot\bp$ model at the $\bK_M$ point at charge neutrality. Generically, the lowest bands of this model have a Dirac fermion spectrum with Fermi velocity $v_M$. In this tripod model approximation, it can be shown (see \cref{sec:tripod-analysis}) that the velocity $v_M$ reaches its minimum (which is generically nonzero unless $\theta_0=0$) near
\begin{equation}\label{eq:alpha-magic}
\alpha^{-1}\approx \sqrt{3},
\end{equation}
given that the energy $E$ at the $\bK_M$ point satisfies $\frac{|E-w_2|}{\hbar v_F|\bK_M|}\ll 1$. Note that the energy $E-w_2$ at the $\bK_M$ point is generically nonzero when $\theta_0$ is nonzero. It is also known that the magic angle condition in \cref{eq:alpha-magic} generically requires $w_0\leq |w_1|$ to avoid hybridization with the remote bands \cite{Bernevig2021a}, and this is also true here (see \cref{fig:small-phi0-moire-band-structures} for examples illustrating this point). By \cref{eq:alpha-def}, we conclude that the first magic angle occurs at
\begin{equation}\label{eq:theta-magic}
\delta\theta=\delta\theta_{\text{magic}}\approx \pm \frac{\sqrt{3}w_1}{\hbar v_F \sqrt{N} |\bK|}\ .
\end{equation}
The tripod model approximation, however, does not give the higher (i.e. second, third, etc.) magic angles.

\cref{fig:velocity-and-bandwidth}\textbf{(a)} and \textbf{(b)} show numerical results for Dirac velocities $v_M$ and the bandwidth of the lowest two moir\'e bands at charge neutrality, near the commensurate configurations with $(m, n) = (1, 0)$ ($\theta_0=0^\circ$) and $(m, n) = (5, 3)$ ($\theta_0\approx38.2^\circ$), respectively. The blue curves show $v_M/v_F$ values computed from the tripod model, and have a minimum around the angle in \cref{eq:theta-magic}. The red curves show the accurate $v_M/v_F$ values computed using $768$ moir\'e bands (see \cref{sec:wilson-and-truncation} and \cref{fig:moire-hopping-lattice}). In both cases, the value of $v_M/v_F$ is computed by numerical differentiation in the $\bq_1$ direction at $\bK_M$. Intriguingly, at $\theta_0 \approx 38.2^\circ$, the accurate Fermi velocity $v_M$ at the first magic angle $\delta\theta_{\text{magic}}$ is almost zero and much smaller than that found in the tripod model. The black curves show the total bandwidth (in units of $\hbar v_F|\bK_M|$) of the lowest two bands at charge neutrality using $768$ moir\'e bands. From the accurate $v_M/v_F$ (red) and bandwidth (black) curves, we clearly see the first magic angle around the value in \cref{eq:theta-magic}. There are higher (i.e. smaller) magic angles near $\theta_0 \approx 38.2^\circ$ as well, where the lowest two bands become flat.

\cref{fig:moire-band-structures}\textbf{(a)} and \textbf{(c)} show the moir\'e band structures at the first magic angle $\delta\theta=\delta\theta_{\text{magic}}$ the commensurate configurations with $(m, n) = (1, 0)$ ($\theta_0=0^\circ$) and $(m, n) = (5, 3)$ ($\theta_0\approx38.2^\circ$), respectively. The band structure with $\theta_0=0^\circ$ shows the usual magic angle moir\'e bands of small angle TBG studied in \cite{Bistritzer2011}. At $\theta_0 \approx 38.2^\circ$, the band structure is clearly not symmetric across the Fermi level, indicating the absence of both particle-hole symmetry $P$ \cite{song_all_2019,song2021TBGII,bultinck_ground_2020} and chiral symmetry $C$ \cite{Tarnopolsky2019} (see definitions in \cref{sec:parameter-signs}). The lowest two moir\'e bands at charge neutrality are still approximately flat near the $\bK_M$ and $-\bK_M$ points, and are energetically shifted close to a remote conduction band. The two bands are however not quite flat near the $\bGamma_M$ point.

It is known that in small angle TBG, lattice relaxation has the effect of slightly reducing the value of $w_0$ \cite{Nam2017,koshino_maximally_2018,Carr2019}. Although we do not here consider relaxation from first principles, it is nonetheless worthwhile to consider the effect of a reduction in $w_0$ on the moir\'e band structure. \cref{fig:moire-band-structures}\textbf{(b)} and \textbf{(d)} show moir\'e band structures using the same parameters as in \cref{fig:moire-band-structures}\textbf{(a)} and \textbf{(c)}, but with $w_0$ reduced by $20\%$. In both cases, we see that the two lowest bands at charge neutrality develop a gap from the higher bands, but are otherwise qualitatively similar. Moir\'e band structures at the first magic angle in \cref{eq:theta-magic} near the other commensurate configurations listed in \cref{tbl:w-parameter-table} are shown in Appendix \cref{fig:additional-moire-band-structures-near-commensuration}. Additionally, other example moir\'e band structures near the first magic angle can be found in \cref{fig:bandstructure-hypermagic,fig:small-phi0-moire-band-structures}.

Appendix \cref{tbl:accurate-parameters} shows the values of $\delta\theta_{\text{magic}}$ for the first six commensurate configurations. Due to the small magnitude of $w_0$ and $w_1$ for nonzero commensurate angles, the corresponding values of $\delta\theta_{\text{magic}}$ are so small that they likely cannot be achieved experimentally. However, we note the possibility that atomic structural reconstructions (e.g. charge density wave orders) may occur in large twist angle TBG and enhance the effective interlayer hoppings $w_0$ and $w_1$. Additionally, lattice relaxation or corrugation or couplings mediated by higher graphene bands could also change these parameters. Provided these perturbations do not break the symmetries of the moir\'e superlattice (translation, $C_{3z}$, $C_{2z}\mathcal{T}$, and $C_{2x}$), the form of effective continuum model will not change, and we may arrive at larger first magic angles in nearly commensurate TBG.

\section{Flat bands in the continuum model parameter space: the hypermagic regime}\label{sec:hypermagic}
Regarding the possibility that the actual model parameters may change due to atomic structural reconstruction, lattice relaxation or corrugation, or couplings mediated by higher graphene bands, we now investigate the band structure of the TBG continuum model near commensuration in \cref{eq:continuum-hamiltonian-complete} with arbitrary parameters. We reveal the existence of a remarkable \emph{hypermagic regime} centered at $\phi_0 = \pm\pi/2$ where many moir\'e bands (often $8$ or more) become extremely flat simultaneously.

\subsection{Model simplification}
We first simplify the continuum model in \cref{eq:continuum-hamiltonian-complete} by applying a unitary transformation of the basis from $\boldsymbol{\ket{\bp}_c}$ to $\boldsymbol{\ket{\bp}'_c}=\boldsymbol{\ket{\bp}_c} U_{\theta_0}$, where 
\begin{equation}\label{eq:rotation-transform}
\begin{split}
U_{\theta_0} &= \begin{pmatrix}
e^{-i(\theta_0/4)\sigma_z} & 0\\
0 & e^{i(\theta_0/4)\sigma_z}
\end{pmatrix}.
\end{split}
\end{equation}
Such a transformation removes the rotation angles $\pm\theta_0/2$ for the Dirac cones, and transforms the Hamiltonian into $\tilde{H} = \int d^2\bp' d^2\bp \boldsymbol{\ket{\bp'}'_c} \mathcal{H}'(\bp', \bp) \boldsymbol{\bra{\bp}'_c}$, where the Hamiltonian matrix is given by
\begin{equation}\label{eq:continuum-hamiltonian-rotate-basis}
\begin{split}
\mathcal{H}'(\bp', \bp)& = w_2 I \delta^2(\bp' -\bp)\\
&+ \hbar v_F\begin{pmatrix}
\bsigma \cdot \bp & 0\\
0 & \bsigma\cdot \bp \\
\end{pmatrix}\delta^2(\bp'-\bp)\\
&+ \sum_{j=1}^3 \begin{pmatrix}
0 & T'_{\bQ_j}\\
0 & 0
\end{pmatrix}\delta^2(\bp' -\bp - \bq_j)\\
&+ \sum_{j=1}^3 \begin{pmatrix}
0 & 0\\
T^{\prime\dagger}_{\bQ_j} & 0
\end{pmatrix}\delta^2(\bp' -\bp + \bq_j).
\end{split}
\end{equation}
Here, $\bsigma = \sigma_x \bhatx + \sigma_y \bhaty$ is a vector of Pauli matrices, and
\begin{equation}
\begin{split}
T'_{\bQ_j} &= e^{i(\theta_0/4)\sigma_z}T_{\bQ_j}e^{i(\theta_0/4)\sigma_z}.
\end{split}
\end{equation}
More explicitly,
\begin{equation}\label{eq:define-T'-matrices}
\begin{split}
T'_{\bQ_j}= w_0 e^{i\phi_0 \sigma_z} +w_1\left(\sigma_x\cos\zeta_j + \sigma_y\sin\zeta_j\right),
\end{split}
\end{equation}
where $\zeta_j=\frac{2\pi(j-1)}{3}$ for $j \in \{1, 2, 3\}$, and we have defined
\begin{equation}\label{eq:def-phi0}
\phi_0 = \chi_0 + \frac{\theta_0}{2}.
\end{equation}
This implies that the angles $\chi_0$ and $\theta_0$ do not have fully independent effects on the band structure. We are left with a single angle variable $\phi_0$ in the continuum model of \cref{eq:continuum-hamiltonian-rotate-basis}, occurring in the matrices $T'_{\bQ_j}$ in \cref{eq:define-T'-matrices}. We note that the angle $\phi_0$ in \cref{eq:def-phi0} also occurs in the expression for the energy gap in the commensurate $AA$ stacking configuration in \cref{eq:comm-gap}. This can also be understood via the transformation in \cref{eq:rotation-transform}.

The model can similarly be written in the transformed real space basis
\begin{equation}\label{eq:transformed-real-space-basis}
\boldsymbol{\ket{\br}'_c} = \boldsymbol{\ket{\br}_c}U_{\theta_0}.
\end{equation}
The Hamiltonian then becomes $\tilde{H} = \int d^2\br \boldsymbol{\ket{\br}'_c} \mathcal{H}'(\br) \boldsymbol{\bra{\br}'_c}$, where the Hamiltonian matrix is given by
\begin{equation}
\mathcal{H}'(\br) = w_2 I + \begin{pmatrix}
-i\hbar v_F \bsigma \cdot \nabla & T'(\br)\\
T^{\prime\dagger}(\br) & -i\hbar v_F \bsigma \cdot \nabla
\end{pmatrix},
\end{equation}
and where we have defined
\begin{equation}
\begin{split}
T'(\br) &= \sum_{j=1}^3 T'_{\bQ_j} e^{i \br \cdot \bq_j}
\end{split}
\end{equation}
in terms of the matrices $T'_{\bQ_j}$ in \cref{eq:define-T'-matrices}.

By the results of \cref{sec:parameter-signs}, we can assume without loss of generality that $s = 1$ (recall that $s$ affects the direction of $\bq_j$, see \cref{eq:define-Q1-Q2-Q3,eq:define-q_j}), and
\begin{equation}
\phi_0\in\left[0, \frac{\pi}{2}\right],\quad w_0\geq0,\quad w_1\geq0,\quad \delta\theta \geq 0.
\end{equation}
In addition, since $w_2$ simply shifts the energy bands globally, we assume $w_2 = 0$ hereafter. As shown in \cref{sec:parameter-signs}, the moir\'e band structures at angle $\phi_0$ and angle $-\phi_0$ are particle-hole transformations of each other, while the moir\'e band structures at angle $\phi_0$ and angle $\pi-\phi_0$ are equivalent.

We note that in the chiral limit $w_0=0$ \cite{Tarnopolsky2019}, the continuum model in \cref{eq:continuum-hamiltonian-rotate-basis} is independent of the angle $\phi_0$. This is revealed as a symmetry of the TBG continuum model in the chiral limit in Ref. \cite{Wang2021}.

\begin{figure}
	\centering
	\includegraphics{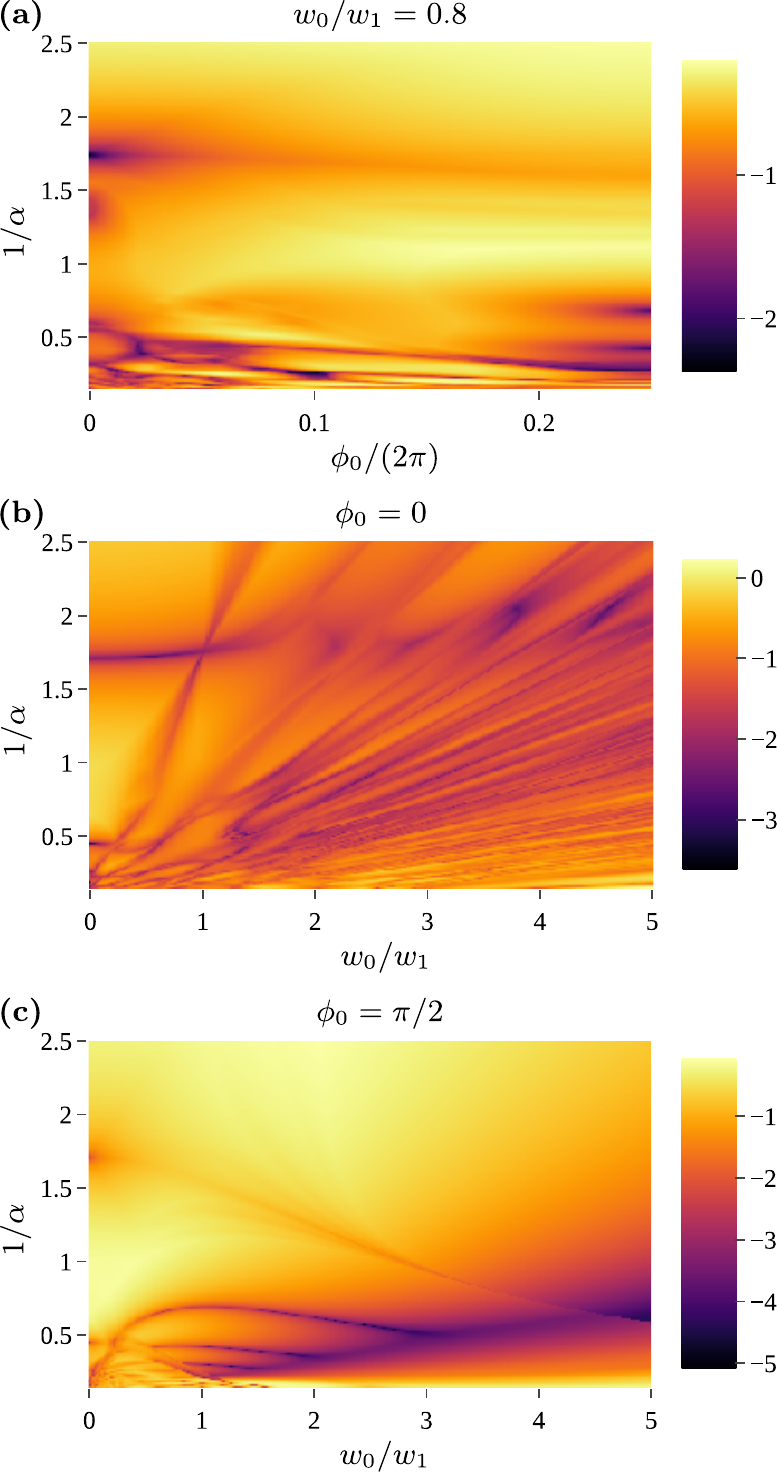}
	\caption{Heatmaps showing the base $10$ logarithm of the bandwidth (in units of $\hbar v_F |\bK_M|$) of the two bands nearest the Fermi level at charge neutrality. As in \cref{fig:velocity-and-bandwidth}, the bandwidth was computed as the largest difference between a conduction energy and a valence energy in the lowest two bands at charge neutrality among the points $\bGamma_M, \bK_M, \bM_M, \bK_M/2, \bM_M/2, -\bM_M/2$ in $\text{BZ}_M$. For this computation, we use the model in \cref{eq:continuum-hamiltonian-rotate-basis} with the quasi-momentum truncation illustrated in Appendix \cref{fig:moire-hopping-lattice}. Panel \textbf{(a)} shows the logarithm of the bandwidth as a function of $\alpha^{-1}$ (defined in \cref{eq:alpha-def}) and $\phi_0/(2\pi)$ while $w_0/w_1$ is fixed at $0.8$. The nearly horizontal dark curve near $\alpha^{-1} = \sqrt{3}$ is part of the first magic manifold (see \cref{sec:first-magic-manifold}). Panels \textbf{(b)} and \textbf{(c)} show the logarithm of the bandwidth as a function of $\alpha^{-1}$ and $w_0/w_1$ while $\phi_0$ is fixed at $0$ and $\pi/2$, respectively. In panel \textbf{(b)}, the nearly horizontal dark curve at $\alpha^{-1}\approx\sqrt{3}$ and $0\leq w_0/w_1\lesssim 1$ corresponds to the first magic manifold of small angle TBG. In panel \textbf{(c)}, the three nearly horizontal dark curves around $\alpha^{-1}= 0.7$, $0.4$, and $0.3$ contain many simultaneous flat bands and are part of the hypermagic regime discussed in \cref{sec:hypermagic-subsection}.}
	\label{fig:heatmap-middle-bands}
\end{figure}

\begin{figure}
	\centering
	\includegraphics{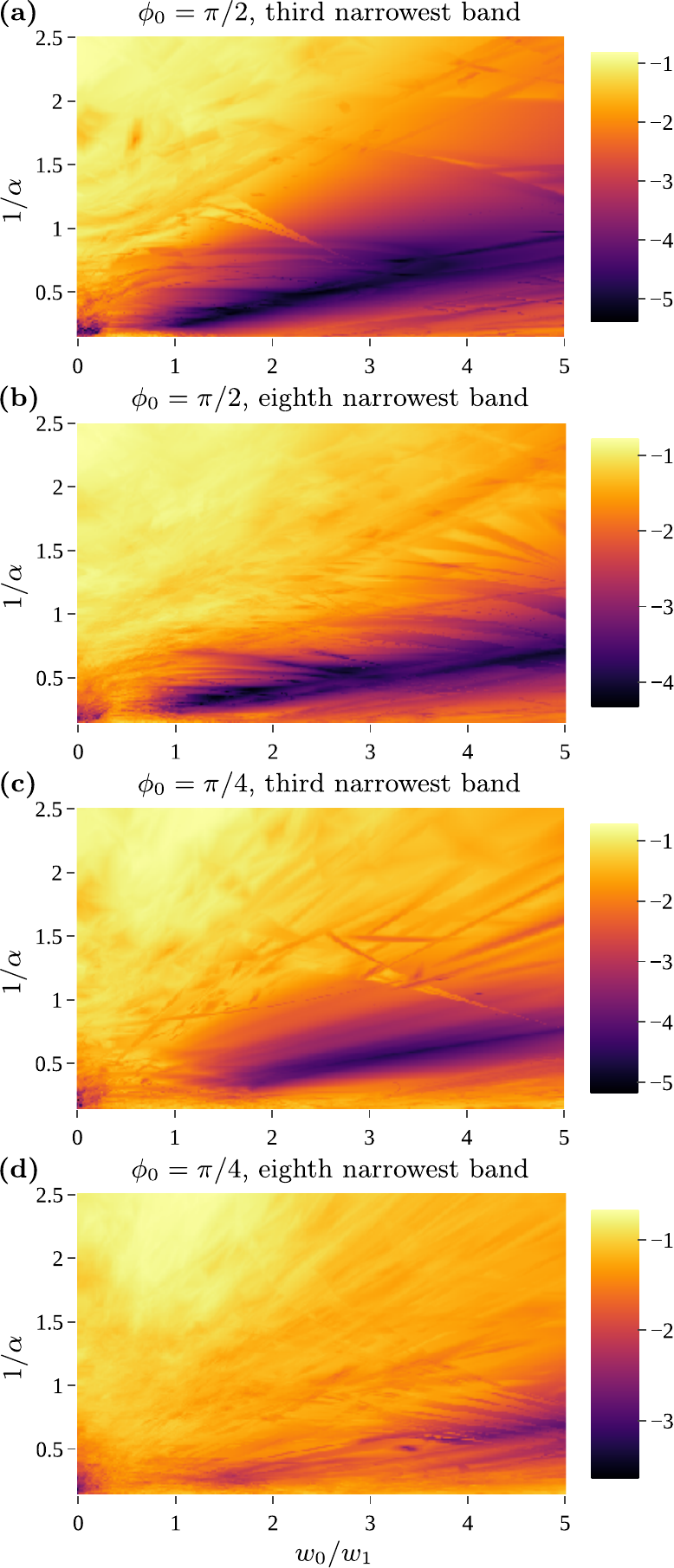}
	\caption{Heatmaps showing the base $10$ logarithm of the bandwidth (in units of $\hbar v_F |\bK_M|$) of the third and eighth narrowest bands among the first $20$ conduction bands and the first $20$ valence bands at charge neutrality for $\phi_0 = \pi/2$ and $\pi/4$. The bandwidth was computed with the points $\bGamma_M, \bK_M, \bM_M, \bK_M/2, \bM_M/2, -\bM_M/2$ in $\text{BZ}_M$. For this computation, we use the model in \cref{eq:continuum-hamiltonian-rotate-basis} with the quasi-momentum truncation illustrated in Appendix \cref{fig:moire-hopping-lattice}. The dark regions indicate parameters in the hypermagic regime discussed in \cref{sec:hypermagic-subsection}. See Appendix \cref{fig:additional-heatmaps} for similar heatmaps with $\phi_0 = 0$, $\pi/8$, and $3\pi/8$.}
	\label{fig:heatmap-hypermagic}
\end{figure}

\begin{figure*}
	\centering
	\includegraphics{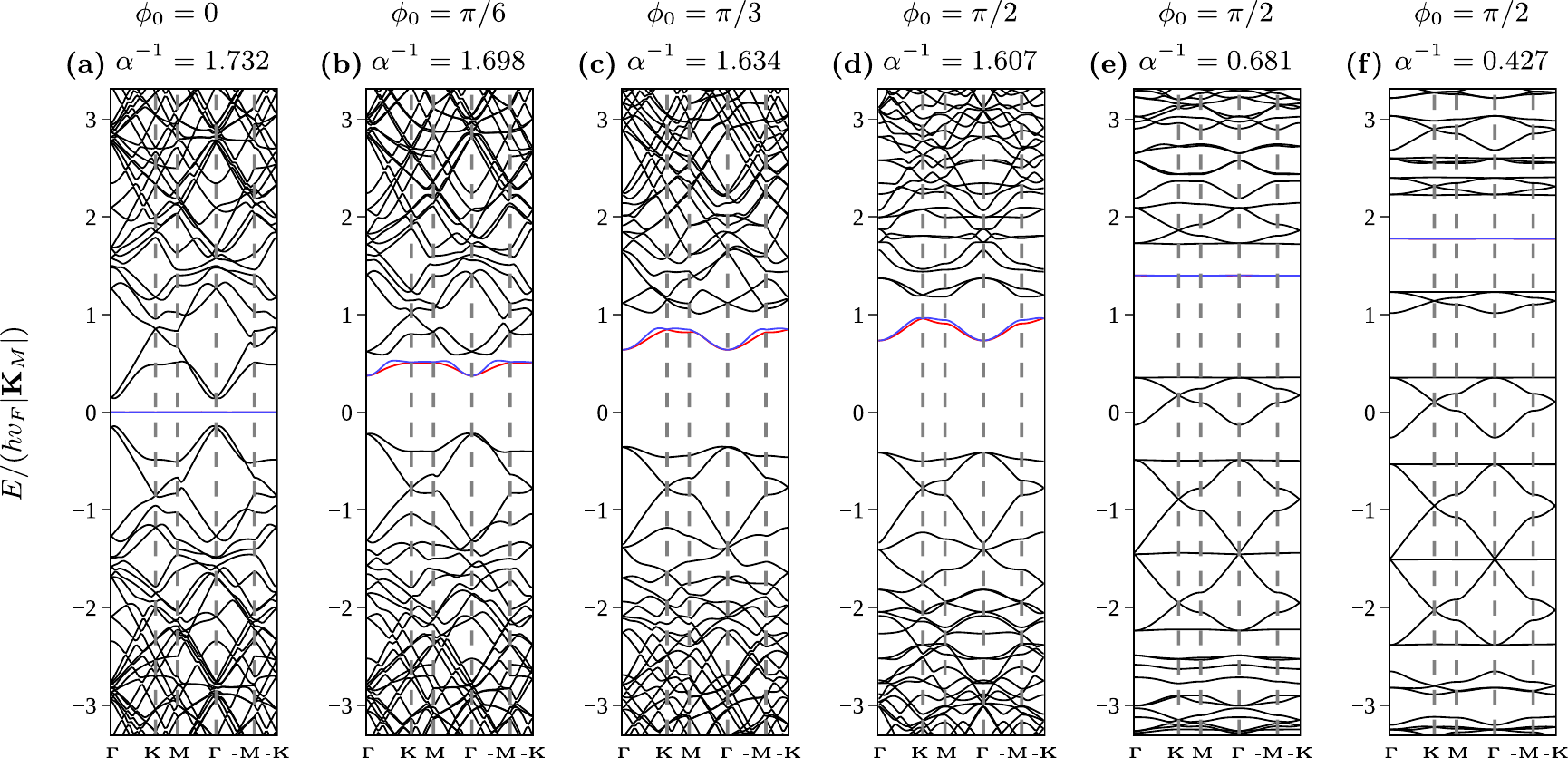}
	\caption{Moir\'e band structures using the model in \cref{eq:continuum-hamiltonian-rotate-basis} with $w_0/w_1 = 0.8$, $w_2=0$, and the quasi-momentum truncation illustrated in Appendix \cref{fig:moire-hopping-lattice}. The horizontal axes follow the moir\'e quasi-momentum trajectory $\bGamma_M \to \bK_M \to \bM_M \to \bGamma_M \to -\bM_M \to -\bK_M$. The two bands nearest charge neutrality are shown in blue and red while all other bands are shown in black. The first four band structures have parameters in the first magic manifold (see \cref{sec:first-magic-manifold}) with varying $\phi_0$. The last two band structures take parameters from the top two dark curves in \cref{fig:bandstructure-hypermagic}\textbf{(c)}.}
	\label{fig:bandstructure-hypermagic}
\end{figure*}

\begin{figure}
	\centering
	\includegraphics{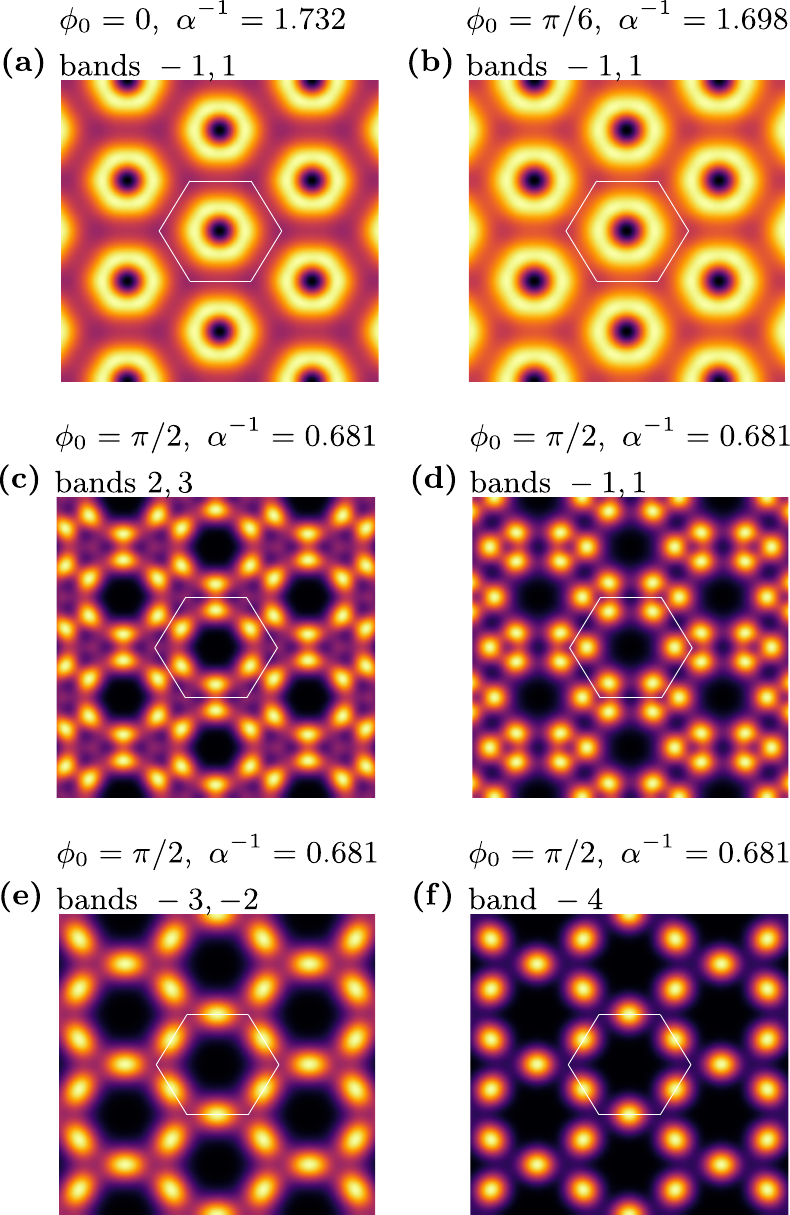}
	\caption{Real space wavefunction plots at $\bGamma_M$ using the model in \cref{eq:continuum-hamiltonian-rotate-basis} with $w_0/w_1 = 0.8$ and the quasi-momentum truncation illustrated in Appendix \cref{fig:moire-hopping-lattice}. Each plot shows the sum of the squares of the norms of the wavefunctions at $\bGamma_M$ in the indicated bands, as a function of space. See \cref{sec:real-space-wavefunctions} for more details. Light colors indicate large values and dark colors indicate small values, but the color scales in each plot are independent. The valence (conduction) bands are denoted with negative (positive) integers, so the highest (lowest) valence (conduction) band is denoted $-1$ ($1$). The white hexagons indicate the hexagonal primitive unit cell of the moir\'e superlattice. Panels \textbf{(a)} and \textbf{(b)} correspond to \cref{fig:bandstructure-hypermagic}\textbf{(a)} and \textbf{(b)} while panels \textbf{(c)}-\textbf{(f)} correspond to \cref{fig:bandstructure-hypermagic}\textbf{(e)}.}
	\label{fig:real-space-wavefunctions}
\end{figure}

\subsection{Changing $\phi_0$ in the first magic manifold}\label{sec:first-magic-manifold}
We first describe the evolution of the flat bands with respect to the angle variable $\phi_0$ defined in \cref{eq:def-phi0} with the magic angle criteria $\alpha^{-1}\approx \sqrt{3}$ and $0\leq w_0/w_1\leq 1$ (see \cref{eq:alpha-magic}). Following Ref. \cite{Bernevig2021a}, we refer to the parameter space satisfying these conditions and minimizing the bandwidth of the lowest two bands at charge neutrality as the \emph{first magic manifold}.

\cref{fig:heatmap-middle-bands} contains three heatmaps showing the base $10$ logarithm of the bandwidth (in units of $\hbar v_F |\bK_M|$) of the two lowest bands at charge neutrality. The first magic manifold appears in \cref{fig:heatmap-middle-bands}\textbf{(b)} (where $\phi_0 = 0$) as a dark nearly horizontal curve on the left of the plot. In the first magic manifold with $\phi_0 = 0$, the lowest two flat bands at charge neutrality are symmetric about zero energy due to an anti-commuting particle-hole symmetry $P$ defined in \cref{sec:parameter-signs} \cite{song_all_2019}. Band structures for parameters in the first magic manifold with $\phi_0 = 0$ can be found in \cref{fig:moire-band-structures}\textbf{(a)}, \textbf{(b)} and \cref{fig:bandstructure-hypermagic}\textbf{(a)}.

For a fixed $w_0/w_1$ and with $\alpha^{-1}\approx\sqrt{3}$, tuning $\phi_0$ away from $0$ shifts the two flat bands at charge neutrality away from zero energy (breaking the particle-hole symmetry $P$), and gradually increases the bandwidth of the flat bands. As shown in \cref{fig:heatmap-middle-bands}\textbf{(a)}, the bandwidth around $\alpha^{-1}=\sqrt{3}$ increases as $\phi_0$ increases from $0$ to $\pi/2$, but still shows a local minimum near $\alpha^{-1}=\sqrt{3}$. The precise value of $\alpha^{-1}$ that minimizes the bandwidth decreases as $\phi_0$ increases. The increase of the bandwidth is mostly due to band curvature at the $\bGamma_M$ point. This can be seen in \cref{fig:moire-band-structures}\textbf{(c)}, \textbf{(d)} and  \cref{fig:bandstructure-hypermagic}\textbf{(b)}-\textbf{(d)}. In particular, the lowest two bands at charge neutrality remain quite flat near the $\bK_M$ and $-\bK_M$ points in the first magic manifold for small $\phi_0$. See Appendix \cref{fig:additional-moire-band-structures-near-commensuration,fig:small-phi0-moire-band-structures} for additional band structures in the first magic manifold.

\cref{fig:real-space-wavefunctions}\textbf{(a)} and \textbf{(b)} show the real space wavefunctions at $\bGamma_M$ corresponding to the flat bands in \cref{fig:bandstructure-hypermagic}\textbf{(a)} and \textbf{(b)}. We see that when $\phi_0$ is increased from $0$ in the first magic manifold, the annular shape of the real space wavefunctions remains unchanged.

\subsection{The hypermagic regime}\label{sec:hypermagic-subsection}
One may have noticed that in the bandwidth plot of \cref{fig:heatmap-middle-bands}\textbf{(a)} (where $w_0/w_1=0.8$) there are three dark spots at $\phi_0=\pi/2$ near $\alpha^{-1}=0.7$, $0.4$, and $0.3$, indicating parameters with very small bandwidths for the lowest two bands at charge neutrality. The situation is identical at $\phi_0=-\pi/2$, which is related to $\phi_0=\pi/2$ by a particle-hole transformation $P$ (see \cref{sec:parameter-signs}).

To investigate what happens to the flat bands at $\phi_0=\pm\pi/2$, we compute the bandwidth of the lowest two bands at charge neutrality at angle $\phi_0=\pi/2$, as a function of $\alpha^{-1}$ and $w_0/w_1$. The result is given in \cref{fig:heatmap-middle-bands}\textbf{(c)}, where we find a small bandwidth region containing three curves with $\alpha^{-1}$ values around $0.7$, $0.4$, and $0.3$ when $0 \leq w_0/w_1 \lesssim 3$. These curves start at $w_0/w_1 = 0$ and extend to at least $w_0/w_1 = 5$. The upper two curves merge around $w_0/w_1=0.2$, $\alpha^{-1}=0.45$ and contain the so-called second magic angle in the chiral limit at $w_0/w_1 = 0$, $\alpha^{-1} = 0.45$ \cite{Tarnopolsky2019}. The third magic angle in the chiral limit at $w_0/w_1 = 0$, $\alpha^{-1} = 0.267$ lies on the lowest curve.

\cref{fig:bandstructure-hypermagic}\textbf{(e)} and \textbf{(f)} show example moir\'e band structures at points on each of the upper two dark curves in \cref{fig:heatmap-middle-bands}\textbf{(c)}. Surprisingly, in both cases, we find several extremely flat bands in addition to the lowest two bands at charge neutrality. In total, there are at least $7$ flat bands in \cref{fig:bandstructure-hypermagic}\textbf{(e)} and $9$ flat bands in \cref{fig:bandstructure-hypermagic}\textbf{(f)}! Additional moir\'e band structures with parameters lying on the curves in \cref{fig:heatmap-middle-bands}\textbf{(c)} are given in \cref{fig:additional-moire-bandstructures}. All of the plots show multiple flat bands, including those in \cref{fig:additional-moire-bandstructures}\textbf{(i)} and \textbf{(j)} which correspond to the second and third magic angles in the chiral limit.

To further investigate this multiple flat band phenomenon, we plot in \cref{fig:heatmap-hypermagic} the bandwidth of the third and eighth narrowest bands among the first 20 conduction bands and the first 20 valence bands at charge neutrality for $\phi_0 = \pi/2$ and $\pi/4$. In \cref{fig:heatmap-hypermagic}\textbf{(a)}, we see that for $\phi_0 = \pi/2$ there is a large region (the dark diagonal band rising from the bottom left of the plot) in which there are $3$ or more flat bands. Additionally, \cref{fig:heatmap-hypermagic}\textbf{(b)} shows that the region in which there are $8$ or more flat bands is nearly as large as the region in which there are $3$ or more flat bands. \cref{fig:heatmap-hypermagic}\textbf{(c)} and \textbf{(d)} show that when $\phi_0$ is decreased to $\pi/4$ the flat bands are often still present though less narrow. Appendix \cref{fig:additional-heatmaps} shows similar heatmaps for the angles $\phi_0 = 0$, $\pi/8$, and $3\pi/8$. At $\phi_0 = 0$, there are very few parameters for which there are more than two flat bands. We call the parameter region centered at $\phi_0 = \pm\pi/2$ in which there are many simultaneous flat bands the \emph{hypermagic regime}.

\begin{figure*}
	\centering
	\includegraphics{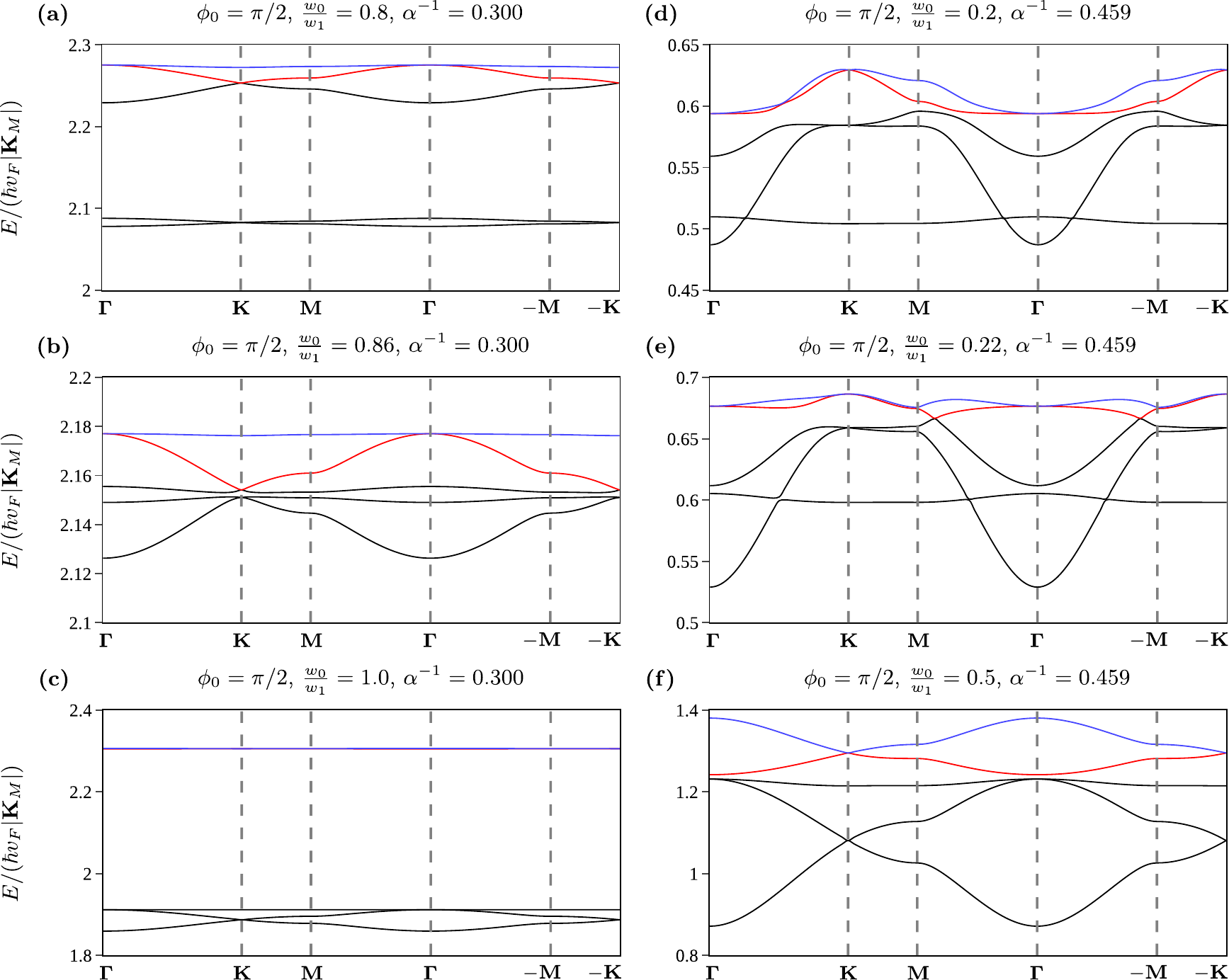}
	\caption{Zoomed plots of moir\'e band structures near charge neutrality using the Hamiltonian in \cref{eq:continuum-hamiltonian-rotate-basis} and the quasi-momentum truncation illustrated in \cref{fig:moire-hopping-lattice}. The horizontal axes follow the moir\'e quasi-momentum trajectory $\bGamma_M \to \bK_M \to \bM_M \to \bGamma_M \to -\bM_M \to -\bK_M$. The two bands nearest charge neutrality are shown in blue and red while all other bands are shown in black. Panels \textbf{(a)}-\textbf{(c)} show a band inversion transition near the lowest dark curve in \cref{fig:heatmap-middle-bands}\textbf{(c)}. In panel \textbf{(a)}, the lowest two moir\'e bands at charge neutrality form two of a group of three connected kagome-like bands. In contrast, in panel \textbf{(c)}, the lowest two moir\'e bands at charge neutrality are gapped from the remote bands. A larger band structure with the same parameters as panel \textbf{(a)} is shown in Appendix \cref{fig:additional-moire-bandstructures-generic}\textbf{(b)}. Panels \textbf{(d)}-\textbf{(f)} show a gap closing transition between the lowest two bands at charge neutrality and the remote bands around the crossing of the upper two dark curves in \cref{fig:heatmap-middle-bands}\textbf{(c)}. At this crossing, the topology of the lowest two bands changes from fragile topological in panel \textbf{(d)} to trivial in panel \textbf{(f)}. The Wilson bands corresponding to panels \textbf{(d)} and \textbf{(f)} are shown in \cref{fig:wilson-bands}\textbf{(g)} and \textbf{(h)}.}
	\label{fig:gap-and-inversion-moire-band-structures}
\end{figure*}

Taking a closer look at the moir\'e bands around charge neutrality in \cref{fig:bandstructure-hypermagic}\textbf{(e)} and \textbf{(f)}, we see groups of three connected bands in which one band is very flat, there are Dirac cones at $\bK_M$ and $-\bK_M$ between the other two bands, and there is a quadratic band touching at $\bGamma_M$ between the flat band and one of the other bands. The second to fourth valence bands at charge neutrality in \cref{fig:bandstructure-hypermagic}\textbf{(e)} and \textbf{(f)} are examples of this pattern. Each such group of three connected bands resembles those of a tight-binding model on the kagome lattice \cite{Bergman2008,tang2011,xu_kagome_2015}. Furthermore, the corresponding real space wavefunctions at $\bGamma_M$ in \cref{fig:real-space-wavefunctions}\textbf{(e)} and \textbf{(f)} show a kagome lattice pattern and the Wilson loop bands in \cref{fig:wilson-bands}\textbf{(f)} are consistent with exponentially localizable Wannier functions \cite{yu_equivalent_2011,Aris2014WL}. Intriguingly, there is a band inversion transition along the lowest dark curve in \cref{fig:heatmap-middle-bands}\textbf{(c)} around $w_0/w_1 = 0.86$, $\alpha^{-1} = 0.3$. For $w_0/w_1$ slightly below $0.86$, the lowest two moir\'e bands at charge neutrality are part of a group of three kagome-like bands while for $w_0/w_1$ just above $0.86$, they form a pair of two isolated bands. This transition is illustrated in \cref{fig:gap-and-inversion-moire-band-structures}\textbf{(a)}-\textbf{(c)}.

In addition to the groups of three connected bands, we also see groups of four connected bands in which the top and bottom bands are very flat, there are Dirac cones at $\bK_M$ and $-\bK_M$ between the middle two bands, and there are two quadratic band touchings at $\bGamma_M$, each involving one flat band. The second to fifth conduction bands at charge neutrality in \cref{fig:bandstructure-hypermagic}\textbf{(e)} and \textbf{(f)} are examples of this pattern. These groups of four bands resemble those of the $p_x,p_y$ 2-orbital honeycomb lattice tight-binding model \cite{Bergman2008,Wu2007}. Furthermore, the corresponding real space wavefunctions at $\bGamma_M$ in \cref{fig:real-space-wavefunctions}\textbf{(c)} show a honeycomb lattice pattern and the Wilson loop bands in \cref{fig:wilson-bands}\textbf{(d)} are consistent with exponentially localizable Wannier functions \cite{yu_equivalent_2011,Aris2014WL}. We note that similar groups of three or four bands were also observed in a recent study of twisted Kitaev bilayers in Ref. \cite{Haskell2022}. Additional moir\'e band structures with parameters in the hypermagic regime including some with $\phi_0 < \pi/2$ can be found in Appendix \cref{fig:additional-moire-bandstructures-generic}.

The continuum model in \cref{eq:continuum-hamiltonian-rotate-basis} (with $w_2=0$) at $\phi_0=\pm\pi/2$ clearly has neither the particle-hole symmetry $P$ nor the chiral symmetry $C$ (see \cref{sec:parameter-signs} for the definitions of these operators), due to the asymmetry between conduction bands and valence bands, for example in \cref{fig:bandstructure-hypermagic}\textbf{(e)} and \textbf{(f)}. As shown in \cref{sec:parameter-signs}, conjugation by $P$ maps the Hamiltonian $\tilde{H}$ at angle $\phi_0$ to $-\tilde{H}$ at angle $-\phi_0$, while keeping the other parameters invariant. In contrast, conjugation by $C$ maps $\tilde{H}$ at angle $\phi_0$ to $-\tilde{H}$ at angle $\phi_0-\pi$, while keeping the other parameters invariant. Therefore, the continuum model at angle $\phi_0=\pm\pi/2$ has a combined $CP$ symmetry:
\begin{equation}
[CP,\tilde{H}]=0\quad \text{when}\quad \phi_0=\pm\frac{\pi}{2}.
\end{equation}
No other values of $\phi_0$ possess this symmetry unless $w_0=0$.

\subsection{Band topology}\label{sec:topology}
Lastly, we discuss the band topology of the lowest two moir\'e bands at charge neutrality. It is known that in the BM model for small angle TBG \cite{Bistritzer2011}, which corresponds to $\phi_0=0$ here (see \cref{eq:def-phi0}), the lowest two moir\'e bands carry a fragile topology protected by $C_{2z}\mathcal{T}$ symmetry, provided the two bands are disconnected from all other bands \cite{po_origin_2018,song_all_2019,po_faithful_2019,ahn_failure_2019,lian2020,po_fragile_2018,cano_fragile_2018,Slager2019WL}. It was further shown in Ref. \cite{song2021TBGII} that in the presence of both $C_{2z}\mathcal{T}$ symmetry and the anti-commuting particle-hole symmetry $P$, the fragile topology becomes stable. See \cref{sec:parameter-signs} for the definition of the $P$ operator and recall that particle-hole symmetry is present only when $\phi_0 = 0$.

\begin{figure}
	\centering
	\includegraphics{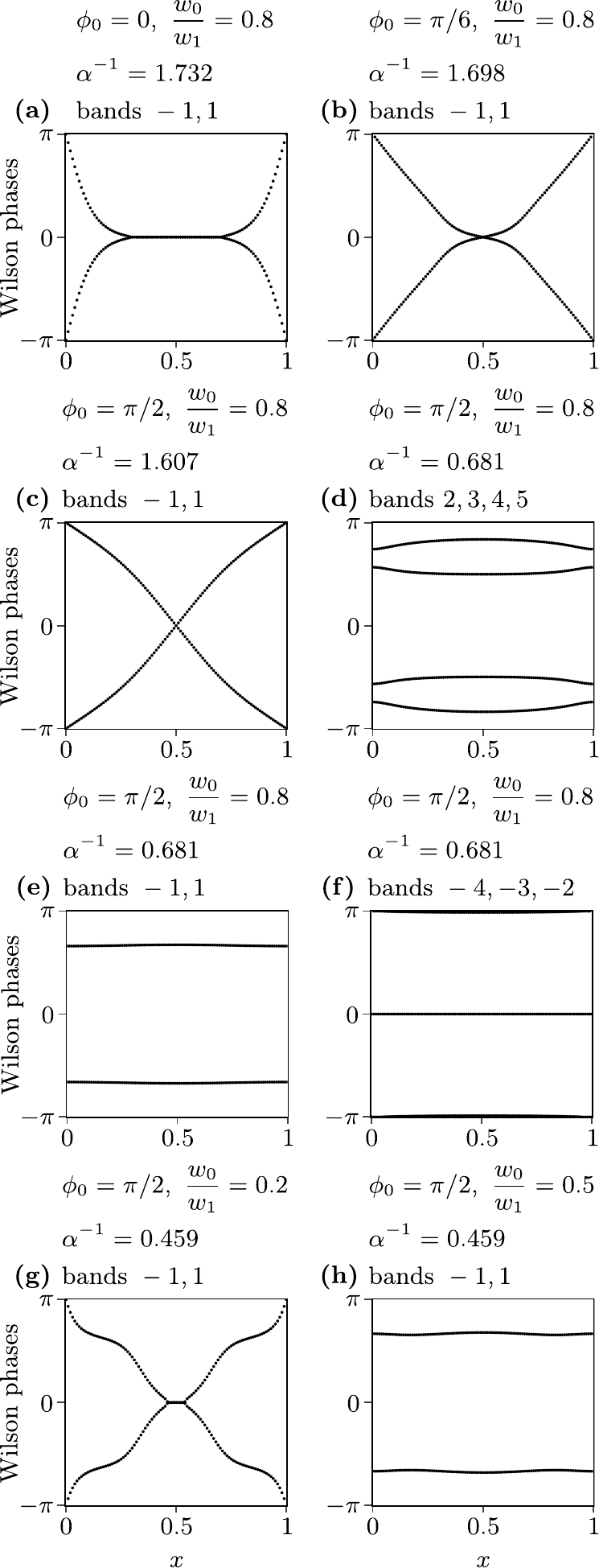}
	\caption{Wilson loop bands for various connected groups of energy bands using the Hamiltonian in \cref{eq:continuum-hamiltonian-rotate-basis} and the quasi-momentum truncation illustrated in \cref{fig:moire-hopping-lattice}. See \cref{sec:wilson-and-truncation} for an explanation of Wilson loop band structure. The valence (conduction) bands are denoted with negative (positive) integers, so the highest (lowest) valence (conduction) band is denoted $-1$ ($1$). The parameters of panels \textbf{(a)}-\textbf{(c)} are the same as those in \cref{fig:bandstructure-hypermagic}\textbf{(a)}, \textbf{(b)}, and \textbf{(d)}. The parameters of panels \textbf{(d)}-\textbf{(f)} are the same as those in \cref{fig:bandstructure-hypermagic}\textbf{(e)}. The parameters of panels \textbf{(g)} and \textbf{(h)} are the same as those in \cref{fig:gap-and-inversion-moire-band-structures}\textbf{(d)} and \textbf{(f)}.}
	\label{fig:wilson-bands}
\end{figure}

The fragile topology in the lowest two moir\'e bands at charge neutrality can be detected by computing their Wilson loop winding number modulo $2$ \cite{song_all_2019,song2021TBGII,yu_equivalent_2011,Aris2014WL}. See \cref{sec:wilson-and-truncation} for an explanation of the Wilson loop matrix and its band structure. \cref{fig:wilson-bands}\textbf{(a)} shows the Wilson loop bands of the lowest two moir\'e bands using parameters corresponding to small angle TBG at the first magic angle. We find a winding number of $1$, indicating non-trivial fragile topology. Away from $\phi_0=0$, the system no longer has particle-hole symmetry $P$, so the fragile topology of the lowest two moir\'e bands can potentially be lost.

We find that the fragile topology of the lowest two moir\'e flat bands at charge neutrality remains robust for any $\phi_0\in[0,\pi/2]$ in the first magic manifold (see \cref{sec:first-magic-manifold}) as long as they are gapped from the remote bands. Two examples of Wilson loop bands in the first magic manifold (with $w_0/w_1=0.8$) are given in \cref{fig:wilson-bands}\textbf{(b)} and \textbf{(c)} and both have a winding number of $1$.

Computing Wilson loop bands in the hypermagic regime, we find that among parameters for which the lowest two bands are gapped from the higher bands, it is possible for the lowest two bands to have either trivial topology or nontrivial fragile topology. In order to transition from one of these possibilities to the other, there must be a gap closing between the lowest two bands and the higher bands. We illustrate one such gap closing in \cref{fig:gap-and-inversion-moire-band-structures}\textbf{(d)}-\textbf{(f)}. The gap closing occurs in \cref{fig:gap-and-inversion-moire-band-structures}\textbf{(e)} near the crossing between the upper two dark curves in \cref{fig:heatmap-middle-bands}\textbf{(c)}. The parameters in \cref{fig:gap-and-inversion-moire-band-structures}\textbf{(d)} are near the second magic angle in the chiral limit and as a result the lowest two bands at charge neutrality have fragile topology \cite{po_origin_2018,song_all_2019,po_faithful_2019,ahn_failure_2019,lian2020}. In contrast, the bands in \cref{fig:gap-and-inversion-moire-band-structures}\textbf{(f)} are topologically trivial and resemble those of a honeycomb lattice tight-binding model. The Wilson loop bands corresponding to \cref{fig:gap-and-inversion-moire-band-structures}\textbf{(d)} and \textbf{(f)} are given in \cref{fig:wilson-bands}\textbf{(g)} and \textbf{(h)} and have Wilson loop winding numbers of $1$ and $0$, respectively. \cref{fig:wilson-bands}\textbf{(e)} shows the Wilson bands corresponding to the lowest two bands at charge neutrality in \cref{fig:bandstructure-hypermagic}\textbf{(e)} which are topologically trivial.

\section{Discussion}\label{sec:discussion}
We have derived an effective low energy continuum model for TBG at angle $\theta=\theta_0+\delta\theta$ near generic commensurate angles $\theta_0$. The model is characterized by complex interlayer hopping amplitudes $w_0 e^{i\chi_0}$ and $w_0e^{-i\chi_0}$ at commensurate $AA$ stackings, a real interlayer hopping amplitude $w_1$ at commensurate $AB/BA$ stackings, and a global energy shift $w_2$. The twist angle $\theta_0$ and the phase $\chi_0$ combine into a single angle parameter $\phi_0=\chi_0+\theta_0/2$ which affects the band structure of the effective continuum model in \cref{eq:continuum-hamiltonian-rotate-basis}. Unless $\theta_0=0$, as in small angle TBG, $\phi_0$ is generically nonzero. Taking the $\delta\theta\rightarrow 0$ limit yields a low-energy model for commensurate TBG, which gives a nonzero charge neutrality gap in the $AA$ stacking case if $\phi_0\neq 0\pmod{\pi}$, and gapless quadratic band touching in the $AB/BA$ stacking cases. For commensurate angle $\theta_0\approx 38.2^\circ$, the gap in the $AA$ stacking case is around $\SI{1.6}{\milli\electronvolt}$ and is therefore experimentally detectable. Away from commensurate angles, we find the first magic angle $\delta\theta_{\text{magic}}$ near a generic commensurate angle $\theta_0$ is still approximately given by $\alpha^{-1}=\sqrt{3}$ with $\alpha$ defined in \cref{eq:alpha-def}. When $\phi_0\neq 0$ at the first magic angle, the lowest two moir\'e bands at charge neutrality are generically flat except in the vicinity of the $\bGamma_M$ point.

We have also revealed a hypermagic parameter regime centered at $\phi_0 = \pm\pi/2$, in which several moir\'e bands (often $8$ or more) become flat simultaneously. The hypermagic regime includes the second and third magic angles in the chiral limit as well as parameters with large $w_0/w_1$. We have identified a gap closing transition in the hypermagic regime between the lowest two bands at charge neutrality and the higher bands, across which the topology of the lowest two bands changes from fragile topological to trivial.

Many of the flat bands in the hypermagic regime belong to disconnected groups of bands which may be understood in terms of effective tight-binding models. Some groups of three bands resemble the kagome lattice tight-binding model which contains a flat band \cite{Bergman2008,tang2011,xu_kagome_2015}. Other groups of four bands resemble the $p_x,p_y$ 2-orbital honeycomb lattice tight-binding model which contains two flat bands \cite{Wu2007,Bergman2008}. 

The lowest two bands at charge neutrality often resemble the honeycomb lattice tight-binding model which can be used to describe monolayer graphene. If such hypermagic parameters can be achieved experimentally, one may expect the strongly interacting physics in the flat bands to be analogous to that in the conventional Hubbard model with trivial single-particle bands. This may allow the occurrence of anti-ferromagnetic states, in contrast to the spin-valley ferromagnetic states in interacting magic angle TBG with $\phi_0=0$ \cite{lian_tbg4_2021,bernevig_tbg5_2021,xie_tbg6_2021,bultinck_ground_2020,kang_strong_2019}.

A practical future concern is how to achieve a continuum model with $\phi_0$ near $\pm \pi/2$ and a sufficiently large energy scale for the parameters $w_0$ and $w_1$ to observe the hypermagic regime in experiment. The effective hopping parameters $w_0$ and $w_1$ at nonzero commensurate angles $\theta_0$ (without lattice relaxation or other effects not considered here) are generically small. For example, $w_0$ and $w_1$ at $\theta_0 \approx 38.2^\circ$ are about $1$ percent of those at $\theta_0=0^\circ$. One idea to enhance $w_0$ and $w_1$ is to explore the possibility of atomic interaction induced structural reconstruction (e.g. charge density waves) or lattice relaxation, which may enhance the moir\'e potential modulation between commensurate $AA$ and $AB/BA$ stackings. In addition, for small twist angles near the untwisted configuration $\theta_0=0$, breaking the mirror symmetry $M_y$ (while preserving the other symmetries) would allow $\chi_0$ to be nonzero, and therefore also $\phi_0$ to be nonzero. Thus, strong $M_y$ breaking perturbations could transform small angle TBG into a large $\phi_0$ model realization. Another interesting question is whether or not there exist other moir\'e models (e.g. involving twisted graphene multilayers or other twisted materials) for which there is a similar hypermagic regime where many bands become simultaneously flat. If other such models exist, it would be interesting to consider their common features and the underlying reasons for the existence of these hypermagic regimes. We leave these ideas and questions for future study.

\begin{acknowledgments}
We thank Jonah Herzog-Arbeitman, Frank Schindler, and B. Andrei Bernevig for valuable discussions. This work is supported by the Alfred P. Sloan Foundation, the National Science Foundation through Princeton University’s Materials Research Science and Engineering Center DMR-2011750, and the National Science Foundation under award DMR-2141966. Additional support is provided by the Gordon and Betty Moore Foundation through Grant GBMF8685 towards the Princeton theory program.
\end{acknowledgments}

\bibliography{bibliography}
\clearpage
\appendix
\onecolumngrid

\begin{center}
    {\bf \large Appendices}    
\end{center}

\section{Microscopic Hamiltonian matrix elements}\label{sec:hamiltonian-matrix-elements}
In this appendix, we derive \cref{eq:intra-matrix-element,eq:inter-matrix-element} for the intra- and interlayer microscopic Hamiltonian matrix elements. Recall that $L$ is the Bravais lattice of monolayer graphene, $P$ is its reciprocal lattice, $\text{BZ}$ is the Brillouin zone, $L_l = R_{-l\theta/2}L$, and $P_l = R_{-l\theta/2}P$. The Bloch states $\ket{\bk, l, \alpha}$ are defined by \cref{eq:Bloch-definition} and satisfy the normalization condition \cref{eq:Bloch-normalization}. We first derive \cref{eq:intra-matrix-element} under the simplifying assumption $\mu = 0$ so that \cref{eq:Hamiltonian-matrix-elements} becomes $\braket{\br', l, \alpha' | H | \br, l, \alpha} = t_+(\br' + \btau_{\alpha'}^{l'} - \br - \btau_\alpha^l)$. Using the identity
\begin{equation}\label{eq:lattice-delta-functions}
\frac{1}{|\text{BZ}|} \sum_{\br \in L} e^{i\bk \cdot \br} = \sum_{\bG \in P} \delta^2(\bk - \bG),
\end{equation}
where $|\text{BZ}|$ is the area of $\text{BZ}$, we compute
\begin{equation}
\begin{split}
\braket{\bk', l, \alpha' | H | \bk, l, \alpha} &= \frac{1}{|\text{BZ}|}\sum_{\br, \br' \in L_l} e^{-i\bk'\cdot (\br' + \btau^l_{\alpha'})}e^{i\bk\cdot (\br + \btau^l_\alpha)}t_+(\br' + \btau^l_{\alpha'} - \br - \btau^l_\alpha)\\
&= \frac{1}{|\text{BZ}|} \sum_{\br' \in L_l} e^{-i\br' \cdot(\bk' - \bk)} \sum_{\br \in L_l} e^{-i\bk' \cdot \btau_{\alpha'}^l} e^{i\bk \cdot (\br -\br' + \btau_\alpha^l)} t_+(\br' -\br + \btau^l_{\alpha'} - \btau^l_\alpha)\\
&= \sum_{\bG_l \in P_l}\delta^2(\bk' - \bk - \bG_l) \sum_{\br \in L_l} e^{-i\bk' \cdot \btau_{\alpha'}^l} e^{i\bk \cdot (-\br + \btau_\alpha^l)} t_+(\br + \btau^l_{\alpha'} - \btau^l_\alpha)\\
&= \sum_{\bG_l \in P_l}\delta^2(\bk' - \bk - \bG_l) e^{-i\bG_l \cdot \btau_{\alpha'}^l}\sum_{\br \in L_l} e^{-i\bk \cdot (\br + \btau_{\alpha'}^l - \btau_\alpha^l)} t_+(\br + \btau^l_{\alpha'} - \btau^l_\alpha)\\
&= \braket{\bk', l, \alpha' | \bk, l, \alpha'} \sum_{\br \in L + \btau_{\alpha'} - \btau_\alpha} e^{-i(R_{l\theta/2}\bk) \cdot \br} t_+(\br).
\end{split}
\end{equation}
Note that we have used the rotational symmetry of the $t_+(\br)$ function in the last step. When $\mu \neq 0$, the Hamiltonian is modified by subtraction of $\mu$ times the identity. As a result, the general form of the matrix element is
\begin{equation}
\braket{\bk', l, \alpha' | H | \bk, l, \alpha} = \braket{\bk', l, \alpha' | \bk, l, \alpha'} \left(-\mu + \sum_{\br \in L + \btau_{\alpha'} - \btau_\alpha} e^{-i(R_{l\theta/2}\bk) \cdot \br} t_+(\br)\right)
\end{equation}
which is \cref{eq:intra-matrix-element}.

Next, we derive \cref{eq:inter-matrix-element}. Using \cref{eq:lattice-delta-functions} and the identities
\begin{equation}
\begin{split}
t_-(\br) &= \int \frac{d^2\bq}{(2\pi)^2} \hat{t}_-(\bq) e^{i\bq \cdot \br}\\
|\Omega||\text{BZ}| &= (2\pi)^2,
\end{split}
\end{equation}
where $|\Omega|$ is the area of the primitive unit cell $\Omega$ of $L$, we compute
\begin{equation}
\begin{split}
\braket{\bk', -l, \alpha' | H | \bk, l, \alpha} &= \frac{1}{|\text{BZ}|}\sum_{\br' \in L_{-l}} \sum_{\br \in L_l} e^{-i\bk'\cdot (\br' + \btau^{-l}_{\alpha'})}e^{i\bk\cdot (\br + \btau^l_\alpha)}t_-(\br' + \btau^{-l}_{\alpha'} - \br - \btau^l_\alpha)\\
&= \frac{1}{|\text{BZ}|} \sum_{\br' \in L_{-l}} \sum_{\br \in L_l} \int \frac{d^2\bq}{(2\pi)^2} \hat{t}_-(\bq) e^{-i\bk'\cdot (\br' + \btau^{-l}_{\alpha'})}e^{i\bk\cdot (\br + \btau^l_\alpha)} e^{i \bq \cdot (\br' + \btau^{-l}_{\alpha'} - \br - \btau^l_\alpha)}\\
&= |\text{BZ}| \int \frac{d^2\bq}{(2\pi)^2} \hat{t}_-(\bq) e^{i\btau^{-l}_{\alpha'} \cdot (\bq - \bk')} e^{i\btau^l_\alpha \cdot (\bk-\bq)}\sum_{\bG_{-l} \in P_{-l}} \delta^2(\bq - \bk' - \bG_{-l}) \sum_{\bG_l \in P_l} \delta^2(\bk - \bq + \bG_l)\\
&= \sum_{\bG_- \in P_-} \sum_{\bG_+ \in P_+} \frac{\hat{t}_-(\bk+\bG_l)}{|\Omega|} e^{i\btau^{-l}_{\alpha'}\cdot \bG_{-l}} e^{-i\btau^l_\alpha \cdot \bG_l}
\delta^2(\bk + \bG_l - \bk'- \bG_{-l})
\end{split}
\end{equation}
which is \cref{eq:inter-matrix-element}.

\section{Dirac cones}\label{sec:dirac-cones}
In this appendix, we derive \cref{eq:dirac-cone}. Since this equation is an approximation of \cref{eq:intra-matrix-element} and both equations depend on the crystal momentum $\bk$ only through $R_{l\theta/2}\bk$, it suffices to consider the case $\theta = 0$. That is, we need to show that the single particle Hamiltonian for monolayer graphene at $\bK + \bp$ takes the form
\begin{equation}\label{eq:dirac-cone-monolayer}
\hbar v_F\bsigma_0 \cdot \bp + O(|\bp|^2)
\end{equation}
when the chemical potential is chosen appropriately. Although this is well known, the most common derivation employs a model of graphene that has only first or second order hopping (for example, see Refs. \cite{Semenoff1984,CastroNeto2009}). We will now give an argument based on symmetry to show that \cref{eq:dirac-cone-monolayer} holds with arbitrary order hopping. This is similar to the symmetry argument given in \cref{sec:symmetry-and-model-parameters} in the case of twisted bilayer graphene near commensuration.

For monolayer graphene, we consider an orthonormal basis of spinless $p_z$ orbitals $\ket{\br, \alpha}$ for $\br \in L$ and $\alpha \in \{A, B\}$ localized at $\br + \btau_\alpha$. We ignore the electron spin because of the weak spin-orbit coupling in graphene \cite{Sichau2019}. The Bloch states are defined by
\begin{equation}
\ket{\bk, \alpha} = \frac{1}{\sqrt{|\text{BZ}|}} \sum_{\br \in L} e^{i\bk\cdot(\br  + \btau_\alpha)}\ket{\br,\alpha}
\end{equation}
for crystal momentum vectors $\bk$, and satisfy the normalization condition
\begin{equation}\label{eq:bloch-normalization-monolayer}
\braket{\bk', \alpha' | \bk, \alpha} = \delta_{\alpha',\alpha}\sum_{\bG \in P}\delta^2(\bk'-\bk - \bG)e^{-i\btau_\alpha \cdot \bG}.
\end{equation}

We consider a microscopic Hamiltonian $H_{\text{mono}}$ with matrix elements
\begin{equation}
\braket{\br', \alpha' | H_{\text{mono}} | \br, \alpha} = t_+(\br' + \btau_{\alpha'} - \br - \btau_\alpha) - \mu \delta_{\br',\br}\delta_{\alpha',\alpha}
\end{equation}
where $\mu$ is a chemical potential and $t_+ : \R^2 \to \R$ is a rotationally symmetric function (i.e. $t_+(\br)$ depends only on $|\br|$). The symmetries of $H_{\text{mono}}$ are generated by the unitary operators $C_{6z}$ (rotation by $\pi/3$ about $\bhatz$), $M_y$ (reflection across the $xz$ plane), and the anti-unitary operator $\mathcal{T}$ (time-reversal). These operators take the form
\begin{equation}
\begin{split}
C_{6z} \ket{\bk, \alpha} &= \ket{R_{\pi/3} \bk, -\alpha}\\
M_y \ket{\bk, \alpha} &= \ket{R^x \bk, -\alpha}\\
\mathcal{T} \ket{\bk, \alpha} &= \ket{-\bk, \alpha}
\end{split}
\end{equation}
where $R^x$ denotes reflection across the $x$ axis. The symmetry subgroup that preserves the high-symmetry crystal momentum $\bK$ is generated by $C_{2z}\mathcal{T}$, $C_{3z}$, and $M_y$, where $C_{2z} = C_{6z}^3$ and $C_{3z} = C_{6z}^2$. Using \cref{eq:bloch-normalization-monolayer} we find
\begin{align}
C_{2z}\mathcal{T}\ket{\bK + \bp, \alpha} &= \ket{\bK + \bp, -\alpha}\label{eq:C_2zT-monolayer}\\
C_{3z}\ket{\bK + \bp, \alpha} &= e^{i(2\pi/3)\alpha}\ket{\bK + R_{2\pi/3}\bp, \alpha}\label{eq:C_3z-monolayer}\\
M_y\ket{\bK + \bp, \alpha} &= \ket{\bK + R^x \bp, -\alpha}.
\end{align}

If we expand the matrix elements of $H_{\text{mono}}$ to second order around $\bK$, we find
\begin{equation}\label{eq:expand-H-mono}
\braket{\bK + \bp', \alpha' | H_{\text{mono}} | \bK + \bp, \alpha} = (\mathcal{H}_{\text{mono}}(\bp)_{\alpha',\alpha} + O(|\bp|^2))\delta^2(\bp'-\bp)
\end{equation}
where $\mathcal{H}_{\text{mono}}(\bp)$ is a Hermitian $2\times 2$ matrix that is linear in $\bp$. Requiring
\begin{equation}
[C_{2z}\mathcal{T}, H_{\text{mono}}] = [C_{3z}, H_{\text{mono}}] = [M_y, H_{\text{mono}}] = 0
\end{equation}
implies
\begin{align}
\mathcal{H}_{\text{mono}}(\bp) &= \sigma_x \overline{\mathcal{H}_{\text{mono}}(\bp)} \sigma_x\label{eq:C_2zT-H-monolayer}\\
&= e^{-i(2\pi/3)\sigma_z}\mathcal{H}_{\text{mono}}(R_{2\pi/3}\bp) e^{i(2\pi/3)\sigma_z}\label{eq:C_3z-H-monolayer}\\
&= \sigma_x \mathcal{H}_{\text{mono}}(R^x \bp) \sigma_x\label{eq:M_xz-H-monolayer}
\end{align}
where we use the notation $\overline{M}$ for the complex conjugate of a matrix $M$. We now expand $\mathcal{H}_{\text{mono}}$ in Pauli matrices as
\begin{equation}
\begin{split}
\mathcal{H}_{\text{mono}}(\bp) &= h_0^0 \sigma_0 + h_0^x \sigma_x + h_0^y \sigma_y + h_0^z \sigma_z\\
&+ (h_x^0 \sigma_0 + h_x^x \sigma_x + h_x^y \sigma_y + h_x^z \sigma_z)p_x\\
&+ (h_y^0 \sigma_0 + h_y^x \sigma_x + h_y^y \sigma_y + h_y^z \sigma_z)p_y
\end{split}
\end{equation}
where the $h$ coefficients are real. First, we choose the value of $\mu$ so that $h_0^0 = 0$. Next, \cref{eq:C_2zT-H-monolayer} implies $h_0^z = h_x^z = h_y^z = 0$ and \cref{eq:C_3z-H-monolayer} implies $h_0^x = h_0^y = h_x^0 = h_y^0 = 0$ and $h_y^x + i h_y^y = i(h_x^x + i h_x^y)$. If we define $v_F$ and $\phi_F$ by $\hbar v_F e^{i\phi_F} = h_x^x + i h_x^y$ we have
\begin{equation}
\mathcal{H}_{\text{mono}}(\bp) = \hbar v_F \bsigma_{\phi_F} \cdot \bp.
\end{equation}
Finally, \cref{eq:M_xz-H-monolayer} implies $\phi_F = 0$ so the Hamiltonian is described by \cref{eq:dirac-cone-monolayer}. We conclude that the $C_{2z}\mathcal{T}$ and $C_{3z}$ symmetries imply that $\mathcal{H}_{\text{mono}}$ takes the form of a Dirac cone and $M_y$ symmetry determines the rotation angle of the Dirac cone.

\section{Equivalent configurations}\label{sec:equivalent-configurations}
Note that the microscopic Hamiltonian in \cref{eq:Hamiltonian-matrix-elements} is uniquely determined up to unitary equivalence by the relative positions of the carbon atoms in the $xy$ plane and their partitioning into two layers. We will therefore consider systems differing only by an isometry of the $xy$ plane and a relabeling of the basis states to be equivalent. This leads to significant redundancy in the specification of bilayer configurations, as we will now show.

With angle and translation parameters $(\theta, \bd)$, the atoms are located at sites
\begin{equation}\label{eq:atom-locations}
\{R_{-\theta/2}(\br + \btau_\alpha) - \bd/2 | \br \in L, \alpha \in \{A, B\}\} \cup \{R_{\theta/2}(\br + \btau_\alpha) + \bd/2 | \br \in L, \alpha \in \{A, B\}\}
\end{equation}
where the two terms indicate the top and bottom layers. Since this set and partitioning is invariant under the mapping $\theta \mapsto -\theta$, $\bd \mapsto -\bd$ (with an interchange of the two layers) the configurations with parameters $(\theta, \bd)$ and $(-\theta, -\bd)$ are equivalent.

Next, consider the configuration with parameters $(\theta + \pi/3, R_{-\pi/6} \bd)$. If we rotate the whole system by the angle $\pi/6$, the bottom layer atoms are located at
\begin{equation}
\{R_{\theta/2 + \pi/3}(\br + \btau_\alpha) + \bd/2 | \br \in L, \alpha \in \{A, B\}\}
\end{equation}
and the top layer atoms are located at
\begin{equation}
\{R_{-\theta/2}(\br + \btau_\alpha) - \bd/2 | \br \in L, \alpha \in \{A, B\}\}.
\end{equation}
Since $R_{\pi/3} L = L$ and $R_{\pi/3}\btau_\alpha - \btau_{-\alpha} \in L$, the bottom layer atoms are equivalently located at
\begin{equation}
\{R_{\theta/2}(\br + \btau_\alpha) + \bd/2 | \br \in L, \alpha \in \{A, B\}\}.
\end{equation}
Since these locations now match \cref{eq:atom-locations}, we see that the configurations with parameters $(\theta, \bd)$ and $(\theta + \pi/3, R_{-\pi/6} \bd)$ are equivalent. As a result of these equivalences, we can restrict $\theta$ to the interval $[0, \pi/3)$ and note that the configurations $(\theta, \bd)$ and $(\pi/3 - \theta, -R_{-\pi/6}\bd)$ are equivalent.

Next, consider the configuration with parameters $(\theta, \bd + \bX)$ for a vector $\bX \in \R^2$. If we translate the whole system by $\bX/2$, the atoms are located at sites
\begin{equation}
\{R_{-\theta/2}(\br + \btau_\alpha) - \bd/2 | \br \in L, \alpha \in \{A, B\}\} \cup \{R_{\theta/2}(\br + \btau_\alpha) + \bd/2 + \bX | \br \in L, \alpha \in \{A, B\}\}.
\end{equation}
If $\bX \in L_-$ then this matches \cref{eq:atom-locations} so the configurations with parameters $(\theta, \bd)$ and $(\theta, \bd + \bX)$ are equivalent. Similarly, if we translate the whole system by $-\bX/2$, we see that when $\bX \in L_+$ the configurations with parameters $(\theta, \bd)$ and $(\theta, \bd + \bX)$ are equivalent. Putting these results together, we see that whenever $\bX \in L_- + L_+$, the configurations with parameters $(\theta, \bd)$ and $(\theta, \bd + \bX)$ are equivalent. 

We show in \cref{sec:commensuration-lattices} that when $\theta$ is a commensurate angle, $L_- + L_+$ is a Bravais lattice whose reciprocal lattice is $P_- \cap P_+$. Furthermore, it follows from the results of \cref{sec:stacking} that for commensurate $\theta$, no set $S$ larger than $L_- + L_+$ has the property that the configurations with parameters $(\theta, \bd)$ and $(\theta, \bd + \bX)$ are equivalent for all $\bd\in\R^2$ and all $\bX \in S$. On the other hand, we show in \cref{sec:density-proof} that when $\theta$ is not a commensurate angle, $L_- + L_+$ is a dense subset of $\R^2$. Since the Hamiltonian depends continuously on $\bd$, it follows that for incommensurate $\theta$ the configurations with parameters $(\theta, \bzero)$ and $(\theta, \bd)$ are equivalent for all $\bd \in \R^2$.

\section{Properties of commensurate configurations}\label{sec:properties-commensurate}
Using a combination of elementary number theory and symmetry arguments, these appendices enumerate and characterize the commensurate configurations of TBG. The approach taken in \cref{sec:enumeration-commensurate,sec:commensuration-lattices} is similar to that in Ref. \cite{Shallcross2010}, but we include detailed derivations for completeness. We follow the notations of \cref{sec:microscopic-Hamiltonian,sec:commensurate-case}.

\subsection{Enumeration of commensurate configurations}\label{sec:enumeration-commensurate}
We first seek to enumerate the commensurate configurations. Recall that $\ba_1, \ba_2$ are primitive vectors for $L$ and $\bb_1, \bb_2$ are primitive vectors for $P$, as illustrated in \cref{fig:lattice-definitions}. Let $a$ and $b$ denote matrices with columns $(\ba_1, \ba_2)$ and $(\bb_1, \bb_2)$ respectively. Explicitly, we have
\begin{equation}
\begin{split}
a &= a_0 \begin{pmatrix}
\sqrt{3} & \sqrt{3}/2\\
0 & -3/2
\end{pmatrix}\\
b &= \frac{2\pi}{a_0}\begin{pmatrix}
\sqrt{3}/3 & 0\\
1/3 & -2/3
\end{pmatrix}.
\end{split}
\end{equation}
Recall from \cref{sec:commensurate-case} that the bilayer system is commensurate when $L_- \cap L_+ \neq \{\bzero\}$, and in this case $L_- \cap L_+$ is the commensuration superlattice. $L_- \cap L_+ \neq \{\bzero\}$ is equivalent to the existence of nonzero integer vectors $\bu_+$ and $\bu_-$ such that
\begin{equation}\label{eq:define-u-plus-u-minus}
\bu_+ = a^{-1}R_\theta a \bu_-.
\end{equation}
Similarly, $P_- \cap P_+ \neq \{\bzero\}$ is equivalent to the existence of nonzero integer vectors $\bv_+$ and $\bv_-$ such that
\begin{equation}
\bv_+ = b^{-1}R_\theta b \bv_-.
\end{equation}
Note that
\begin{align}
a^{-1}R_\theta a &= \begin{pmatrix}
x_0 + y_0 & 2y_0\\
-2y_0 & x_0 - y_0
\end{pmatrix}\label{eq:a-inv-R-a}\\
b^{-1}R_\theta b &= \begin{pmatrix}
x_0 - y_0 & 2y_0\\
-2y_0 & x_0 + y_0
\end{pmatrix}\label{eq:b-inv-R-b}
\end{align}
where $x_0 = \cos\theta$, $y_0 = \frac{1}{\sqrt{3}}\sin\theta$. It follows that the bilayer system is commensurate if and only if $x_0$ and $y_0$ are both rational, which is equivalent to the $L_- \cap L_+ \neq \{\bzero\}$ and $P_- \cap P_+ \neq \{\bzero\}$.

From here on, we will use $\theta_0$ in place of $\theta$ when we assume the system is commensurate in order to match the notation of \cref{sec:commensurate-case}. If the system is commensurate, then $(x_0, y_0)$ is a rational point on the ellipse $x^2 + 3y^2 = 1$. Unless $(x_0, y_0) = (1,0)$, the line through $(x_0, y_0)$ and $(1,0)$ intersects the $y$ axis at a rational point $(0, m/n)$ where $m,n$ are relatively prime integers with $n > 0$. Solving $x^2 + 3y^2 = 1$ simultaneously with $x = -\frac{n}{m}y + 1$ yields
\begin{equation}
\begin{split}
x_0 &= \frac{3m^2-n^2}{3m^2+n^2}\\
y_0 &= \frac{2mn}{3m^2 + n^2}.
\end{split}
\end{equation}
The special case $(x_0, y_0) = (1,0)$ corresponds to $(m,n) = (1,0)$. By the results of \cref{sec:equivalent-configurations}, we can restrict $\theta_0 \in [0, \pi/3)$ so that $m > n \geq 0$ and $\theta_0 = \cos^{-1}(x_0)$.

\subsection{Commensuration lattices}\label{sec:commensuration-lattices}
We now determine the primitive vectors and reciprocal lattices of $L_- \cap L_+$ and $P_- \cap P_+$ assuming $\theta_0$ is a commensurate angle. We have
\begin{align}
a^{-1}R_{\theta_0} a &= \frac{1}{N}\begin{pmatrix}
\alpha & \beta\\
-\beta & \gamma
\end{pmatrix}\label{eq:a-R-theta_0}\\
b^{-1}R_{\theta_0} b &= \frac{1}{N}\begin{pmatrix}
\gamma & \beta\\
-\beta & \alpha
\end{pmatrix}\label{eq:b-R-theta_0}\\
\alpha &= (m+n)(3m-n)/d_0\\
\beta &= 4mn/d_0 = \alpha - \gamma\\
\gamma &= (m-n)(3m+n)/d_0\\
N &= (3m^2+n^2)/d_0\label{eq:define-N}
\end{align}
where $d_0$ is the greatest common divisor of the numerators of $\alpha, \beta, \gamma, N$. Note that $\alpha$ should not be confused with the model parameter defined in \cref{eq:alpha-def} and used in the main text. If $3 \nmid n$ (i.e. $3$ does not divide $n$) then the numerator of $N$ is $1 \pmod 3$ so $3 \nmid d_0$. On the other hand, if $3 | n$ then $3 | d_0$ but $9 \nmid d_0$ since $3 \nmid m$. In either case $3 \nmid N$. If one of $m$ and $n$ is even, then the numerator of $N$ is odd so $d_0$ is odd. On the other hand, if $m$ and $n$ are both odd then $4 | d_0$, but considering the numerator of $\beta$ we see that $8 \nmid d_0$. If $p$ is a prime divisor of $d_0$ other than $2$ then considering the numerator of $\beta$, we see that $p | m$ or $p | n$ but not both. Considering the numerator of $N$, we see that $p = 3$. We conclude
\begin{equation}\label{eq:define-d_0}
d_0 = \gcd(4mn, 3m^2 + n^2) = \begin{cases}
1 & \text{if } (2 | n \text{ or } 2 | m) \text{ and } 3 \nmid n\\
3 & \text{if } (2 | n \text{ or } 2 | m) \text{ and } 3 | n\\
4 & \text{if } 2 \nmid n \text{ and } 2\nmid m \text{ and } 3 \nmid n\\
12 & \text{if } 2 \nmid n \text{ and } 2\nmid m \text{ and } 3 | n
\end{cases}
\end{equation}
so that $\gcd(\beta, N) = 1$. Also, since $\det(a^{-1}R_{\theta_0}a) = 1$ we have $\alpha \gamma + \beta^2 = N^2$.

Assume for now that $(m, n) \neq (1, 0)$ so that $\beta \neq 0$. Writing $\bu_-$ in components as $\bu_- = x \bhatx + y\bhaty$ for integers $x, y$, \cref{eq:define-u-plus-u-minus} becomes
\begin{equation}
\frac{1}{N}\begin{pmatrix}
\alpha & \beta\\
-\beta & \gamma
\end{pmatrix} \begin{pmatrix}
x \\ y
\end{pmatrix} \in \Z^2
\end{equation}
which is in turn equivalent to the pair of congruences
\begin{align}
\alpha x + \beta y &\equiv 0 \pmod{N}\label{eq:solve-u-plus-1}\\
-\beta x + \gamma y &\equiv 0 \pmod{N}.\label{eq:solve-u-plus-2}
\end{align}
Since $\beta \neq 0$ and $\gcd(\beta, N) = 1$, we can multiply \cref{eq:solve-u-plus-2} through by $\beta$. However since $-\beta^2 \equiv \alpha\gamma \pmod{N}$, we see that this equation is implied \cref{eq:solve-u-plus-1}. Furthermore, \cref{eq:solve-u-plus-1} can be solved as
\begin{equation}
\begin{split}
x &= n_1\\
y &= n_1(-\beta^{-1}\alpha)  + n_2N
\end{split}
\end{equation}
for integers $n_1, n_2$, where $\beta^{-1}$ is the smallest non-negative integer such that $\beta^{-1}\beta = 1 \pmod{N^2}$. As a result, the set of integer vectors $\bu$ such that $a^{-1}R_{\theta_0} a \bu$ is an integer vector forms a Bravais lattice with primitive vectors
\begin{equation}
\begin{split}
\bu_1^- &= \bhatx -\beta^{-1}\alpha \bhaty\\
\bu_2^- &= N \bhaty.
\end{split}
\end{equation}
In the case $(m, n) = (1, 0)$, $a^{-1}R_{\theta_0} a = I$, $N = 1$, and $\beta^{-1} = 0$ so this result still holds. The image of this lattice under $a^{-1}R_{\theta_0} a$ is also a Bravais lattice with corresponding primitive vectors
\begin{equation}
\begin{split}
\bu_1^+ &= a^{-1}R_{\theta_0} a \bu_1^- = -\alpha\rho\bhatx + (\beta\rho - \beta^{-1}N)\bhaty\\
\bu_2^+ &= a^{-1}R_{\theta_0} a \bu_2^- = \beta\bhatx + \gamma\bhaty\\
\rho &= (\beta^{-1}\beta - 1)/N \in N\Z.
\end{split}
\end{equation}
We conclude that the commensuration superlattice takes the form
\begin{equation}
L_- \cap L_+ = \{R_{-\theta_0/2}a(n_1 \bu_1^+ + n_2 \bu_2^+) | n_1, n_2 \in \Z\} = \{R_{\theta_0/2}a(n_1 \bu_1^- + n_2 \bu_2^-) | n_1, n_2 \in \Z\}.
\end{equation}
Note that the unit cell of $L_- \cap L_+$ has area $N |\Omega|$.

We can use this result to compute the reciprocal lattice of $L_- \cap L_+$. Let this reciprocal lattice be called $\tilde{P}$ and note that primitive vectors for $\tilde{P}$ can be given by
\begin{equation}\label{eq:define-bu}
\begin{split}
\tilde{\bu}_1 &= \frac{1}{N}R_{-\theta_0/2}b (\gamma \bhatx -\beta \bhaty) = R_{\theta_0/2} b \bhatx\\
\tilde{\bu}_2 &= R_{-\theta_0/2} b [(-\beta \rho/N + \beta^{-1})\bhatx -(\alpha \rho/N)\bhaty] = \frac{1}{N}R_{\theta_0/2}b (\beta^{-1}\alpha\bhatx + \bhaty).
\end{split}
\end{equation}
Since $\rho/N$ is an integer, $\tilde{\bu}_1 \in P_-, \tilde{\bu}_2 \in P_+$ so $\tilde{P} \subset P_- + P_+$. However, by the definition of the reciprocal lattice $P_- + P_+ \subset \tilde{P}$ so that $\tilde{P} = P_- + P_+$. Note that the unit cell for $\tilde{P}$ has area $|\text{BZ}|/N$.

Since \cref{eq:a-R-theta_0,eq:b-R-theta_0} are related by the interchange of $\alpha$ and $\gamma$, corresponding results for the reciprocal lattices can be obtained by interchanging $\alpha$ and $\gamma$. The set of integer vectors $\bv$ such that $b^{-1}R_{\theta_0} b \bv$ is an integer vector forms a Bravais lattice with primitive vectors
\begin{equation}\label{eq:define-v^-}
\begin{split}
\bv_1^- &= \bhatx -\beta^{-1}\gamma\bhaty\\
\bv_2^- &= N\bhaty
\end{split}
\end{equation}
and the image of this lattice under $b^{-1}R_{\theta_0} b$ is also a Bravais lattice with corresponding primitive vectors
\begin{equation}\label{eq:define-v^+}
\begin{split}
\bv_1^+ &= b^{-1}R_{\theta_0} b \bv_1^- = -\gamma\rho\bhatx + (\beta\rho - \beta^{-1}N)\bhaty\\
\bv_2^+ &= b^{-1}R_{\theta_0} b \bv_2^- = \beta\bhaty + \alpha\bhaty.
\end{split}
\end{equation}
We conclude
\begin{equation}
P_- \cap P_+ = \{R_{-\theta_0/2}b(n_1 \bv_1^+ + n_2 \bv_2^+) | n_1, n_2 \in \Z\} = \{R_{\theta_0/2}b(n_1 \bv_1^- + n_2 \bv_2^-) | n_1, n_2 \in \Z\},
\end{equation}
the reciprocal lattice of $P_- \cap P_+$ is $L_- + L_+$, and the unit cell of $P_- \cap P_+$ has area $N |\text{BZ}|$.

\subsection{Equivalences between top and bottom $\bK$ and $\bK'$ points}\label{sec:K-K'-equivalences}
We will now derive \cref{eq:K_+ - K_-,eq:K_+ - K'_-,eq:K_l - K'_l} starting with \cref{eq:K_l - K'_l}. By \cref{eq:define-high-symmetry-points},
\begin{equation}
\begin{split}
\bK_l &= R_{-l\theta_0/2} \bK\\
&= R_{-l\theta_0/2}(2\bb_1 + \bb_2)/3\\
&= R_{-l\theta_0/2}b(2\bhatx + \bhaty)/3
\end{split}
\end{equation}
and similarly $\bK'_l = R_{-l\theta_0/2}b(\bhatx + 2\bhaty)$, so that $\bK_l - \bK'_l = R_{-l\theta_0/2}(\bhatx - \bhaty)/3$. Examining the primitive vectors $\tilde{\bu}_1$ and $\tilde{\bu}_2$ for $P_- + P_+$ in \cref{eq:define-bu}, we see that if $R_{-l\theta_0/2}b \bv \in P_- + P_+$ where $\bv$ is a rational vector then the denominators of $\bv \cdot \bhatx$ and $\bv \cdot \bhaty$ must divide $N$. Since $3 \nmid N$, it follows that $\bK_l - \bK'_l \not\in P_- + P_+$, which is \cref{eq:K_l - K'_l}.

Next, since $3 \nmid N$ there is an integer $k \in \{0, 1, 2\}$ such that
\begin{equation}\label{eq:define-integer-k}
kN = 2 + \beta^{-1}\gamma \pmod{3}
\end{equation}
so that by \cref{eq:define-v^-} we have
\begin{equation}
\bv_1^- + k \bv_2^- = \bhatx + 2\bhaty \pmod{3}.
\end{equation}
Recalling that $\bv_j^+ = b^{-1}R_{\theta_0} b \bv_j^-$ for $j=1,2$ we then have
\begin{equation}
\begin{split}
\bK_- &= -R_{\theta_0/2}b(\bv_1^- + k\bv_2^-)/3 + \bG_-\\
&= -R_{-\theta_0/2}b(\bv_1^+ + k\bv_2^+)/3 + \bG_-\\
\bK'_- &= R_{\theta_0/2}b(\bv_1^- + k\bv_2^-)/3 + \bG'_-\\
&= R_{-\theta_0/2}b(\bv_1^+ + k\bv_2^+)/3 + \bG'_-
\end{split}
\end{equation}
for some $\bG_-, \bG'_- \in P_-$. Multiplying these equations by $R_{-\theta_0}$, we find
\begin{equation}
\begin{split}
\bK_+ &= -R_{-\theta_0/2}b(\bv_1^- + k\bv_2^-)/3 + \bG_+\\
\bK'_+ &= R_{-\theta_0/2}b(\bv_1^- + k\bv_2^-)/3 + \bG'_+
\end{split}
\end{equation}
where $\bG_+ = R_{-\theta_0}\bG_- \in P_+$ and $\bG'_+ = R_{-\theta_0}\bG'_+ \in P_+$. Note that
\begin{equation}
\bK_+ - \bK_- = R_{-\theta_0/2}b(\bv_1^+ - \bv_1^- + k(\bv_2^+ - \bv_2^-))/3 + \bG_+ - \bG_-
\end{equation}
so $\bK_+ - \bK_- \in P_- + P_+$ if and only if $R_{-\theta_0/2}b(\bv_1^+ - \bv_1^- + k(\bv_2^+ - \bv_2^-))/3 \in P_- + P_+$. By the same argument as before, we see that $\bK_+ - \bK_- \in P_- + P_+$ if and only if
\begin{equation}
\bv_1^+ + k\bv_2^+ = \bhatx + 2\bhaty \pmod{3}
\end{equation}
in which case we also have $\bK'_+ - \bK'_- \in P_- + P_+$. Similarly, $\bK_+ - \bK'_- \in P_- + P_+$ if and only if
\begin{equation}
\bv_1^+ + k\bv_2^+ = 2\bhatx + \bhaty \pmod{3}
\end{equation}
in which case we also have $\bK'_+ - \bK_- \in P_- + P_+$.

Using \cref{eq:define-v^+,eq:define-integer-k}, $\alpha \gamma + \beta^2 = N^2 = 1 \pmod{3}$, and $\beta = \alpha - \gamma$, one can show
\begin{equation}
\bv_1^+ + k\bv_2^+ = N(\alpha + \gamma)(2\bhatx + \bhaty) \pmod{3}.
\end{equation}
Additionally, using \cref{eq:define-d_0} we find
\begin{equation}
\begin{split}
N(\alpha + \gamma) &= 2(9m^4 - n^4)/d_0^2\\
&= \begin{cases}
1 \pmod{3} & \text{if } 3 \nmid n\\
2 \pmod{3} & \text{if } 3 | n.
\end{cases}
\end{split}
\end{equation}
Let $(\bJ_-, \bJ'_-)$ denote $(\bK_-, \bK'_-)$ when $3 | n$ and $(\bK'_-, \bK_-)$ when $3 \nmid n$. We then conclude $\bK_+ - \bJ_-, \bK'_+ - \bJ'_- \in P_- + P_+$, which is equivalent to \cref{eq:K_+ - K_-,eq:K_+ - K'_-}.

\subsection{Pairs of complementary commensurate configurations}\label{sec:pi/3-complementarity}
It follows from \cref{sec:equivalent-configurations} that when $\theta_0$ is a commensurate angle, $\pi/3-\theta_0$ is also a commensurate angle. We will now prove this statement another way and consider an important relationship between the two configurations that is used in \cref{sec:commensurate-case}.

Returning to the notation of \cref{sec:enumeration-commensurate}, let
\begin{equation}
\begin{split}
x_1 &= \cos(\pi/3-\theta_0) = \frac{1}{2}(x_0 + 3y_0)\\
y_1 &= \frac{1}{\sqrt{3}}\sin(\pi/3-\theta_0) = \frac{1}{2}(x_0 - y_0).
\end{split}
\end{equation}
Since $\theta_0$ is a commensurate angle, $x_0$ and $y_0$ are rational, and therefore $x_1$ and $y_1$ are rational. It follows that $\pi/3 - \theta_0$ is also commensurate. If $(x_0, y_0)$ corresponds to the integer pair $(m_0, n_0)$ and $(x_1, y_1)$ corresponds to the integer pair $(m_1, n_1)$ then
\begin{equation}
\frac{m_1}{n_1} = \frac{y_1}{1-x_1} = \frac{3m_0^2-2m_0n_0-n_0^2}{3(m_0-n_0)^2}.
\end{equation}
If $3 | n_0$ then $3\nmid m_0$ so the denominator of this fraction is divisible by $3$ exactly once. However, the numerator is also divisible by $3$ so we conclude $3 \not | n_1$. On the other hand, suppose $3 \not | n_0$. It is straightforward to see that the largest power of $3$ dividing the numerator is the same as the largest power of $3$ dividing $m_0 - n_0$, so we conclude $3 | n_1$. As a result, in one of the commensurate configurations corresponding to $\theta_0$ and $\pi/3-\theta_0$ we have $(\bJ_-, \bJ'_-) = (\bK_- , \bK'_-)$, and in the other we have $(\bJ_-, \bJ'_-) = (\bK'_- , \bK_-)$.

\subsection{The lattices $\mathcal{Q}_+$ and $\mathcal{Q}_0$}\label{sec:minimal-norm}
In this section, we prove \cref{eq:Q-lattices-form}, find the minimal norm elements of $(\bK_- + P_-) \cap (\bK_+ + P_+)$, and derive the forms of $L_- \cap L_+$, $P_- \cap P_+$, $L_- + L_+$, and $P_- + P_+$. As explained in \cref{sec:commensurate-case}, we assume $3 | n$ so that $\bJ_- = \bK_-$. Since $(\bK_- + P_-) \cap (\bK_+ + P_+)$ is closed under addition by elements of $P_- \cap P_+$ and has the property that the difference of any two elements is in $P_- \cap P_+$, we must have
\begin{equation}
(\bK_- + P_-) \cap (\bK_+ + P_+) = \bk_0 + P_- \cap P_+
\end{equation}
for some vector $\bk_0$. Since $P_- \cap P_+$ is a triangular lattice, $\bk_0 + P_- \cap P_+$ has at most three elements of minimal norm. However, since $(\bK_- + P_-) \cap (\bK_+ + P_+)$ has $3$-fold rotational symmetry and does not contain $0$, it must have exactly three elements of minimal norm. Since $(\bK_- + P_-) \cap (\bK_+ + P_+)$ additionally has symmetry under reflection across the vector $\bK$, one of the elements of minimal norm must be proportional to $\bK$. Since the unit cell of $P_- \cap P_+$ has area $N |\text{BZ}|$, we conclude $P_- \cap P_+ = \sqrt{N}P$ and the element of minimal norm proportional to $\bK$ must be $\bQ_1 = s\sqrt{N}\bK$ where $s$ is $1$ or $-1$. The other two elements of minimal norm are $\bQ_2 = R_{2\pi/3}\bQ_1$ and $\bQ_3 = R_{4\pi/3}\bQ_1$, and we can write
\begin{equation}
(\bK_- + P_-) \cap (\bK_+ + P_+) = s\sqrt{N}\bK + P_- \cap P_+.
\end{equation}
Recalling from \cref{sec:commensuration-lattices} that the reciprocal lattice of $P_- \cap P_+$ is $L_- + L_+$, it follows that $L_- + L_+ = L/\sqrt{N}$. Applying the same argument to the real space lattices, we see that $L_- \cap L_+ = \sqrt{N}L$ so that $P_- + P_+ = P/\sqrt{N}$.

We will now determine the sign $s$. We have $s\sqrt{N}\bK - \bK_l \in P_l$ or equivalently $(s\sqrt{N}R_{l\theta_0/2} - I)\bK \in P$. Using the half-angle formulas and the results of \cref{sec:enumeration-commensurate,sec:commensuration-lattices} we find
\begin{align}
\cos(\theta_0/2) &= \frac{m\sqrt{3}}{\sqrt{d_0 N}}\label{eq:cos-theta_0/2}\\
\sin(\theta_0/2) &= \frac{n}{\sqrt{d_0 N}}\label{eq:sin-theta_0/2}\\
(s\sqrt{N}R_{l\theta_0/2} - I)\bK &= \frac{4\pi\sqrt{3}}{9a_0}\left((sm\sqrt{3/d_0} - 1)\bhatx + (sln/\sqrt{d_0})\bhaty\right).
\end{align}
For comparison,
\begin{equation}
n_1 \bb_1 + n_2 \bb_2 = \frac{4\pi\sqrt{3}}{9a_0}\left((3n_1/2)\bhatx + (n_1/2 - n_2)\sqrt{3}\bhaty\right).
\end{equation}
By \cref{eq:define-d_0}, when $m \pm n$ is odd, we have $d_0 = 3$ so the equation $(s\sqrt{N}R_{l\theta_0/2}-I)\bK = n_1\bb_1 + n_2\bb_2$ has a solution if and only if $sm = 1 \pmod{3}$. When $m \pm n$ is even, we have $d_0 = 12$ so the same equation now has a solution if and only if $sm = 2 \pmod{3}$. We summarize both cases by saying
\begin{equation}\label{eq:determine-s}
s = \frac{m \pm n}{\sqrt{d_0/3}} \pmod{3} \text{ and } s = \pm 1.
\end{equation}

\subsection{$AA$, $AB$, $BA$ stacking commensurate configurations}\label{sec:stacking}
\begin{figure}
	\centering
	\includegraphics{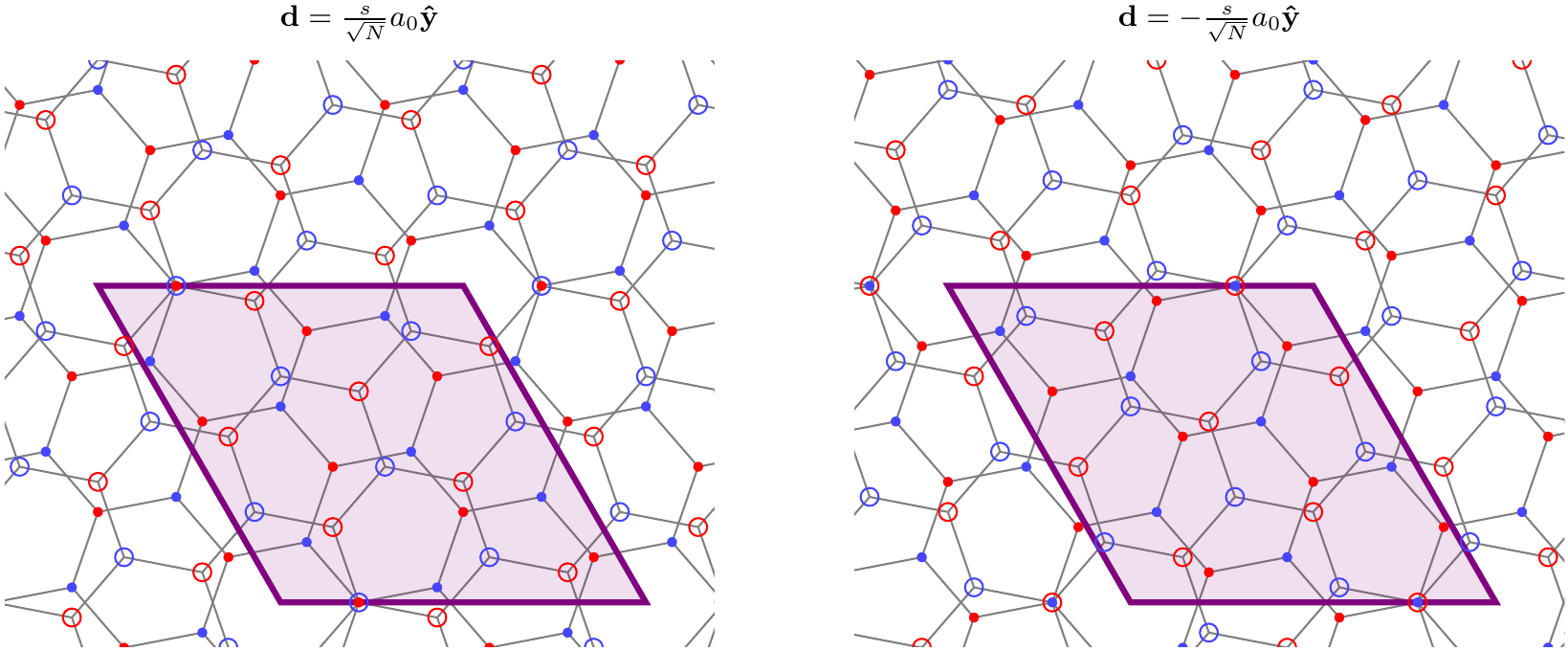}
	\caption{The real space structure of commensurate twisted bilayer graphene as in \cref{fig:commensurate-lattice-AA} but now with nonzero $\bd$. The left (right) plot corresponds to $AB$ ($BA$) stacking. The top (bottom) atoms are represented by dots (circles), the $A$ ($B$) sublattices in each layer are colored blue (red), and the purple rhombus is a primitive unit cell for $L_- \cap L_+$.}
	\label{fig:commensurate-lattice-AB}
\end{figure}
We say that a commensurate configuration has $AA$ stacking if there is an $A$ sublattice atom on the top layer that is directly above some $A$ sublattice atom on the bottom layer. Similarly we say that a commensurate configuration has $AB$ ($BA$) stacking if there is a $B$ ($A$) sublattice atom on the top layer that is directly above some $A$ ($B$) sublattice atom on the bottom layer. For the commensurate configuration with $\theta_0 = 0$, it is clear that $\bd = \bzero$ (i.e. no interlayer translation) corresponds to $AA$ stacking, $\bd = a_0 \bhaty$ corresponds to $AB$ stacking, and $\bd = -a_0 \bhaty$ corresponds to $BA$ stacking. We will now derive a generalization of this correspondence for commensurate configurations with $3 | n$.

We first consider $AA$ stacking. In this case, there is a pair of vectors $\br_+ \in L_+$ and $\br_- \in L_-$ such that $\br_+ + \btau_A^+ = \br_- + \btau_A^-$. Equivalently, we have $\btau_A^+ - \btau_A^- \in L_- + L_+$. Using \cref{eq:sin-theta_0/2}, we have
\begin{equation}
\begin{split}
\btau_A^+ - \btau_A^- &= (R_{-\theta_0/2} \btau_A - \bd/2) - (R_{\theta_0/2} \btau_A + \bd/2)\\
&= -(R_{\theta_0/2} - R_{-\theta_0/2})(a_0 \bhaty) - \bd\\
&= 2 \sin(\theta_0/2) (a_0\bhatx) - \bd\\
&= \frac{2n}{\sqrt{d_0N}} a_0\bhatx - \bd\\
&= n' \ba_1/\sqrt{N} - \bd
\end{split}
\end{equation}
where $n' = 2n/\sqrt{3d_0}$ is an integer since $3 | n$ and $d_0 \in \{3, 12\}$ by \cref{eq:define-d_0}. Since we found in \cref{sec:minimal-norm} that $L_- + L_+ = L/\sqrt{N}$ we see that $\btau_A^+ - \btau_A^- \in L_- + L_+$ if and only if $\bd \in L_- + L_+$. We conclude that $AA$ stacking corresponds to $\bd \in L_- + L_+$. Since $\btau_B = R_{-\pi/3} \btau_A$ and $\ba_2 = R_{-\pi/3}\ba_1$ we have
\begin{equation}
\btau_B^+ - \btau_B^- = n' \ba_2/\sqrt{N} - \bd
\end{equation}
so that $\btau_B^+ - \btau_B^- \in L_- + L_+$ if and only if $\bd \in L_- + L_+$. It follows that $AA$ stacking can equivalently be defined by saying that there is a $B$ sublattice atom on the top layer that is directly above some $B$ sublattice atom on the bottom layer. A commensurate configuration with $AA$ stacking is shown in \cref{fig:commensurate-lattice-AA}.

Next, we consider $AB$ and $BA$ stacking. In $AB$ stacking, there are vectors $\br_l \in L_l$ such that $\br_+ + \btau_B^+ = \br_- + \btau_A^-$, or equivalently $\btau_B^+ - \btau_A^- \in L_- + L_+$. Similarly, $BA$ stacking is equivalent to $\btau_A^+ - \btau_B^- \in L_- + L_+$. Using \cref{eq:cos-theta_0/2,eq:sin-theta_0/2}, we have
\begin{equation}
\begin{split}
\btau_A^l - \btau_B^{-l} &= -(R_{l\theta_0/2}\btau_B - R_{-l\theta_0/2}\btau_A) - l\bd\\
&= -R_{-\pi/6}(R_{(l\theta_0-\pi/3)/2} - R_{-(l\theta_0-\pi/3)/2})\btau_A - l\bd\\
&= 2\sin((l\theta_0-\pi/3)/2)R_{-\pi/6}(a_0\bhatx) - l\bd\\
&= (l\sin(\theta_0/2) \sqrt{3} - \cos(\theta_0/2))(\btau_B - \btau_A) - l\bd\\
&= -m' (\btau_B - \btau_A)/\sqrt{N} - l\bd
\end{split}
\end{equation}
where $m' = (m - ln)/\sqrt{d_0/3}$ is an integer with $m' = s\pmod{3}$ by \cref{eq:determine-s}. It follows that $\btau_A^l - \btau_B^{-l} \in L_- + L_+$ if and only if $\bd \in -\frac{ls a_0}{\sqrt{N}}\bhaty + L_- + L_+$ so that $AB$ stacking corresponds to $\bd \in \frac{s}{\sqrt{N}} a_0 \bhaty + L_- + L_+$ and $BA$ stacking corresponds to $\bd \in -\frac{s}{\sqrt{N}} a_0 \bhaty + L_- + L_+$. Commensurate configurations with $AB$ and $BA$ stacking are shown in \cref{fig:commensurate-lattice-AB}.

\section{$L_- + L_+$ and $P_- + P_+$ are dense for incommensurate $\theta$}\label{sec:density-proof}
Suppose $\theta$ is an incommensurate angle. Recall from \cref{sec:enumeration-commensurate} that this implies $x_0$ and $y_0$ are not both rational. It follows from \cref{eq:a-inv-R-a} that both columns and both rows of the matrix $a^{-1}R_\theta a$ contain an irrational value. It is well known that for any irrational number $z$, the set of fractional parts of integer multiples of $z$ is dense in the interval $[0, 1)$. Equivalently, the set of integer linear combinations of $1$ and $z$ is dense in $\R$. It follows that the set of integer linear combinations of $a^{-1}R_\theta a \bhatx$, $a^{-1}R_\theta a \bhaty$, $\bhatx$, $\bhaty$ is dense in $\R^2$. Since the linear map $R_{-\theta/2} a$ is continuous and density is preserved under continuous maps, we conclude that $L_- + L_+$ is dense in $\R^2$. A similar argument using \cref{eq:b-inv-R-b} shows that $P_- + P_+$ is dense in $\R^2$ as well.

\section{Properties of the distance function $d$}\label{sec:distance-function}
\begin{enumerate}
\item We consider \cref{item:distance-0} in \cref{sec:incommensurate-case} which claims $d(\bk, l, \bk, l) = 0$. If we write $\bk + \bG_l = \bk + \bG_{-l}$ where $\bG_l = \bG_{-l} = \bzero$, we then have $d(\bk, l, \bk, l) = |\bG_{-l}| = 0$.

\item We consider \cref{item:distance-symmetry} in \cref{sec:incommensurate-case} which claims $d(\bk, l, \bk', l') = d(\bk', l', \bk, l)$. If $\bk' - \bk \not\in P_- + P_+$ then $d(\bk, l, \bk', l') = \infty = d(\bk', l', \bk, l)$. Otherwise, suppose $\bk + \bG_l = \bk' + \bG_{-l}$ for some $\bG_- \in P_-$, $\bG_+ \in P_+$. There are two cases to consider:
\begin{itemize}
\item If $l' = -l$ then $\bk' + \bG_{l'} = \bk + \bG_{-l'}$ so that
\begin{equation}
d(\bk, l, \bk', l') = |\bk + \bG_l| = |\bk' + \bG_{l'}| = d(\bk', l', \bk, l).
\end{equation}

\item If $l' = l$ then $\bk' - \bG_{l'} = \bk - \bG_{-l'}$ so that
\begin{equation}
d(\bk, l, \bk', l') = |\bG_{-l}| = |-\bG_{-l'}| = d(\bk', l', \bk, l).
\end{equation}
\end{itemize}

\item We consider \cref{item:distance-triangle-inequality} in \cref{sec:incommensurate-case} which claims
\begin{equation}
d(\bk, l, \bk'', l'') \leq d(\bk, l, \bk', l') + d(\bk', l', \bk'', l'').
\end{equation}
When either term on the right hand side is $\infty$, the inequality is trivially satisfied. If the left hand side is $\infty$ then $\bk'' - \bk \not\in P_- + P_+$. This implies that at least one of the terms on the right hand side must be $\infty$ as well, so the inequality is satisfied.

Otherwise, suppose
\begin{equation}
\begin{split}
\bk + \bG_l &= \bk' + \bG_{-l}\\
\bk' + \bG'_{l'} &= \bk'' + \bG'_{-l'}
\end{split}
\end{equation}
for some $\bG_-, \bG'_- \in P_-$ and $\bG_+, \bG'_+ \in P_+$. It follows that
\begin{equation}
\bk + \bG_l + \bG'_{l'} = \bk'' + \bG_{-l} + \bG'_{-l'}.
\end{equation}
We now consider three cases:
\begin{enumerate}
\item Suppose $l = l' = l''$ and without loss of generality we take $l = l' = l'' = +$. Then $d(\bk, l, \bk', l') =  |\bG_-|$, $d(\bk', l', \bk'', l'') = |\bG'_-|$ and $\bk + \bG''_l = \bk'' + \bG''_{-l}$ where
\begin{equation}
\begin{split}
\bG''_- &= \bG_- + \bG'_-\\
\bG''_+ &= \bG_+ + \bG'_+.
\end{split}
\end{equation}
We then have
\begin{equation}
\begin{split}
d(\bk, l, \bk'', l'') &= |\bG''_-|\\
&= |\bG_- + \bG'_-|\\
&\leq d(\bk, l, \bk', l') + d(\bk', l', \bk'', l'').
\end{split}
\end{equation}

\item Suppose $l = l' \neq l''$ and without loss of generality we take $l = l' = +$, $l'' = -$. Then $d(\bk, l, \bk', l') = |\bG_-|$, $d(\bk', l', \bk'', l'') = |\bk'' + \bG'_-|$ and $\bk + \bG''_l = \bk'' + \bG''_{-l}$ where
\begin{equation}
\begin{split}
\bG''_- &= \bG_- + \bG'_-\\
\bG''_+ &= \bG_+ + \bG'_+.
\end{split}
\end{equation}
We then have
\begin{equation}
\begin{split}
d(\bk, l, \bk'', l'') &= |\bk'' + \bG''_-|\\
&= |\bG_- + (\bk'' + \bG'_-)|\\
&\leq d(\bk, l, \bk', l') + d(\bk', l', \bk'', l'').
\end{split}
\end{equation}\label{item:l=l'-neq-l''}

\item Suppose $l = l'' \neq l'$ and without loss of generality we take $l = l'' = +$, $l' = -$. Then $d(\bk, l, \bk', l') = |\bk' + \bG_-|$, $d(\bk', l', \bk'', l'') = |\bk' + \bG'_-|$ and $\bk + \bG''_l = \bk'' + \bG''_{-l}$ where
\begin{equation}
\begin{split}
\bG''_- &= \bG_- - \bG'_-\\
\bG''_+ &= \bG_+ - \bG'_+.
\end{split}
\end{equation}
We then have
\begin{equation}
\begin{split}
d(\bk, l, \bk'', l'') &= |\bG''_-|\\
&= |\bG_- - \bG'_-|\\
&= |(\bk' + \bG_-) - (\bk' + \bG'_-)|\\
&\leq d(\bk, l, \bk', l') + d(\bk', l', \bk'', l'').
\end{split}
\end{equation}
\end{enumerate}
The last case in which $l \neq l' = l''$ follows from the symmetry of $d$ and the case $l = l' \neq l''$.
\end{enumerate}

\section{Level sets of $d$}\label{sec:distance-level-sets}
In this section we prove the characterization of $d$ described in \cref{sec:low-energy-model}. Recall that $P^0_\pm = R_{-l\theta_0/2}P$ and that $\theta = \theta_0 + \delta\theta$ is an incommensurate angle, where $\theta_0$ is a commensurate angle and $\delta\theta$ is small. Let $\bk \in \R^2$, $l \in \{+, -\}$, and let $\bk_0 = R_{l\delta\theta/2}\bk$. Suppose that $\bk', l'$ satisfy $d(\bk, l, \bk', l') < \infty$ so that we can write $\bk + \bG_l = \bk' + \bG_{-l}$ for unique vectors $\bG_- \in P_-$, $\bG_+ \in P_+$. Define $\bG^0_\pm = R_{\pm \delta\theta/2} \bG_\pm \in P^0_\pm$ and $\bk'_0 = \bk_0 + \bG^0_l - \bG^0_{-l} \in \bk_0 + P^0_- + P^0_+$. We then have
\begin{equation}
\begin{split}
\bk' - R_{l\delta\theta/2}\bk'_0 &= (\bk + \bG_l - \bG_{-l}) - R_{l\delta\theta/2}(\bk_0 + \bG^0_l - \bG^0_{-l})\\
&= R_{-l\delta\theta/2}(\bk_0 + \bG^0_l) - R_{l\delta\theta/2}(\bk_0 + \bG^0_l)\\
&= -lD(\delta\theta)\bQ_{-l}
\end{split}
\end{equation}
where $D(\delta\theta)$ is defined by \cref{eq:define-D-delta-theta} and
\begin{equation}
\bQ_{-l} = \bk_0 + \bG_l^0 = \bk'_0 + \bG_{-l}^0 \in (\bk_0 + P^0_l) \cap (\bk'_0 + P^0_{-l}) = \mathcal{Q}(\bk_0, l, \bk'_0, -l).
\end{equation}
Similarly,
\begin{equation}
\begin{split}
\bk' - R_{-l\delta\theta/2}\bk'_0 &= (\bk + \bG_l - \bG_{-l}) - R_{-l\delta\theta/2}(\bk_0 + \bG^0_l - \bG^0_{-l})\\
&= R_{-l\delta\theta/2}\bG^0_{-l} - R_{l\delta\theta/2} \bG^0_{-l}\\
&= -lD(\delta\theta)\bQ_l
\end{split}
\end{equation}
where
\begin{equation}
\bQ_l = \bG^0_{-l} = \bk_0 - \bk'_0 + \bG^0_l \in P^0_{-l} \cap (\bk_0 - \bk'_0 + P^0_l) = \mathcal{Q}(\bk_0, l, \bk'_0, l).
\end{equation}
It follows that
\begin{equation}
\bk' = R_{-l'\delta\theta/2}\bk'_0 - lD(\delta\theta)\bQ_{l'}
\end{equation}
where $\bQ_{l'} \in \mathcal{Q}(\bk_0, l, \bk'_0, l')$. Furthermore, the vectors $\bk'_0$ and $\bQ_{l'}$ are uniquely determined because the vectors $\bG_-$ and $\bG_+$ are uniquely determined. Additionally, since $|\bQ_{-l}| = |\bk_0 + \bG^0_l| = |\bk + \bG_l|$ and $|\bQ_l| = |\bG^0_l| = |\bG_l|$ we have $d(\bk, l, \bk', l') = |\bQ_{l'}|$. The converse statement can be proved simply by tracing the above argument backwards.

\section{Equivalence of small rotations and spatially varying translations}\label{sec:rotation-translation}
We now derive \cref{eq:T-rotation-translation,eq:S-rotation-translation} which relate the $T$ and $S^l$ potentials in commensurate and incommensurate configurations. In this section, we denote continuum states $\ket{\bp, l, \alpha}_c$ and $\ket{\br, l, \alpha}_c$ with twist angle $\theta = \theta_0 + \delta\theta$ and translation vector $\bd$ by $\ket{\bp, l, \alpha, \delta\theta, \bd}_c$ and $\ket{\br, l, \alpha, \delta\theta, \bd}_c$, respectively. Since $\ket{\bp, l, \alpha, 0, \bd}_c$ is a state with crystal momentum $\bK_l + \bp$ which has been shifted by $-l\bd/2$ we must have
\begin{equation}
\ket{\bp, l, \alpha, 0, \bd}_c = e^{-il(\bK_l + \bp) \cdot \bd/2}\ket{\bp, l, \alpha, 0, \bzero}_c.
\end{equation}
Similarly, since $\ket{\bp, l, \alpha, \delta\theta, \bzero}_c$ is a momentum state that has been rotated by $-l\delta\theta/2$, we must have
\begin{equation}
\ket{\bp, l, \alpha, \delta\theta, \bzero}_c = \ket{R_{l\delta\theta/2}\bp, l, \alpha, 0, \bzero}_c.
\end{equation}
By \cref{eq:define-real-continuum-states}, we then have
\begin{equation}
\begin{split}
\ket{\br, l, \alpha, 0, \bd} &= e^{-il \bK_l \cdot \bd/2} \ket{\br + l\bd/2, l, \alpha, 0, \bzero}\\
\ket{\br, l, \alpha, \delta\theta, \bzero} &= \ket{R_{l\delta\theta/2}\br, l, \alpha, 0, \bzero}
\end{split}
\end{equation}
so that \cref{eq:rotation-translation-equiv-simple} implies
\begin{equation}
\ket{\br, l, \alpha, \delta\theta, \bzero} = e^{il\bK_l \cdot D(\delta\theta)\br/2}\ket{\br, l, \alpha, 0, D(\delta\theta)\br} + O(\delta\theta^2).
\end{equation}

Next, let the continuum Hamiltonian $\tilde{H}$ with twist angle $\theta = \theta_0 + \delta\theta$ and translation vector $\bd$ be denoted $\tilde{H}(\delta\theta, \bd)$. Since the pattern of atoms near position $\br$ with $\theta = \theta_0 + \delta\theta$ and $\bd = \bzero$ is the same to first order in $\delta\theta$ as the pattern with $\theta = \theta_0$ and $\bd = D(\delta\theta)\br$, we must have
\begin{equation}
\bra{\br', l', \alpha', \delta\theta, \bzero}_c \tilde{H}(\delta\theta, \bzero) \ket{\br, l, \alpha, \delta\theta, \bzero}_c = \bra{\br', l', \alpha', \delta\theta, \bzero}_c \tilde{H}(0, D(\delta\theta)\br) \ket{\br, l, \alpha, \delta\theta, \bzero}_c + O(\delta\theta^2)
\end{equation}
It follows that
\begin{equation}
\begin{split}
T(\br, \delta\theta, \bzero) &= \begin{pmatrix}
\bra{\br, +, A, \delta\theta, \bzero}_c & \bra{\br, +, B, \delta\theta, \bzero}_c
\end{pmatrix}
\tilde{H}(\delta\theta, \bzero)
\begin{pmatrix}
\ket{\br, -, A, \delta\theta, \bzero}_c\\ \ket{\br, -, B, \delta\theta, \bzero}_c
\end{pmatrix}\\
&= e^{-i(\bK_- + \bK_+)\cdot D(\delta\theta)\br/2} \begin{pmatrix}
\bra{\br, +, A, 0, D(\delta\theta)\br}_c & \bra{\br, +, B, 0, D(\delta\theta)\br}_c
\end{pmatrix}
\tilde{H}(0, D(\delta\theta)\br)
\begin{pmatrix}
\ket{\br, -, A, 0, D(\delta\theta)\br}_c\\ \ket{\br, -, B, 0, D(\delta\theta)\br}_c
\end{pmatrix}\\
&+ O(\delta\theta^2)\\
&= e^{-i\cos(\theta/2)\bK \cdot D(\delta\theta)\br} T(\br, 0, D(\delta\theta)\br) + O(\delta\theta^2)
\end{split}
\end{equation}
which is equivalent to \cref{eq:T-rotation-translation}. \cref{eq:S-rotation-translation} follows from a similar calculation.

\section{Symmetry representations}\label{sec:symmetry-representations}
In this section, we give the representations of the unitary and anti-unitary symmetries of twisted bilayer graphene referred to in \cref{sec:symmetry-and-model-parameters}. For $\theta \neq 0$, the spinless symmetries of the full Hamiltonian are generated by the unitary operators $C_{6z}$ (rotation by $\pi/3$ about $\bhatz$), $C_{2x}$ (rotation by $\pi$ about $\bhatx$), and the anti-unitary operator $\mathcal{T}$ (time-reversal). These operators take the form
\begin{equation}
\begin{split}
C_{6z} \ket{\bk, l, \alpha} &= \ket{R_{\pi/3} \bk, l, -\alpha}\\
C_{2x} \ket{\bk, l, \alpha} &= -\ket{R^x \bk, -l, -\alpha}\\
\mathcal{T} \ket{\bk, l, \alpha} &= \ket{-\bk, l, \alpha}
\end{split}
\end{equation}
where $R^x$ denotes reflection across the $x$ axis. The minus sign for $C_{2x}$ reflects the fact that $\ket{\br, l, \alpha}$ are $p_z$ orbitals. The symmetry subgroup preserving valley is generated by $C_{2z}\mathcal{T}$, $C_{3z}$, and $C_{2x}$, where $C_{2z} = C_{6z}^3$ and $C_{3z} = C_{6z}^2$. Using \cref{eq:Bloch-normalization}, we find
\begin{equation}
\begin{split}
C_{2z}\mathcal{T}\ket{\bK_l + \bp, l, \alpha} &= \ket{\bK_l + \bp, l, -\alpha}\\
C_{3z}\ket{\bK_l + \bp, l, \alpha} &= e^{i(2\pi/3)\alpha} \ket{\bK_l + R_{2\pi/3}\bp, l, \alpha}\\
C_{2x}\ket{\bK_l + \bp, l, \alpha} &= -\ket{\bK_{-l} + R^x \bp, -l, -\alpha}.
\end{split}
\end{equation}
As a result, the appropriate representations on the $\ket{\bp, l, \alpha}_c$ space are
\begin{equation}
\begin{split}
C_{2z}\mathcal{T}\boldsymbol{\ket{\bp}_c} &= \boldsymbol{\ket{\bp}_c}\begin{pmatrix}
\sigma_x & 0\\
0 & \sigma_x
\end{pmatrix}\\
C_{3z}\boldsymbol{\ket{\bp}_c} &= \boldsymbol{|}R_{2\pi/3}\bp\boldsymbol{\rangle}_{\textbf{c}}\begin{pmatrix}
e^{i(2\pi/3)\sigma_z} & 0\\
0 & e^{i(2\pi/3)\sigma_z}
\end{pmatrix}\\
C_{2x}\boldsymbol{\ket{\bp}_c} &= \boldsymbol{|}R^x\bp\boldsymbol{\rangle}_{\textbf{c}}\begin{pmatrix}
0 & -\sigma_x\\
-\sigma_x & 0
\end{pmatrix}
\end{split}
\end{equation}
where $\boldsymbol{\ket{\bp}_c}$ is defined in \cref{eq:define-p-row-vector}.

In the case $\theta = 0$ there is an additional valley preserving unitary symmetry $M_y$ (reflection across the $xz$ plane). This operator has representations
\begin{equation}
\begin{split}
M_y \ket{\bk, l, \alpha} &= \ket{R^x \bk, l, -\alpha}\\
M_y \ket{\bK + \bp, l, \alpha} &= \ket{\bK + R^x \bp, l, -\alpha}\\
M_y \boldsymbol{\ket{\bp}_c} &= \boldsymbol{|}R^x\bp\boldsymbol{\rangle}_{\textbf{c}}\begin{pmatrix}
\sigma_x & 0\\
0 & \sigma_x
\end{pmatrix}.
\end{split}
\end{equation}
For $\theta$ near $0$, $M_y$ can be considered an approximate symmetry.

\section{Determining the model parameters when $\delta\theta = 0$}\label{sec:model-parameters}
Recall from \cref{sec:real-space} that in the commensurate case, the continuum Hamiltonian approximates the four bands of $H$ nearest the Fermi level at charge neutrality. Explicitly, this model takes the form of a $\bp$ dependent $4\times 4$ matrix as shown in \cref{eq:commensurate-continuum-model}. In order to determine the parameters for this model, we will now describe a method to determine an effective Hamiltonian for these four bands directly from the microscopic Hamiltonian $H$.

Recall from \cref{sec:commensurate-case} that for commensurate configurations, $H$ is block diagonal with blocks of dimension $4N$. Let $H(\bp)$ be the Hamiltonian block containing Bloch states $\ket{\bK_l + \bp, l, \alpha}$ for $l \in \{+, -\}$, $\alpha \in \{A, B\}$. In practice, the $4N \times 4N$ matrix representation for $H(\bp)$ can be computed accurately from \cref{eq:intra-matrix-element,eq:inter-matrix-element} with finitely many terms for each sum as long as the hopping functions $t_+(\br)$ and $\hat{t}_-(\bk)$ decay rapidly enough. We diagonalize $H(\bp)$ as
\begin{equation}
H(\bp) = \sum_{j=1}^{4N} E_j(\bp) \ket{\bp, j} \bra{\bp, j}
\end{equation}
for real eigenvalues $E_1(\bp) \leq E_2(\bp) \leq \cdots \leq E_{4N}(\bp)$ and orthonormal eigenvectors $\ket{\bp, j}$. The indices $j$ from $2N-1$ to $2N+2$ correspond to the four bands described by the continuum Hamiltonian.

Define the projection operators
\begin{equation}
\begin{split}
P_0(\bp) &= \sum_{l = \pm} \sum_{\alpha = \pm} \ket{\bK_l + \bp, l, \alpha}\bra{\bK_l + \bp, l, \alpha}\\
P_1(\bp) &= \sum_{j=2N-1}^{2N+2} \ket{\bp, j}\bra{\bp, j}.
\end{split}
\end{equation}
Since the states $\ket{\bp, j}$ are almost completely supported on the states $\ket{\bK_l + \bp, l, \alpha}$, the operators $P_0(\bp)$ and $P_1(\bp)$ are nearly the same. It follows that there is a canonical unitary operator $U(\bp)$ called the direct rotation that satisfies
\begin{equation}\label{eq:direct-rotation-property}
U(\bp)P_1(\bp)U^\dagger(\bp) = P_0(\bp)
\end{equation}
and minimizes the Frobenius norm of $U(\bp) - I$ over all unitary operators satisfying \cref{eq:direct-rotation-property} \cite{Bravyi2011}. The only condition upon which this theorem is dependent is $||P_0(\bp) - P_1(\bp)||_{op} < 1$, which is satisfied in practice. Here, we use the notation $||M||_{op}$ to denote the operator norm of $M$. The direct rotation is given explicitly by
\begin{equation}
U(\bp) = \sqrt{(I - 2P_0(\bp))(I - 2P_1(\bp))}
\end{equation}
where $\sqrt{M}$ denotes the operator square root of $M$ and is defined using a branch cut of the function $z \mapsto \sqrt{z}$ along the negative real axis in the complex plane, with $\sqrt{1} = 1$. The operator
\begin{equation}
H_{\text{eff}}(\bp) = \sum_{j=2N-1}^{2N+2} E_j(\bp) U(\bp)\ket{\bp, j} \bra{\bp, j}U^\dagger(\bp)
\end{equation}
is the result of projecting $H(\bp)$ onto the four bands of interest and then applying the direct rotation into the subspace spanned by the Bloch states $\ket{\bK_l + \bp, l, \alpha}$. Under the mapping $\ket{\bK_l + \bp, l, \alpha} \mapsto \ket{\bp, l, \alpha}_c$, $H_\text{eff}(\bp)$ maps to an operator that should be considered the correct continuum Hamiltonian.

Let $\mathcal{H}_{\text{eff}}(\bp)$ be the $4\times 4$ matrix representation of $H_{\text{eff}}(\bp)$ with respect to the basis $\ket{\bK_l + \bp, l, \alpha}$ so that $\mathcal{H}_{\text{eff}}(\bp)$ is directly comparable to the matrix $\mathcal{H}_0(\bp)$ defined in \cref{eq:commensurate-continuum-model}. These two matrices are explicitly dependent on $\bp$, but also implicitly dependent on the translation vector $\bd$. Recall from \cref{sec:stacking} that $\bd = \bzero$ corresponds to $AA$ stacking, $\bd = \frac{s}{\sqrt{N}} a_0 \bhaty$ corresponds to $AB$ stacking, and $\bd = -\frac{s}{\sqrt{N}} a_0\bhaty$ corresponds to $BA$ stacking. By \cref{eq:T-S-matrices,eq:truncated-T_0}, we have
\begin{equation}
\begin{split}
T_0(\bzero) &= 3w_0 e^{i\chi_0\sigma_z}\\
T_0(\pm \bd_{AB}) &= \frac{3}{2}w_1(\sigma_x \mp i \sigma_y)
\end{split}
\end{equation}
where $\bd_{AB} = \frac{s}{\sqrt{N}} a_0\bhaty$ so that $w_0, \chi_0, w_2$ determine $\mathcal{H}_0(\bp)$ for $AA$ stacking configurations, while $w_1, w_2$ determine $\mathcal{H}_0(\bp)$ for $AB$ and $BA$ stacking configurations. Furthermore, to determine the model parameters, it suffices to compare $\mathcal{H}_{\text{eff}}(\bp)$ and $\mathcal{H}_0(\bp)$ at $\bp = \bzero$ and a single generic $\bd$ value. For simplicity, we instead use $\bp = \bzero$ and both $\bd = \bzero$ and $\bd = \bd_{AB}$ to determine the model parameters shown in \cref{tbl:w-parameter-table,tbl:accurate-parameters}. These computations are performed using the hopping functions $t_\pm(\br)$ given in \cref{sec:t-functions}.

\begin{figure}
	\centering
	\includegraphics{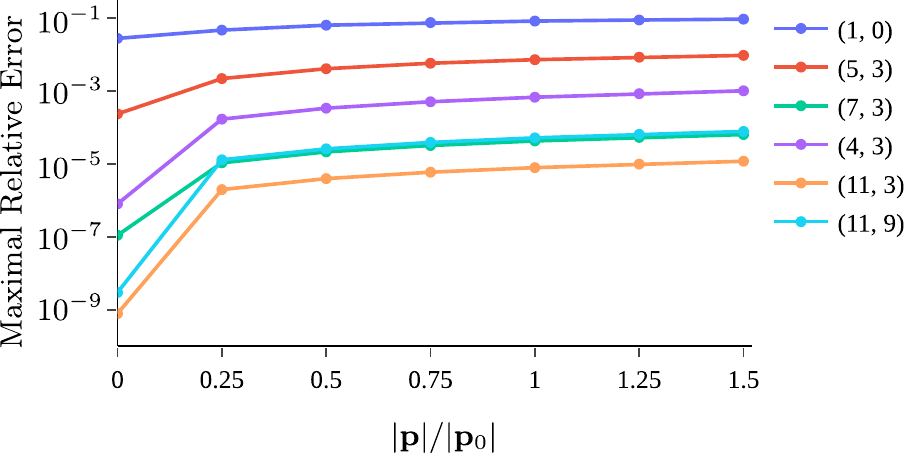}
	\caption{The maximal relative error between $\mathcal{H}_{\text{eff}}(\bp)$ and $\mathcal{H}_0(\bp)$ as a function of $|\bp|/|\bp_0|$ where $\hbar v_F |\bp_0| = 3|w_0|$ (see \cref{eq:relative-error}). The maximum is taken over $\bd$ in a $10\times 10$ discretization of a unit cell of $2L/\sqrt{N}$ and five values of $\bp$ with a given norm.}
	\label{fig:relative-error}
\end{figure}

\begin{figure}
	\centering
	\includegraphics{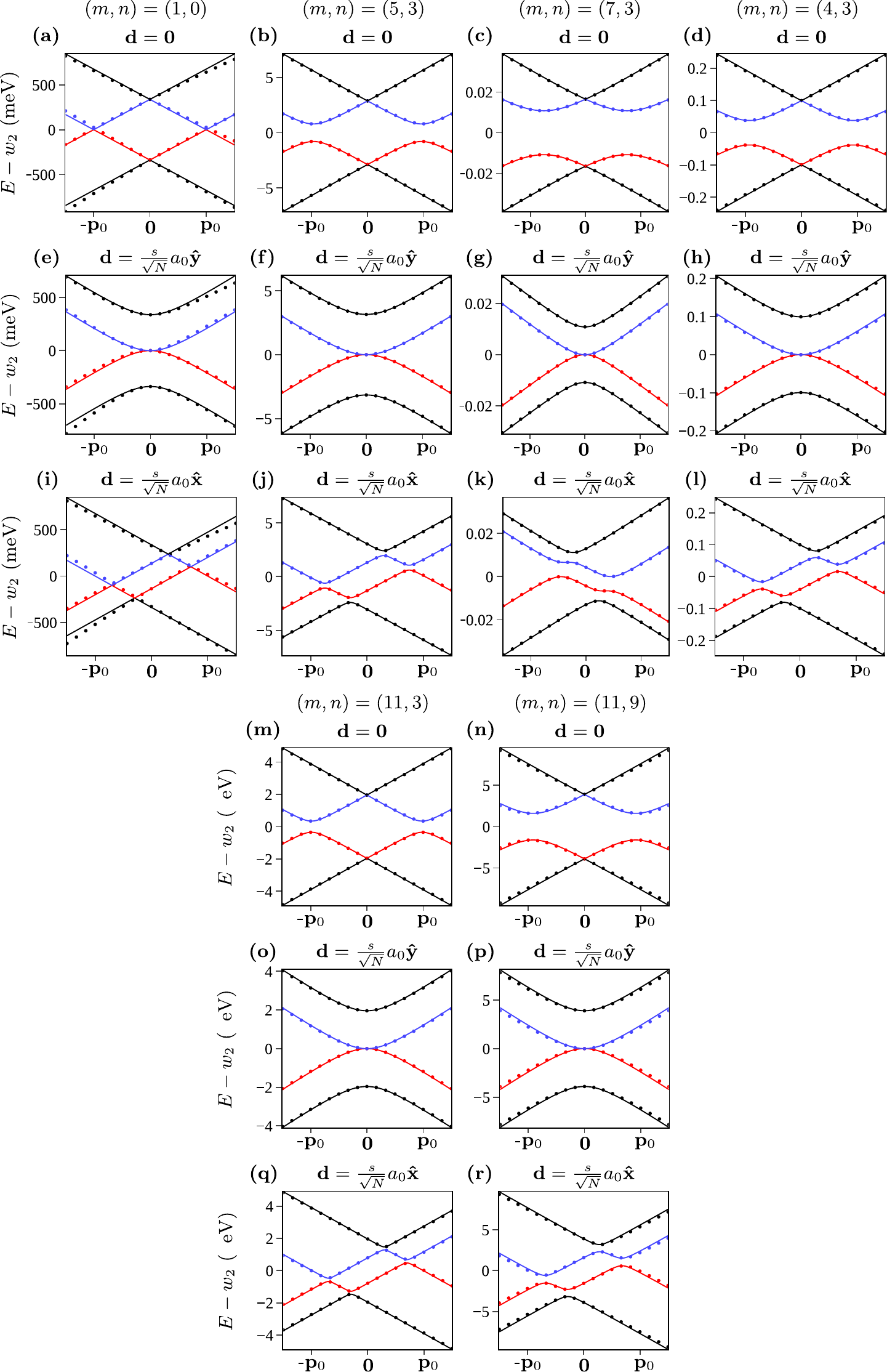}
	\caption{Commensurate band structures. The lines use the model in \cref{eq:commensurate-continuum-model,eq:truncated-T_0} with parameters in \cref{tbl:accurate-parameters} whereas the dots use the microscopic Hamiltonian in \cref{eq:intra-matrix-element,eq:inter-matrix-element}. The vector $\bp$ ranges linearly from $-3\bp_0/2$ to $3\bp_0/2$ where $\hbar v_F \bp_0 = 3|w_0|\bhatx$. Recall that $\bd = \bzero$ and $\bd = \frac{s}{\sqrt{N}} a_0 \bhaty$ correspond to $AA$ and $AB$ stacking, respectively.}
	\label{fig:additional-commensurate-bandstructures}
\end{figure}

To validate the accuracy of these results, we compute the relative error
\begin{equation}\label{eq:relative-error}
\frac{||\mathcal{H}_{\text{eff}}(\bp) - \mathcal{H}_0(\bp)||}{||\mathcal{H}_{\text{eff}}(\bp)||}
\end{equation}
using the parameters in \cref{tbl:accurate-parameters}, where $||M||$ denotes the Frobenius norm of $M$. We compute this relative error as a function of $\bd$ and $\bp$, where $\bd$ varies over a unit cell of $2L/\sqrt{N}$ (recalling from \cref{sec:real-space} that both $H$ and $\tilde{H}$ are periodic up to unitary equivalence with respect to $L/\sqrt{N}$), and $|\bp|$ varies from $0$ to $3 |\bp_0|/2$ where $\hbar v_F |\bp_0| = 3|w_0|$ (see \cref{fig:commensurate-band-structures}). Specifically, for each value of $|\bp|$, we compute the maximal relative error for $\bd$ in a $10\times 10$ discretization of a unit cell of $2L/\sqrt{N}$ and for five values of $\bp$ with the given magnitude. The results are shown in \cref{fig:relative-error} for the first $6$ commensurate configurations. The relative errors for all configurations other than $(m, n) = (1, 0)$ (and $\theta_0 = 0^\circ$) are less than $10^{-2}$ for all $|\bp|$ values considered and are less than $10^{-3.5}$ for $\bp = \bzero$. The relative errors for $(m, n) = (1, 0)$ are larger but still bounded by $10^{-1}$ for all $|\bp|$ values considered, and the relative error at $\bp = \bzero$ is less than $0.03$. We conclude that $\tilde{H}$ is an accurate model for the four bands of $H$ nearest the Fermi level at charge neutrality for all $\bd$ and small $\bp$. \cref{fig:additional-commensurate-bandstructures} compares the eigenvalues of $\mathcal{H}_{\text{eff}}(\bp)$ and $\mathcal{H}_0(\bp)$ for each commensurate configuration in \cref{tbl:accurate-parameters} as a function of $\bp$ for three values of $\bd$.

\begin{table}[h]
	\centering
	\bgroup
	\def\arraystretch{1.2}
	\begin{tabular}{|c|c|c|c|c|c|c|}
		\hline $(m, n)$ & $\theta_0$ & $N$ & $s$ & $\chi_0$ & $(w_0, w_1, w_2)$ in $\si{\micro\electronvolt}$ & $\delta\theta_{\text{magic}}$ in microdegrees\\
		\hline $(1, 0)$ & $0^{\circ}$ & $1$ & $1$ & $0.0^{\circ}$ & $(112682.504, 112682.504, 0.0)$ & $1255782.99$ \\
		\hline $(5, 3)$ & $38.2132107^{\circ}$ & $7$ & $1$ & $-3.09972641^{\circ}$ & $(958.62462, 1051.57009, -4444.6652)$ & $4421.48495$\\
		\hline $(7, 3)$ & $27.7957725^{\circ}$ & $13$ & $-1$ & $125.164435^{\circ}$ & $(5.5027, 3.61749, -4431.53104)$ & $9.96025$\\
		\hline $(4, 3)$ & $46.8264489^{\circ}$ & $19$ & $1$ & $-0.993893031^{\circ}$ & $(33.1618, 33.19161, -4320.05111)$ & $83.97407$\\
		\hline $(11, 3)$ & $17.8965511^{\circ}$ & $31$ & $1$ & $1.23811361^{\circ}$ & $(0.65302, 0.65326, -4426.40937)$ & $1.31881$\\
		\hline $(11, 9)$ & $50.5699921^{\circ}$ & $37$ & $1$ & $-0.861668226^{\circ}$ & $(1.29978, 1.30022, -4026.88676)$ & $2.34902$\\
		\hline
	\end{tabular}
	\egroup
	\caption{Numerically determined model parameters reported with nine significant figures.}
	\label{tbl:accurate-parameters}
\end{table}

\section{$t_\pm(\br)$ functions}\label{sec:t-functions}
Following Ref. \cite{Moon2013}, we take
\begin{equation}
\begin{split}
t_+(\br) &= A_0e^{(a_0 - |\br|)/\delta_0}\\
t_-(\br) &= A_0e^{\left(a_0 - \sqrt{|\br|^2 + r_z^2}\right)/\delta_0}\frac{|\br|^2}{|\br|^2 + r_z^2} + B_0e^{\left(r_z - \sqrt{|\br|^2 + r_z^2}\right)/\delta_0}\frac{r_z^2}{|\br|^2 + r_z^2}
\end{split}
\end{equation}
where $A_0 = -\SI{2.7}{\electronvolt}$ and $B_0 = \SI{0.48}{\electronvolt}$ are transfer integrals, $r_z = 2.36 a_0$ is the interlayer spacing, and $\delta_0 = 0.318 a_0$ is chosen so that $t_+(\ba_1) \approx t_+(\btau_B - \btau_A)/10$. Using \cref{eq:intra-matrix-element,eq:dirac-cone} we find
\begin{equation}
\hbar v_F/a_0 \approx \SI{3.68423316}{\electronvolt}.
\end{equation}

\section{Signs of the parameters and discrete symmetries}\label{sec:parameter-signs}
We now consider the continuum Hamiltonian $\tilde{H}$ in \cref{eq:continuum-hamiltonian-rotate-basis} as a function $\tilde{H}(\phi_0, w_0, w_1, w_2, \delta\theta, s)$ of the shown parameters. By \cref{eq:explicit-q_j}, we have
\begin{equation}\label{eq:negate-deltatheta-s}
\tilde{H}(\phi_0, w_0, w_1, w_2, -\delta\theta, s) = \tilde{H}(\phi_0, w_0, w_1, w_2, \delta\theta, -s) = -\tilde{H}(\phi_0, -w_0, -w_1, -w_2, \delta\theta, s).
\end{equation}
Similarly, by \cref{eq:define-T'-matrices}, we have
\begin{equation}\label{eq:negate-w0}
\tilde{H}(\phi_0 + \pi, w_0, w_1, w_2, \delta\theta, s) = \tilde{H}(\phi_0, -w_0, w_1, w_2, \delta\theta, s).
\end{equation}
Next, we consider the particle hole operator $P$, first chiral operator $C$ (which is often simply called the ``chiral operator" \cite{Tarnopolsky2019} when there is no ambiguity), and second chiral operator $C'$ defined in Ref. \cite{Bernevig2021b}
\begin{equation}
\begin{split}
P \boldsymbol{\ket{\bp}'_c} &= \boldsymbol{\ket{-\bp}'_c} \begin{pmatrix}
0 & -I\\
I & 0
\end{pmatrix}\\
C \boldsymbol{\ket{\bp}'_c} &= \boldsymbol{\ket{\bp}'_c} \begin{pmatrix}
\sigma_z & 0\\
0 & \sigma_z
\end{pmatrix}\\
C' \boldsymbol{\ket{\bp}'_c} &= \boldsymbol{\ket{\bp}'_c} \begin{pmatrix}
\sigma_z & 0\\
0 & -\sigma_z
\end{pmatrix}.
\end{split}
\end{equation}
These operators act within the $\bK$ valley and the origin of quasi-momentum $\bp$ is the $\bGamma_M$ point of the moir\'e Brillouin zone. These operators are unitary and satisfy
\begin{equation}\label{eq:P-C-C'-action}
\begin{split}
P\tilde{H}(\phi_0, w_0, w_1, w_2, \delta\theta, s)P^{-1} &= -\tilde{H}(-\phi_0, w_0, w_1, -w_2, \delta\theta, s)\\
C\tilde{H}(\phi_0, w_0, w_1, w_2, \delta\theta, s)C^{-1} &= -\tilde{H}(\phi_0, -w_0, w_1, -w_2, \delta\theta, s)\\
C'\tilde{H}(\phi_0, w_0, w_1, w_2, \delta\theta, s){C'}^{-1} &= -\tilde{H}(\phi_0, w_0, -w_1, -w_2, \delta\theta, s).
\end{split}
\end{equation}
It follows that $\tilde{H}$ is always equivalent up to a sign and a unitary change of basis from the case in which $s = 1$, $0 \leq \phi_0 \leq \pi/2$, and $w_0, w_1, \delta\theta \geq 0$, so it is sufficient to restrict the parameters in these ranges in calculations.

In particular, we have
\begin{equation}\label{eq:CP-action}
CP\tilde{H}(\phi_0, w_0, w_1, w_2, \delta\theta, s)(CP)^{-1}=\tilde{H}(\pi-\phi_0, w_0, w_1, w_2, \delta\theta, s)\ .
\end{equation}
Therefore, when $\phi_0=\pi/2$, the system has a combined $CP$ symmetry,
although neither $C$ nor $P$ is a symmetry. Moreover, noting that the $CP$ operator map momentum $\bp$ to $-\bp$, $CP$ symmetry implies that the energy spectrum at $\phi_0=\pi/2$ is symmetric between $\bp$ and $-\bp$, as can be seen in \cref{fig:bandstructure-hypermagic}\textbf{(d)}-\textbf{(f)}).

\section{Wilson loops and quasi-momentum truncation}\label{sec:wilson-and-truncation}
In order to make the moir\'e translation symmetry of the Hamiltonian in \cref{eq:continuum-hamiltonian-rotate-basis} more explicit, we now reparametrize the states $\boldsymbol{\ket{\bp}'_c}$ defined in \cref{eq:rotation-transform}, following the approach of \cite{Bernevig2021a}. Note that we can write the moir\'e quasi-momentum $\bp + l \bq_1$ uniquely in the form $\bq + \bg_0$ where $\bq \in \text{BZ}_M$ and $\bg_0 \in D(\delta\theta)\mathcal{Q}_0$. We then have $\bp = \bq - \bg$ where $\bg = l\bq_1 - \bg_0 \in D(\delta\theta)\mathcal{Q}_l$. With this motivation, for $\bq \in \R^2$, $\bg \in D(\delta\theta)\mathcal{Q}_l$, $l \in \{+, -\}$, and $\alpha \in \{A, B\}$ we define
\begin{equation}\label{eq:define-M-states}
\ket{\bq, \bg, \alpha}_M = \ket{\bq - \bg, l, \alpha}'_c
\end{equation}
where the row vector of states $\boldsymbol{\ket{\bp}'_c}$ is given in components by
\begin{equation}\label{eq:define-transformed-momentum-states}
\boldsymbol{\ket{\bp}'_c} = \begin{pmatrix}
\ket{\bp, +, A}'_c & \ket{\bp, +, B}'_c & \ket{\bp, -, A}'_c & \ket{\bp, -, B}'_c
\end{pmatrix}.
\end{equation}
Although the states $\ket{\bq, \bg, \alpha}_M$ for $\bq \in \text{BZ}_M$ form a continuous basis for the Hilbert space, it is useful to define the
overcomplete set of states $\ket{\bq, \bg, \alpha}_M$ for $\bq \in \R^2$.

Using this notation, the continuum Hamiltonian can be written
\begin{equation}\label{eq:continuum-hamiltonian-expanded-matrix}
\tilde{H} = \int_{\text{BZ}_M} d^2\bq \sum_{\substack{\bg', \bg \in \\ D(\delta\theta)(\mathcal{Q}_- \cup \mathcal{Q}_+)}} \sum_{\substack{\alpha', \alpha \in \\ \{A, B\}}} \ket{\bq, \bg', \alpha'}_M \mathcal{H}(\bq)_{(\bg', \alpha'), (\bg, \alpha)}\bra{\bq, \bg, \alpha}_M
\end{equation}
where the infinite dimensional Hamiltonian matrix $\mathcal{H}(\bq)$ has elements
\begin{equation}\label{eq:continuum-hamiltonian-matrix}
\mathcal{H}(\bq)_{(\bg', \alpha'), (\bg, \alpha)} = w_2 \delta_{\bg', \bg} \delta_{\alpha', \alpha} + \hbar v_F (\bsigma \cdot (\bq - \bg))_{\alpha', \alpha}\delta_{\bg', \bg} + \sum_{j=1}^3 (T'_{\bQ_j})_{\alpha', \alpha} \delta_{\bg', \bg-\bq_j} + (T^{\prime\dagger}_{\bQ_j})_{\alpha', \alpha} \delta_{\bg', \bg+\bq_j}.
\end{equation}
For $\bg_0 \in D(\delta\theta)\mathcal{Q}_0$, we have
\begin{equation}
\mathcal{H}(\bq + \bg_0)_{(\bg', \alpha'), (\bg, \alpha)} = \mathcal{H}(\bq)_{(\bg' - \bg_0, \alpha'), (\bg - \bg_0, \alpha)}
\end{equation}
so that
\begin{equation}\label{eq:moire-translation-embedding-matrix}
\mathcal{H}(\bq + \bg_0) = V(\bg_0) \mathcal{H}(\bq) V^\dagger(\bg_0)
\end{equation}
where the unitary matrix $V(\bg_0)$ has elements
\begin{equation}
V(\bg_0)_{(\bg', \alpha'), (\bg, \alpha)} = \delta_{\bg', \bg + \bg_0}\delta_{\alpha', \alpha}
\end{equation}
and is called the embedding matrix.

Consider some set of $N_b \geq 1$ bands of $\mathcal{H}(\bq)$ that are disconnected from all other bands throughout $\text{BZ}_M$. Let $U(\bq)$ be a matrix whose columns form an orthonormal basis for this set of bands. Importantly, we require
\begin{equation}\label{eq:embedding-matrix-periodicity}
U(\bq + \bg_0) = V(\bg_0) U(\bq)
\end{equation}
for $\bg_0 \in D(\delta\theta)\mathcal{Q}_0$. We define the non-Abelian Berry connection
\begin{equation}
\mathcal{A}(\bq) = U^\dagger(\bq) \nabla_\bq U(\bq).
\end{equation}
Although $U$ is not actually periodic, \cref{eq:embedding-matrix-periodicity} implies
\begin{equation}
\mathcal{A}(\bq + \bg_0) = \mathcal{A}(\bq)
\end{equation}
for $\bg_0 \in D(\delta\theta)\mathcal{Q}_0$. As a result, $\mathcal{A}$ is a well defined $U(N_b)$ gauge connection on the torus $\T = \R^2/D(\delta\theta)\mathcal{Q}_0$.

For any closed loop $\bGamma$ in $\T$, we define the gauge covariant Wilson loop unitary
\begin{equation}
W(\bGamma) = \mathcal{P}\exp\left[-\int_{\bGamma} \mathcal{A}(\bq) \cdot d\bq\right]
\end{equation}
where $\mathcal{P}$ indicates path ordering. For each $0 \leq x < 1$, we define the loop $\bGamma_x(t) = x(\bq_3 - \bq_2) + t(\bq_1 - \bq_2)$ for $0 \leq t \leq 1$. Following \cite{Neupert2018}, we compute
\begin{equation}\label{eq:compute-wilson-loop}
\begin{split}
W(\bGamma_x) &= \mathcal{P}\exp\left[-\int_{\bGamma_x} \mathcal{A}(\bq) \cdot d\bq\right]\\
&= \lim_{N_q \to \infty} \prod_{j=N_q-1}^0 \exp[-\mathcal{A}(\bGamma_x(t_{j+1})) \cdot (\bGamma_x(t_{j+1}) - \bGamma_x(t_j))]\\
&= \lim_{N_q \to \infty} \prod_{j=N_q-1}^0 I-\mathcal{A}(\bGamma_x(t_{j+1})) \cdot (\bGamma_x(t_{j+1}) - \bGamma_x(t_j))\\
&= \lim_{N_q \to \infty} \prod_{j=N_q-1}^0 I- U^\dagger(\bGamma_x(t_{j+1})) (U(\bGamma_x(t_{j+1})) - U(\bGamma_x(t_{j})))\\
&= \lim_{N_q \to \infty} \prod_{j=N_q-1}^0 U^\dagger(\bGamma_x(t_{j+1})) U(\bGamma_x(t_{j}))
\end{split}
\end{equation}
where $I$ is the identity matrix and $t_j = j/N_q$. Since $W(\bGamma_x)$ is gauge covariant, its spectrum is gauge invariant. We will refer to the spectrum of $-i\ln(W(\bGamma_x))$ as a function of $x$ as the Wilson loop band structure.

\begin{figure}
	\centering
	\includegraphics{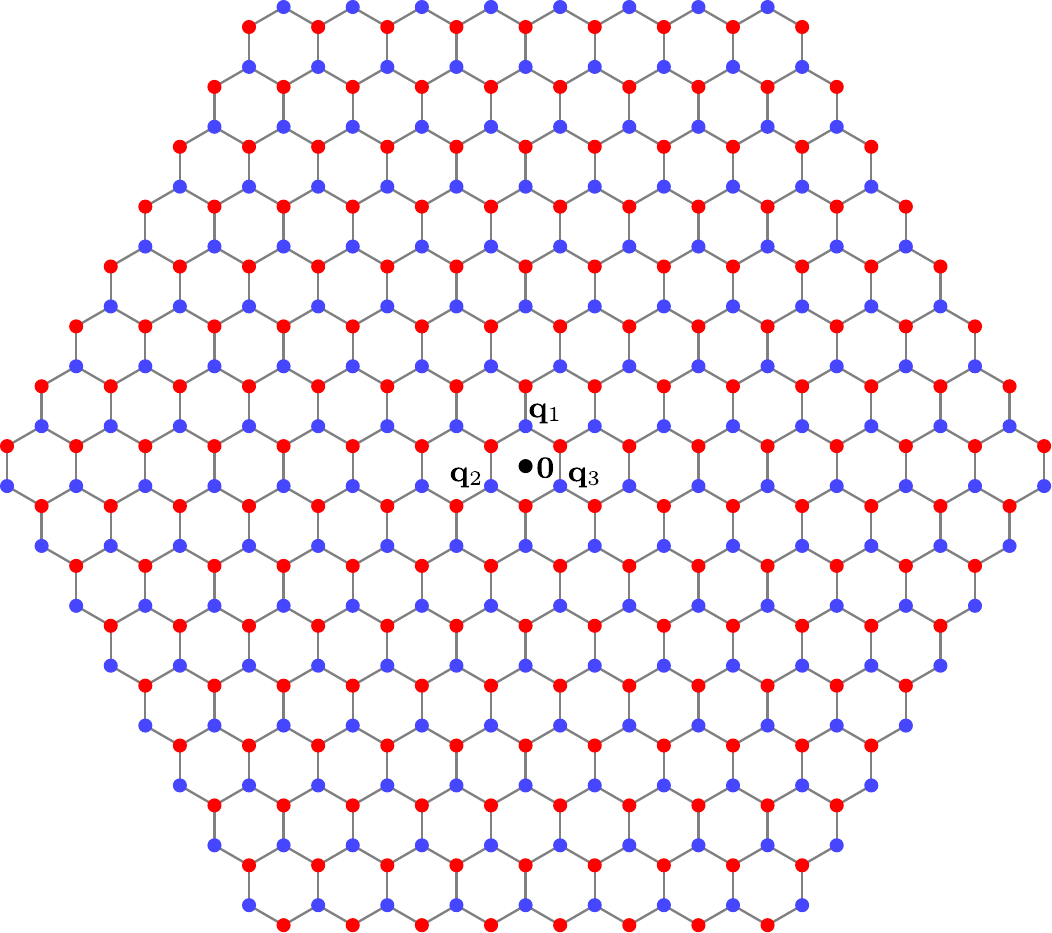}
	\caption{The truncated lattices $D(\delta\theta)\tilde{\mathcal{Q}}_l$ where $\tilde{\mathcal{Q}_l}$ is given by \cref{eq:Q-truncation} with $M = 15$. The blue (red) dots indicate $l = + (-)$, the black dot indicates the origin, and the gray lines correspond to nonzero interlayer matrix elements in \cref{eq:continuum-hamiltonian-matrix}. Each of $\tilde{\mathcal{Q}}_\pm$ has $192$ points so this truncation corresponds to a model with $768$ bands.}
	\label{fig:moire-hopping-lattice}
\end{figure}

When we numerically compute the energy or Wilson loop band structure of $\tilde{H}$, we must truncate the infinite dimensional matrices $\mathcal{H}(\bq)$, $V(\bg_0)$, and $U(\bq)$ to a finite number of dimensions. Since the infinite dimensional nature of $\mathcal{H}(\bq)$ comes from the infinite size of $\mathcal{Q}_l$, we equivalently need to choose a truncation of the lattices $\mathcal{Q}_l$. In order to make the symmetry operators $C_{3z}$, $C_{2x}$, and $P$ well defined (see \cref{sec:symmetry-representations,sec:parameter-signs}), we need a truncation $\tilde{\mathcal{Q}}_l$ of $\mathcal{Q}_l$ satisfying
\begin{equation}\label{eq:Q-truncation-conditions}
\begin{split}
\tilde{\mathcal{Q}}_- &= -\tilde{\mathcal{Q}}_+\\
R_{2\pi/3} \tilde{\mathcal{Q}}_\pm &= R^x\tilde{\mathcal{Q}}_\pm = \tilde{\mathcal{Q}}_\pm.
\end{split}
\end{equation}
One such truncation is given explicitly by
\begin{equation}\label{eq:Q-truncation}
\tilde{\mathcal{Q}}_l = \{n_1 \bQ_1 + n_2 \bQ_2 + n_3 \bQ_3 | n_1 + n_2 + n_3 = l, |n_1| + |n_2| + |n_3| \leq M\}
\end{equation}
for some $M \geq 1$. See \cref{fig:moire-hopping-lattice} for an illustration of $D(\delta\theta)\tilde{\mathcal{Q}}_l$ as defined by \cref{eq:Q-truncation} with $M=15$. This truncation is equivalent to the ``$\bGamma_M$-centered model" in Ref. \cite{Bernevig2021a}. As long as \cref{eq:Q-truncation-conditions} is satisfied, the finite dimensional truncated Hamiltonian retains exact $C_{2z}\mathcal{T}$, $C_{3z}$, and $C_{2x}$ symmetries, and \cref{eq:negate-deltatheta-s,eq:negate-w0,eq:P-C-C'-action,eq:CP-action} hold as well. However, it should be noted that the moir\'e translation symmetry in \cref{eq:moire-translation-embedding-matrix} is exact only when $M = \infty$.

\section{Real space wavefunctions}\label{sec:real-space-wavefunctions}
In this section, we derive the form of the real space wavefunctions shown in \cref{fig:real-space-wavefunctions}. Suppose $v$ is an eigenvector of $\mathcal{H}(\bq)$ where $\mathcal{H}$ is given in \cref{eq:continuum-hamiltonian-matrix}. In that case,
\begin{equation}
\ket{\psi_v} = \sum_{\bg \in D(\delta\theta)(\mathcal{Q}_- \cup \mathcal{Q}_+)} \sum_{\alpha \in \{A, B\}} v_{(\bg,\alpha)} \ket{\bq, \bg, \alpha}_M
\end{equation}
is the corresponding eigenvector of the continuum Hamiltonian $\tilde{H}$. Suppose the row vector of states $\boldsymbol{\ket{\br}'_c}$ defined in \cref{eq:transformed-real-space-basis} can be written in components as
\begin{equation}
\boldsymbol{\ket{\br}'_c} = \begin{pmatrix}
\ket{\br, +, A}'_c & \ket{\br, +, B}'_c & \ket{\br, -, A}'_c & \ket{\br, -, B}'_c
\end{pmatrix}.
\end{equation}
Note that $\ket{\br, l, \alpha}'_c$ differs from $\ket{\br, l, \alpha}_c$ in \cref{eq:define-real-continuum-states} only by a phase and satisfies
\begin{equation}
\ket{\br, l, \alpha}'_c = \frac{1}{2\pi}\int d^2\bp e^{-i\bp\cdot \br}\ket{\bp, l, \alpha}'_c
\end{equation}
where the states $\ket{\bp, l, \alpha}'_c$ are defined in \cref{eq:define-transformed-momentum-states}.

Using \cref{eq:define-M-states}, we have
\begin{equation}
\braket{\br, l, \alpha |'_c \psi_v} = \frac{e^{i\bq \cdot \br}}{2\pi}\sum_{\bg \in D(\delta\theta)\mathcal{Q}_l} v_{(\bg,\alpha)} e^{-i\bg\cdot \br}.
\end{equation}
The plots in \cref{fig:real-space-wavefunctions} show
\begin{equation}
\sum_{l \in \{+, -\}} \sum_{\alpha \in \{A, B\}} |\braket{\br, l, \alpha |'_c \psi_v}|^2
\end{equation}
summed over one or more eigenvectors $v$ with $\bq = \bGamma_M$. Importantly, this quantity is invariant under unitary mixing of the eigenvectors involved.

\section{Tripod model approximation for magic angle conditions}\label{sec:tripod-analysis}
In this section, we use \cref{eq:continuum-hamiltonian-expanded-matrix,eq:continuum-hamiltonian-matrix} to approximate the condition under which the bands of the continuum Hamiltonian nearest the Fermi level at charge neutrality become flat near the $\bK_M$ point in the moir\'e Brillouin zone. In order to make the model analytically tractable, we use the truncation
\begin{equation}
\begin{split}
\tilde{Q}_+ &= \{\bQ_1\}\\
\tilde{Q}_- &= \bQ_1 + \{\bQ_1, \bQ_2, \bQ_3\}
\end{split}
\end{equation}
which produces a model called the ``tripod Hamiltonian" \cite{Bernevig2021a,Bistritzer2011}. Although this truncation does not satisfy \cref{eq:Q-truncation-conditions}, it nonetheless enables a simple calculation of the magic angle in small angle TBG.

We now consider the eigenvalue problem for the tripod Hamiltonian near the $\bK_M$ point. We decompose the eigenvector $\psi$ in the form
\begin{equation}
\psi = \begin{pmatrix}
\ket{\bK_M + \bp, \bq_1, A}_M & \ket{\bK_M + \bp, \bq_1, B}_M
\end{pmatrix} \psi_0
+ \sum_{j=1}^3 \begin{pmatrix}
\ket{\bK_M + \bp, \bq_1 + \bq_j, A}_M & \ket{\bK_M + \bp, \bq_1 + \bq_j, B}_M
\end{pmatrix} \psi_j
\end{equation}
where $\psi_0$, $\psi_1$, $\psi_2$, $\psi_3$ are two dimensional complex column vectors, $\bp$ is a small vector, and the states $\ket{\bq, \bg, \alpha}_M$ are defined in \cref{eq:define-M-states}. The eigenvalue problem then takes the form
\begin{equation}
\begin{split}
(w_2 I + \hbar v_F \bsigma_0 \cdot \bp)\psi_0 + \sum_{j=1}^3 T'_{\bQ_j} \psi_j &= E \psi_0\\
T^{\prime\dagger}_{\bQ_j} \psi_0 + (w_2 I + \hbar v_F \bsigma_0\cdot (\bp - \bq_j))\psi_j &= E\psi_j
\end{split}
\end{equation}
where $E$ is the energy and $T'_{\bQ_j}$ is given by \cref{eq:define-T'-matrices}. Subtracting the $w_2$ terms and multiplying by
\begin{equation}
\lambda = \frac{s}{\hbar v_F |\bK_M|} = \frac{s}{2\hbar v_F \sqrt{N}|\bK| \sin(\delta\theta/2)},
\end{equation}
the eigenvalue problem takes the dimensionless form
\begin{align}
\bsigma_0 \cdot \bp' \psi_0 + \sum_{j=1}^3 \tilde{T}'_{\bQ_j} \psi_j &= E' \psi_0\label{eq:tripod-1}\\
\tilde{T}^{\prime\dagger}_{\bQ_j} \psi_0 + \bsigma_0\cdot (\bp' - \bq'_j))\psi_j &= E'\psi_j\label{eq:tripod-2}
\end{align}
where $E' = \lambda(E - w_2)$, $\bp' = s\bp/|\bK_M|$, $\tilde{T}'_{\bQ_j} = \lambda T'_{\bQ_j} = e^{i(\theta_0/4)\sigma_z}(\lambda T_{\bQ_j})e^{i(\theta_0/4)\sigma_z}$, $\bq'_j = s\bq_j/|\bK_M| = R^{j-1}_{2\pi/3}\bhaty$. We first solve \cref{eq:tripod-2} for $\psi_j$
\begin{align}
\psi_j &= (E' I - \bsigma_0 \cdot (\bp' - \bq'_j))^{-1} \tilde{T}^{\prime\dagger}_{\bQ_j} \psi_0\\
&= \frac{E' I + \bsigma_0 \cdot (\bp' - \bq'_j)}{E'^2 - |\bp' - \bq'_j|^2} \tilde{T}^{\prime\dagger}_{\bQ_j} \psi_0
\end{align}
assuming that $E'^2 - |\bp' - \bq'_j|^2 \neq 0$. Next, we use use \cref{eq:tripod-1} to find
\begin{equation}\label{eq:tripod-combined}
\left(E' I - \bsigma_0 \cdot \bp' + \sum_{j=1}^3 \tilde{T}'_{\bQ_j} \frac{E' I + \bsigma_0 \cdot (\bp' - \bq'_j)}{|\bp' - \bq'_j|^2 - E'^2} \tilde{T}^{\prime\dagger}_{\bQ_j}\right) \psi_0 = 0.
\end{equation}

We first consider the case $\bp' = \bzero$. \cref{eq:tripod-combined} then becomes
\begin{equation}\label{eq:tripod-combined-0}
\begin{split}
0 &= \left(E' I + \sum_{j=1}^3 \tilde{T}'_{\bQ_j} \frac{E' I - \bsigma_0 \cdot \bq'_j}{1 - E'^2} \tilde{T}^{\prime\dagger}_{\bQ_j}\right) \psi_0\\
&= \left(E' I + \frac{E'}{1-E'^2}\sum_{j=1}^3 \tilde{T}'_{\bQ_j} \tilde{T}^{\prime\dagger}_{\bQ_j} - \frac{1}{1 - E'^2} \sum_{j=1}^3 \tilde{T}'_{\bQ_j} (\bsigma_0 \cdot \bq'_j) \tilde{T}^{\prime\dagger}_{\bQ_j}\right) \psi_0.
\end{split}
\end{equation}
Using the identities
\begin{equation}\label{eq:tripod-identities-1}
\begin{split}
\sum_{j=1}^3 \tilde{T}'_{\bQ_j} \tilde{T}^{\prime\dagger}_{\bQ_j} &= 3\lambda^2(w^2_0 + w^2_1) I\\
\sum_{j=1}^3 \tilde{T}'_{\bQ_j} (\bsigma_0\cdot \bq'_j) \tilde{T}^{\prime\dagger}_{\bQ_j} &= 6\lambda^2 w_0 w_1 \sin(\phi_0) I,
\end{split}
\end{equation}
\cref{eq:tripod-combined-0} becomes
\begin{equation}
\left(E' + \frac{E'}{1-E'^2} 3\lambda^2(w^2_0 + w^2_1) - \frac{1}{1-E'^2} 6\lambda^2 w_0w_1\sin(\phi_0)\right)\psi_0 = 0.
\end{equation}
Since $\psi_0 \neq 0$, we conclude
\begin{equation}\label{eq:tripod-cubic}
E'^3 - E' [1 + 3\lambda^2(w^2_0 + w^2_1)] + 6\lambda^2 w_0 w_1 \sin(\phi_0) = 0.
\end{equation}

Note that when $E' = 1$, the cubic polynomial in \cref{eq:tripod-cubic} takes the value
\begin{equation}
-3\lambda^2(w^2_0 + w^2_1) + 6\lambda^2 w_0w_1 \sin(\phi_0) \leq -3\lambda^2(w^2_0 + w^2_1) + 6\lambda^2|w_0||w_1| = -3\lambda^2(|w_0| - |w_1|)^2 \leq 0
\end{equation}
and when $E' = -1$, it takes the value
\begin{equation}
3\lambda^2(w^2_0 + w^2_1) + 6\lambda^2 w_0w_1 \sin(\phi_0) \geq 3\lambda^2(w^2_0 + w^2_1) - 6\lambda^2|w_0||w_1| = 3\lambda^2(|w_0| - |w_1|)^2 \geq 0.
\end{equation}
There is therefore some solution $E' = E'_0$ of \cref{eq:tripod-cubic} with $E'_0$ in the interval $[-1, 1]$. Additionally, when $|\sin(\phi_0)| < 1$ or $|w_0| \neq |w_1|$, we can take $E'_0$ in the interval $(-1, 1)$. In this case, we can approximate $E'^3_0 \approx 0$ and find
\begin{equation}
E'_0 \approx \frac{6 \lambda^2 w_0 w_1 \sin(\phi_0)}{1 + 3\lambda^2 (w^2_0 + w^2_1)} \in (-1, 1).
\end{equation}
When $|\sin(\phi_0)| = 1$ and $|w_0| = |w_1|$, it is possible that there is no solution of \cref{eq:tripod-cubic} in $(-1, 1)$. We will not consider this case further.

Next, we consider \cref{eq:tripod-combined} with $\bp' \neq \bzero$. We take $E' = E'_0 + \delta E'$ and expand to first order in $\delta E'$ and $|\bp'|$. Using $|\bp' - \bq'_j|^2 \approx 1 - 2\bp' \cdot \bq'_j$, $E'^2 \approx E'^2_0 + 2E'_0 \delta E'$, and the fact that $E'_0$ solves \cref{eq:tripod-cubic}, we find
\begin{equation}\label{eq:tripod-combined-nonzero}
\begin{split}
0 &\approx \left((E'_0 + \delta E') I - \bsigma_0 \cdot \bp' + \sum_{j=1}^3 \tilde{T}'_{\bQ_j} \frac{(E'_0 + \delta E') I + \bsigma_0 \cdot (\bp' - \bq'_j)}{1 - E'^2_0 - 2\bp' \cdot \bq'_j - 2 E'_0 \delta E'} \tilde{T}^{\prime\dagger}_{\bQ_j}\right) \psi_0\\
&= \left((E'_0 + \delta E') I - \bsigma_0 \cdot \bp' + \sum_{j=1}^3 \tilde{T}'_{\bQ_j} \frac{(E'_0 + \delta E') I + \bsigma_0 \cdot (\bp' - \bq'_j)}{1 - E'^2_0}\tilde{T}^{\prime\dagger}_{\bQ_j} \frac{1}{1 - \frac{2\bp' \cdot \bq'_j + 2 E'_0 \delta E'}{1 - E'^2_0}} \right) \psi_0\\
&\approx ((E'_0 + \delta E') I - \bsigma_0 \cdot \bp')\psi_0\\
&+ \left(\sum_{j=1}^3 \tilde{T}'_{\bQ_j} \frac{E'_0 I - \bsigma_0 \cdot \bq'_j}{1 - E'^2_0}\tilde{T}^{\prime\dagger}_{\bQ_j} + \sum_{j=1}^3 \tilde{T}'_{\bQ_j} \frac{\delta E' I + \bsigma_0 \cdot \bp'}{1 - E'^2_0}\tilde{T}^{\prime\dagger}_{\bQ_j}\right)\left(1 + \frac{2\bp' \cdot \bq'_j + 2 E'_0 \delta E'}{1 - E'^2_0}\right) \psi_0\\
&\approx \left(\delta E' I -\bsigma_0 \cdot \bp' + \sum_{j=1}^3 \tilde{T}'_{\bQ_j} \frac{E'_0 I - \bsigma_0 \cdot \bq'_j}{1 - E'^2_0}\tilde{T}^{\prime\dagger}_{\bQ_j} \frac{2\bp' \cdot \bq'_j + 2 E'_0 \delta E'}{1 - E'^2_0} + \sum_{j=1}^3 \tilde{T}'_{\bQ_j} \frac{\delta E' I + \bsigma_0 \cdot \bp'}{1 - E'^2_0}\tilde{T}^{\prime\dagger}_{\bQ_j}\right)\psi_0.
\end{split}
\end{equation}
Using the identities in \cref{eq:tripod-identities-1} as well as
\begin{equation}
\begin{split}
\sum_{j=1}^3 \tilde{T}'_{\bQ_j} (\bsigma_0 \cdot \bq'_j)(2\bp' \cdot \bq'_j)\tilde{T}^{\prime\dagger}_{\bQ_j} &= -3\lambda^2 w_1^2 \bsigma_0 \cdot \bp' + 3\lambda^2 w_0^2 \bsigma_{-2\phi_0} \cdot \bp'\\
\sum_{j=1}^3 \tilde{T}'_{\bQ_j}(2\bp' \cdot \bq'_j)\tilde{T}^{\prime\dagger}_{\bQ_j} &= 6\lambda^2 w_0w_1 \bsigma_{-\phi_0-\pi/2} \cdot \bp'\\
\sum_{j=1}^3 \tilde{T}'_{\bQ_j} (\bsigma_0 \cdot \bp')\tilde{T}^{\prime\dagger}_{\bQ_j} &= 3\lambda^2 w_0^2 \bsigma_{-2\phi_0} \cdot \bp',
\end{split}
\end{equation}
\cref{eq:tripod-combined-nonzero} becomes
\begin{equation}\label{eq:tripod-combined-nonzero-simplified}
\left[\left(1 - \frac{2 E'^3_0}{1 - E'^2_0} + \frac{3\lambda^2(w^2_0 + w^2_1)}{1 - E'^2_0}\right)\delta E' I + \left(-1 + \frac{3\lambda^2 w^2_1}{1-E'^2_0} \right) \bsigma_0 \cdot \bp' + \frac{E'_0}{1-E'^2_0}(6\lambda^2 w_0w_1 \bsigma_{-\phi_0-\pi/2} \cdot \bp')\right]\psi_0 = 0.
\end{equation}

We are interested in the conditions under which the terms proportional to $\bp'$ in \cref{eq:tripod-combined-nonzero-simplified} vanish so that $\delta E' = 0$ to first order in $|\bp'|$. If $E'_0 = 0$, this condition is equivalent to $3\lambda^2 w^2_1 = 1$ or
\begin{equation}\label{eq:magic-angle-condition-small}
\lambda = \pm\frac{1}{|w_1| \sqrt{3}}.
\end{equation}
Since $E'_0 = 0$ when $w_0 = 0$ or $\phi_0 = 0$, we recognize this as the magic angle condition identified in Refs. \cite{Bernevig2021a,Bistritzer2011}, which is realizable in small angle TBG. However, there is another solution with $\phi_0 = \pm \pi/2$ and
\begin{equation}
\frac{3\lambda^2 w^2_1}{1-E'^2_0} = 1 \pm \frac{E'_0}{1-E'^2_0}6\lambda^2 w_0 w_1
\end{equation}
or equivalently
\begin{equation}\label{eq:tripod-magic-angle-quadratic}
E'^2_0 \mp 6\lambda^2 w_0w_1 E'_0 + 3\lambda^2 w^2_1 -1 = 0.
\end{equation}
By \cref{eq:tripod-cubic}, $E'_0$ also satisfies
\begin{equation}\label{eq:tripod-magic-angle-qubic}
E'^3_0 - E'_0 [1 + 3\lambda^2(w^2_0 + w^2_1)] \pm 6 \lambda^2 w_0 w_1 = 0
\end{equation}
so we need to find when these two equations have a common solution for $E'_0$ in the interval $(-1, 1)$. Assuming $|w_0| \neq |w_1|$, we can take the approximation $E'^3_0 \approx 0$ so that \cref{eq:tripod-magic-angle-quadratic,eq:tripod-magic-angle-qubic} become
\begin{equation}\label{eq:magic-angle-condition-large}
27w_1^2(w^4_1 - 2 w^2_0 w^2_1 - 3 w^4_0)\lambda^6 + 9(w^4_1 - w^4_0)\lambda^4 - 3(w^2_1 + 2 w^2_0)\lambda^2 - 1 = 0.
\end{equation}
It is easy to see that this equation has real solutions for $\lambda$ if and only if $|w_0| < |w_1|$.

We conclude that the magic angle conditions for the tripod Hamiltonian are \cref{eq:magic-angle-condition-small} when either $w_0 = 0$ or $\phi_0 = 0$, and \cref{eq:magic-angle-condition-large} when $\phi_0 = \pm \pi/2$ and $|w_0| < |w_1|$.

\section{Additional heatmaps and moir\'e band structures}\label{sec:additional-band}
\begin{figure}
	\centering
	\includegraphics{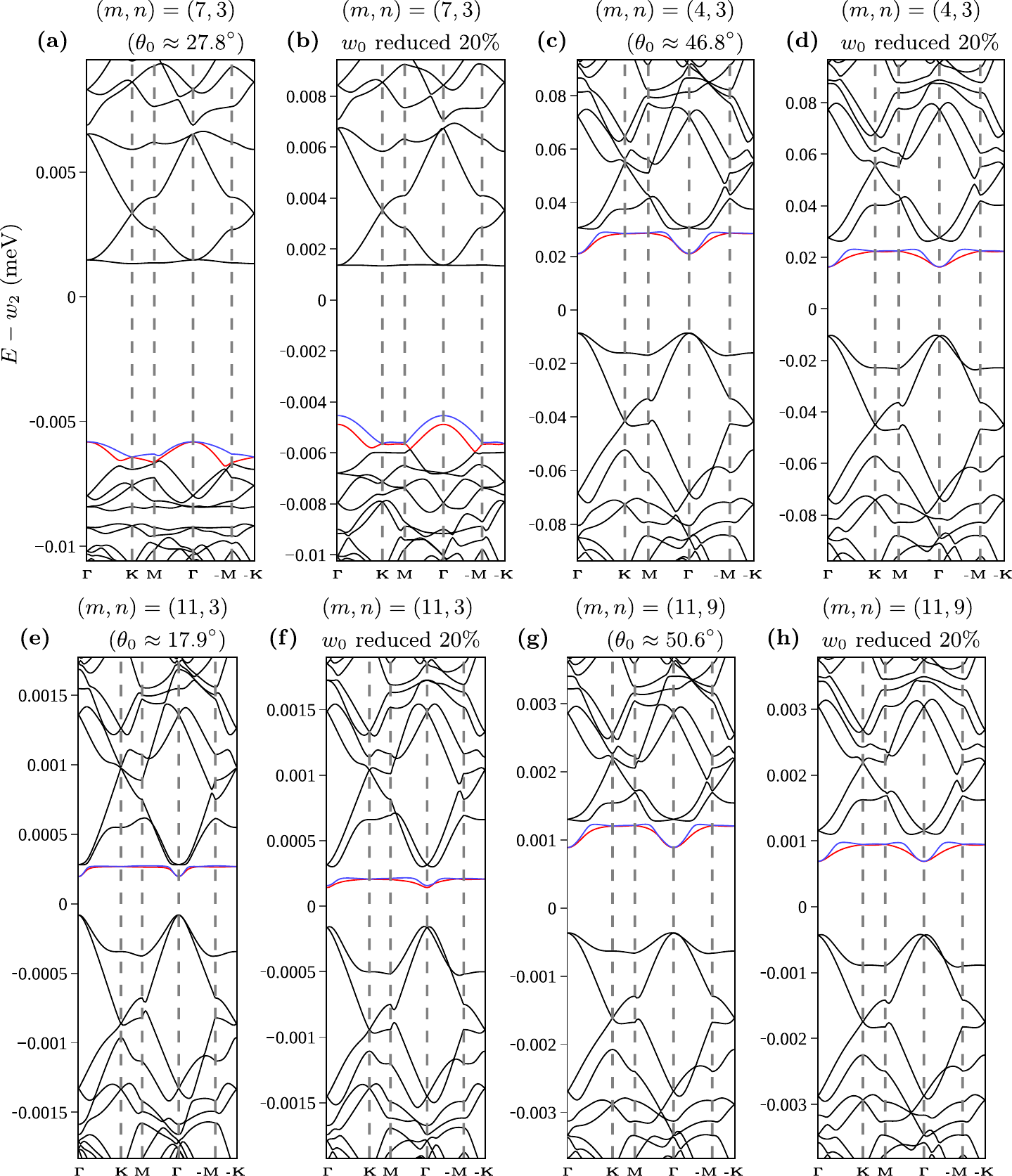}
	\caption{Moir\'e band structures at the first magic angle $\delta\theta = \theta - \theta_0 = \delta\theta_{\text{magic}}$ near the latter four $(m,n)$ commensurate configurations in \cref{tbl:accurate-parameters}. The band structures were computed with the Hamiltonian in  \cref{eq:continuum-hamiltonian-complete} and the quasi-momentum truncation illustrated in Appendix \cref{fig:moire-hopping-lattice}. The horizontal axes follow the moir\'e quasi-momentum trajectory $\bGamma_M \to \bK_M \to \bM_M \to \bGamma_M \to -\bM_M \to -\bK_M$. The two bands nearest charge neutrality are shown in blue and red while all other bands are shown in black. The parameters for panels \textbf{(a)}, \textbf{(c)}, \textbf{(e)}, and \textbf{(g)} are taken from \cref{tbl:accurate-parameters}. The parameters for panels \textbf{(b)}, \textbf{(d)}, \textbf{(f)}, and \textbf{(h)} are the same as those for \textbf{(a)}, \textbf{(c)}, \textbf{(e)}, and \textbf{(g)} except with the values of $w_0$ reduced by $20\%$.}
	\label{fig:additional-moire-band-structures-near-commensuration}
\end{figure}

\begin{figure}
	\centering
	\includegraphics{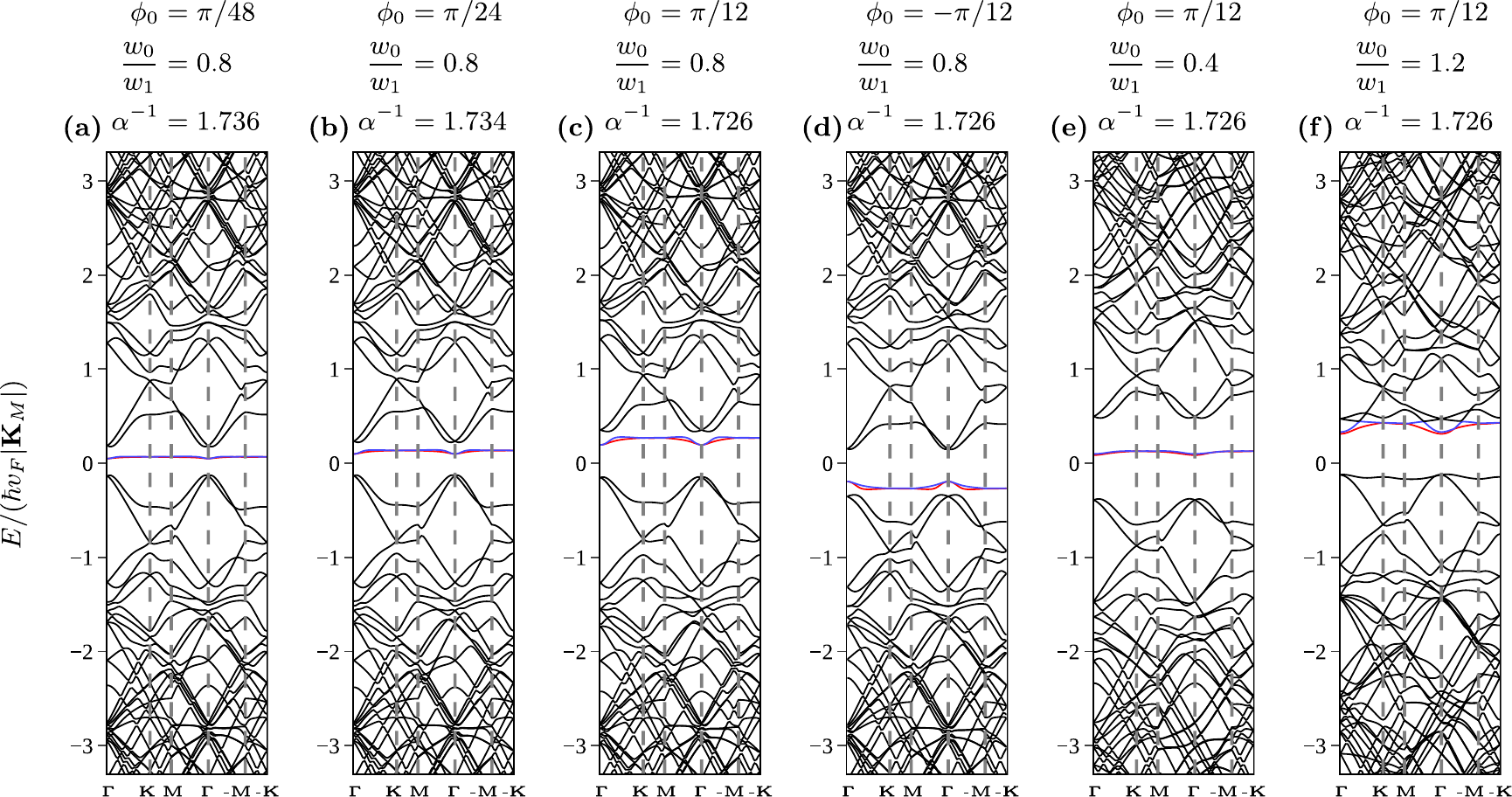}
	\caption{\textbf{(a)}-\textbf{(c)} Moir\'e band structures in the first magic manifold (see \cref{sec:first-magic-manifold}) for four different small values of $\phi_0$. \textbf{(d)}-\textbf{(f)} Variations on panel \textbf{(c)} in which the the value of $\phi_0$ is negated or the value of $w_0/w_1$ is changed.  Note that in panel \textbf{(f)} where $w_0/w_1$ is large, the lowest two bands at charge neutrality are no longer isolated from the higher bands. All panels use the model of \cref{eq:continuum-hamiltonian-rotate-basis} with $w_2=0$ and the quasi-momentum truncation illustrated in Appendix \cref{fig:moire-hopping-lattice}. The horizontal axes follow the moir\'e quasi-momentum trajectory $\bGamma_M \to \bK_M \to \bM_M \to \bGamma_M \to -\bM_M \to -\bK_M$. The two bands nearest charge neutrality are shown in blue and red while all other bands are shown in black.}
	\label{fig:small-phi0-moire-band-structures}
\end{figure}

\begin{figure}
	\centering
	\includegraphics{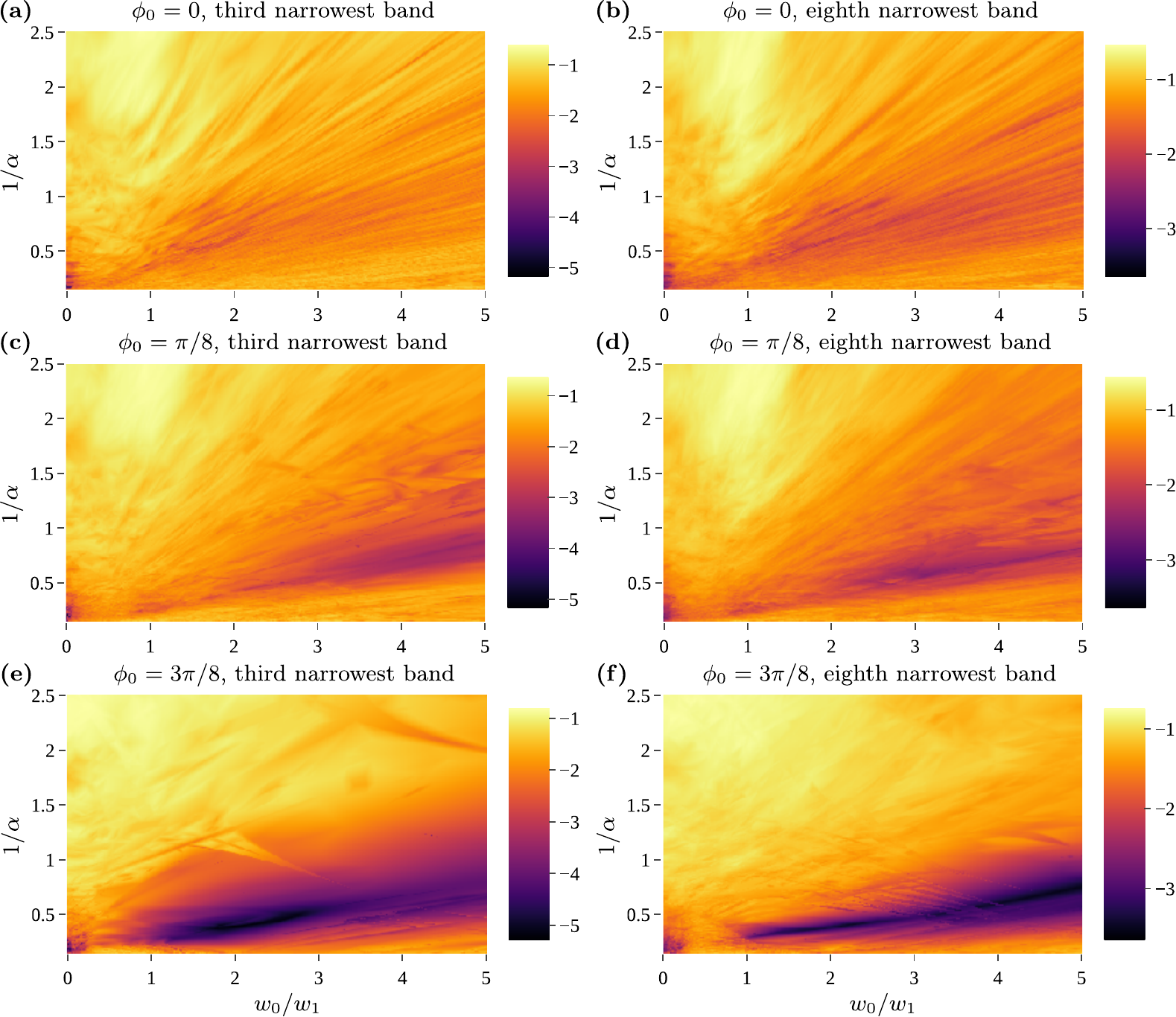}
	\caption{Heatmaps showing the base $10$ logarithm of the bandwidth (in units of $\hbar v_F |\bK_M|$) of the third and eighth narrowest bands among the first $20$ conduction bands and the first $20$ valence bands at charge neutrality for $\phi_0 = 0$, $\pi/8$, and $3\pi/8$. The bandwidth was computed with the points $\bGamma_M, \bK_M, \bM_M, \bK_M/2, \bM_M/2, -\bM_M/2$ in $\text{BZ}_M$. For this computation, we use the model in \cref{eq:continuum-hamiltonian-rotate-basis} with the quasi-momentum truncation illustrated in Appendix \cref{fig:moire-hopping-lattice}. The dark regions indicate parameters in the hypermagic regime discussed in \cref{sec:hypermagic-subsection}. See \cref{fig:heatmap-hypermagic} for similar heatmaps with $\phi_0 = \pi/4$ and $\pi/2$.}
	\label{fig:additional-heatmaps}
\end{figure}

\begin{figure}
	\centering
	\includegraphics{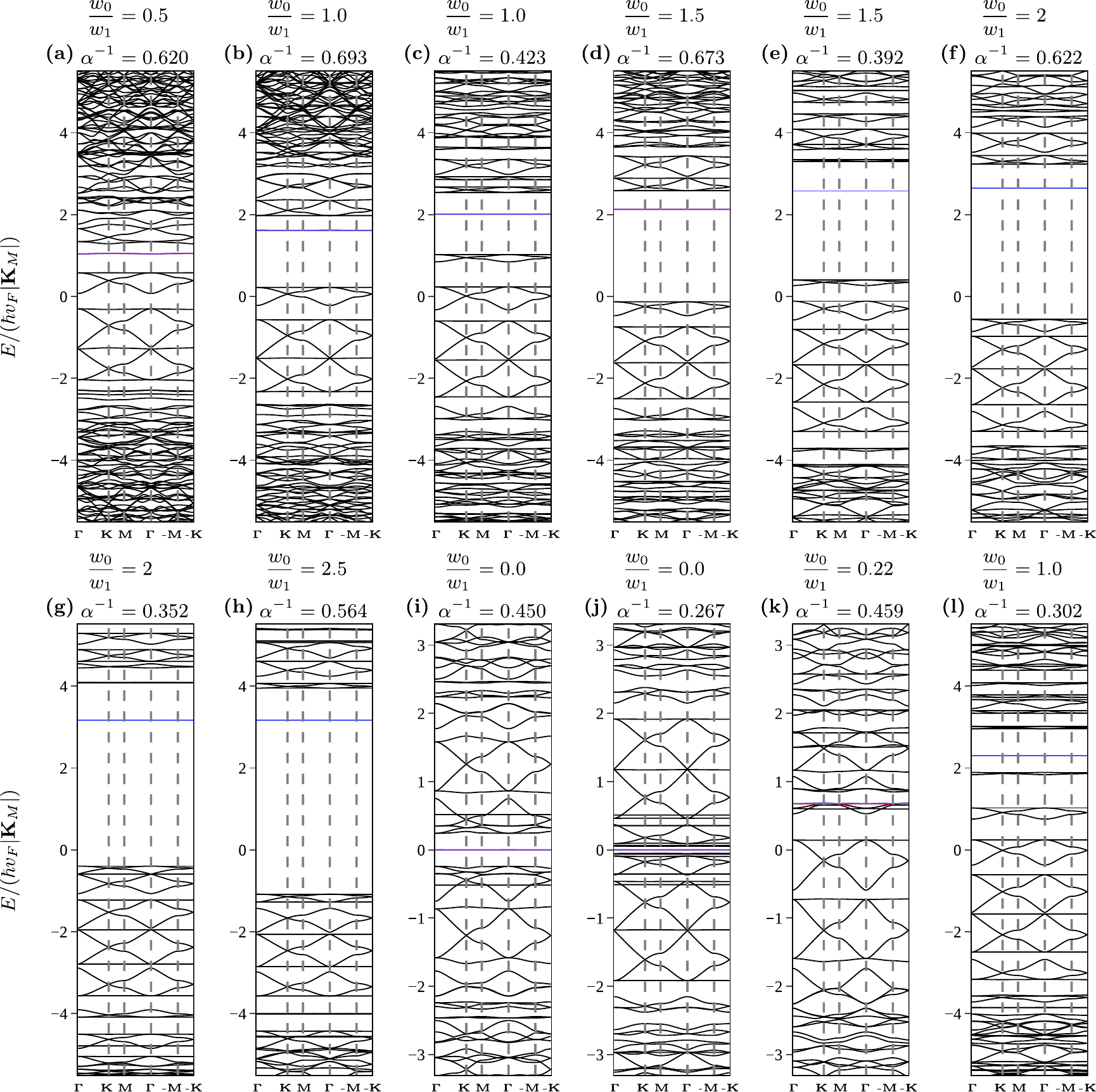}
	\caption{Moir\'e band structures using the model in \cref{eq:continuum-hamiltonian-rotate-basis} with $w_2 = 0$, $\phi_0 = \pi/2$, and various parameters $w_0/w_1$ and $\alpha^{-1}$ located in the three dark curves in \cref{fig:heatmap-middle-bands}\textbf{(c)}. Since each panel has many simultaneous flat bands, these parameters all belong to the hypermagic regime discussed in \cref{sec:hypermagic-subsection}. For this computation, we use the quasi-momentum truncation illustrated in \cref{fig:moire-hopping-lattice}. The horizontal axes follow the moir\'e quasi-momentum trajectory $\bGamma_M \to \bK_M \to \bM_M \to \bGamma_M \to -\bM_M \to -\bK_M$. The two bands nearest charge neutrality are shown in blue and red while all other bands are shown in black. Note that panels \textbf{(i)} and \textbf{(j)} are in the chiral limit $w_0=0$ so that $\phi_0$ does not affect their band structure. The $\alpha^{-1}$ values for panels \textbf{(i)} and \textbf{(j)} correspond to the second and third magic angles in the chiral limit, respectively \cite{Tarnopolsky2019}.}
	\label{fig:additional-moire-bandstructures}
\end{figure}

\begin{figure}
	\centering
	\includegraphics{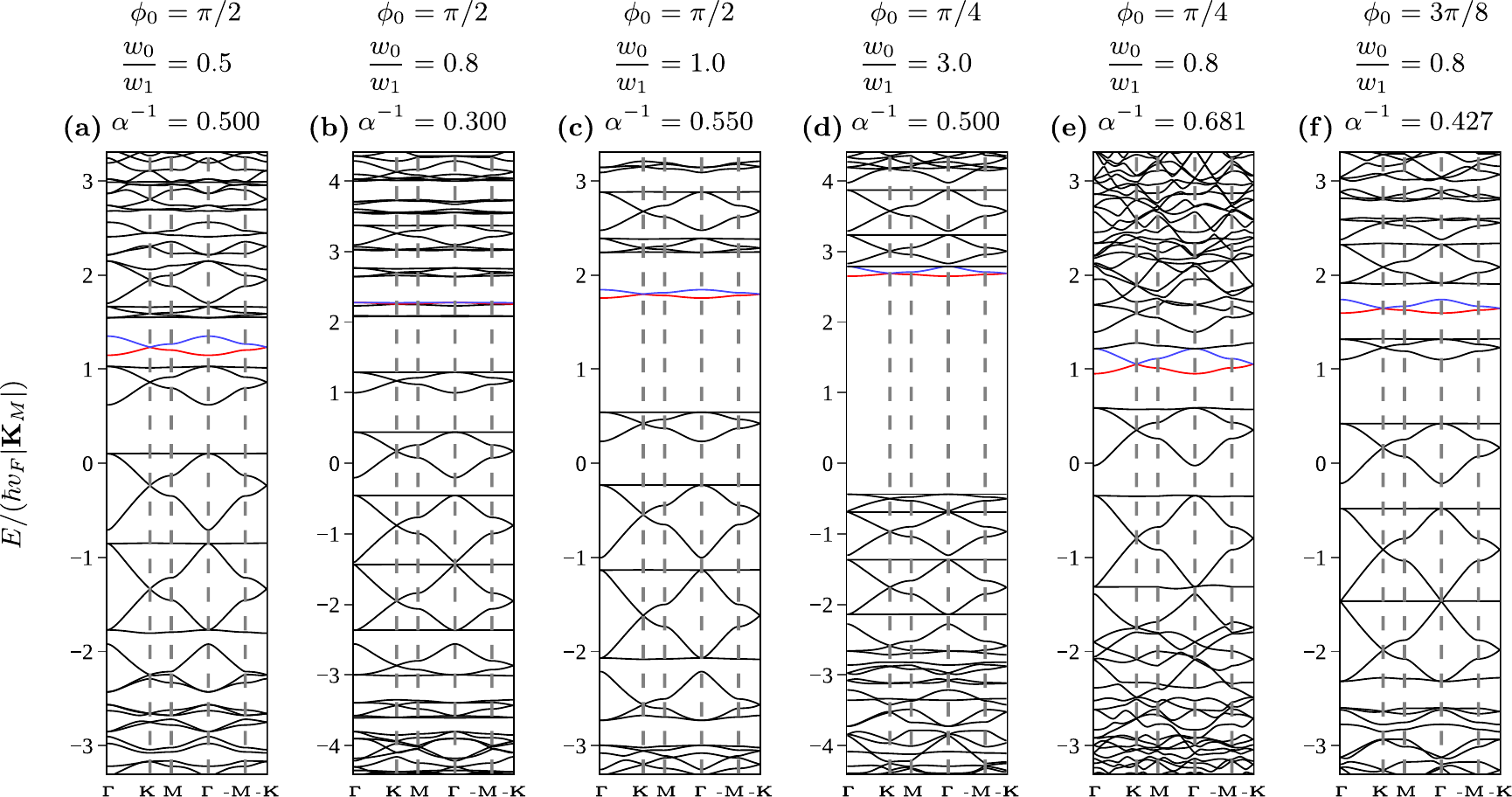}
	\caption{Moir\'e band structures using the model in \cref{eq:continuum-hamiltonian-rotate-basis} with $w_2 = 0$ and various values of $\phi_0$, $w_0/w_1$, and $\alpha^{-1}$ in the hypermagic regime discussed in \cref{sec:hypermagic-subsection} but not in the three dark curves in \cref{fig:heatmap-middle-bands}\textbf{(c)}. For this computation, we use the quasi-momentum truncation illustrated in \cref{fig:moire-hopping-lattice}. The horizontal axes follow the moir\'e quasi-momentum trajectory $\bGamma_M \to \bK_M \to \bM_M \to \bGamma_M \to -\bM_M \to -\bK_M$. The two bands nearest charge neutrality are shown in blue and red while all other bands are shown in black. The lowest bands in panels \textbf{(b)}, \textbf{(d)}, and \textbf{(e)} form kagome-like groups of three as discussed in \cref{sec:hypermagic-subsection}. The lowest bands in panel \textbf{(b)} are also shown in \cref{fig:gap-and-inversion-moire-band-structures}\textbf{(a)}. In contrast, the lowest two bands in panels \textbf{(a)}, \textbf{(c)}, and \textbf{(f)} are gapped from higher bands and resemble those of graphene. Panels \textbf{(a)}-\textbf{(c)} have $\phi_0 = \pi/2$ while panels \textbf{(d)}-\textbf{(f)} have smaller values of $\phi_0$. The $w_0/w_1$ and $\alpha^{-1}$ parameters for panels \textbf{(e)} and \textbf{(f)} are the same as those for \cref{fig:bandstructure-hypermagic}\textbf{(e)} and \textbf{(f)}.}
	\label{fig:additional-moire-bandstructures-generic}
\end{figure}

\cref{fig:additional-moire-band-structures-near-commensuration,fig:small-phi0-moire-band-structures,fig:additional-heatmaps,fig:additional-moire-bandstructures,fig:additional-moire-bandstructures-generic} show some additional heatmaps and moir\'e band structures (see captions for details).

\end{document}